\documentclass[a4paper,english,envcountsame]{llncs}
\usepackage{pslatex}
\usepackage{amsfonts}
\usepackage{amsmath}
\usepackage{amssymb}
\usepackage{paralist}
\usepackage{stmaryrd}
\usepackage{stackrel}
\usepackage{proof}
\usepackage{xspace}

\usepackage[colorlinks]{hyperref}

\newif\ifLongVersion\LongVersiontrue

 \usepackage{stmaryrd}
 \usepackage{xspace}
 \usepackage{amsmath}
 \usepackage{txfonts}
 \usepackage{mfirstuc}
 \usepackage{graphicx}
 \usepackage{paralist}
 \usepackage{ifthen}
 \usepackage{scalerel}

\renewcommand{\vec}[1]{\boldsymbol{#1}}   

\usepackage[final]{commenting}
\declareauthor{ri}{Radu}{blue}
\declareauthor{np}{Nicolas}{red}
\declareauthor{me}{Mnacho}{brown!70!black}

\newtheorem{assumption}{Assumption}
\newtheorem{fact}{Fact}

\newcommand{\bigO}{\mathcal{O}} 
\newcommand{\np}{$\mathsf{NP}$}

\newcommand{\twoexptime}{$2$-$\mathsf{EXPTIME}$}

\newcommand{\set}[1]{\left\{ #1 \right\}}

\newcommand{\tuple}[1]{\left\langle #1 \right\rangle}
\renewcommand{\vec}[1]{\mathbf #1}

\newcommand{\setof}[2]{\left\{#1\,\middle|\:#2\right\}}

\newcommand{\len}[1]{{|{#1}|}}
\newcommand{\card}[1]{{||{#1}||}}

\newcommand{\nat}{{\bf \mathbb{N}}}

\renewcommand{\paragraph}[1]{\noindent{\bf #1}}

\let\Asterisk\undefined
\newcommand{\Asterisk}{\mathop{\scalebox{1.7}{\raisebox{-0.2ex}{$\ast$}}}}%

\newif\ifLongVersion\LongVersiontrue
\renewcommand{\proof}[1]{\ifLongVersion \noindent\emph{Proof}: {#1} \else\fi}

\newcommand{\teq}{\approx}

\newcommand{\locs}{\mathbb{L}}
\newcommand{\nil}{\mathsf{nil}}
\newcommand{\emp}{\mathsf{emp}}
\newcommand{\wand}{
 \mathrel{\mbox{$\hspace*{-0.03em}\mathord{-}\hspace*{-0.66em}
 \mathord{-}\hspace*{-0.36em}\mathord{*}$\hspace*{-0.005em}}}} 
\newcommand{\seplog}{\mathsf{SL}}

\newcommand{\seplogk}[1]{\seplog^{\!\scriptstyle{#1}}}
\newcommand{\shk}[1]{\mathsf{SH}^{\!\scriptstyle{#1}}}

\newcommand{\fv}[1]{\mathsf{fv}({#1})}
\newcommand{\cst}[1]{\mathsf{cst}({#1})}
\newcommand{\term}[1]{\mathsf{trm}({#1})}

\newcommand{\dom}{\mathrm{dom}}

\newcommand{\interv}[2]{\llbracket#1\mathrel{{.}\,{.}}\nobreak#2\rrbracket}

\newcommand{\img}{\mathrm{img}}

\newcommand{\size}[1]{\mathrm{size}(#1)}
\newcommand{\probwidth}[1]{\mathrm{width}({#1})}
\newcommand{\vars}{\mathbb{V}}
\newcommand{\preds}{\mathbb{P}}
\newcommand{\const}{\mathbb{C}}
\newcommand{\constset}{\mathcal{C}}
\newcommand{\constpart}{\mathfrak{C}}

\renewcommand{\int}{\mathsf{\scriptscriptstyle{i}}}

\newcommand{\astore}{\mathfrak{s}}
\newcommand{\aheap}{\mathfrak{h}}
\newcommand{\aaheap}{\overline{\aheap}}

\renewcommand{\iff}{\Leftrightarrow}

\newcommand{\alloc}[2]{\mathsf{alloc}_{#2}({#1})}

\newcommand{\allocpar}[2]{\mathsf{palloc}_{#2}({#1})}
\newcommand{\alloconst}[2]{\mathsf{calloc}_{#2}({#1})}

\newcommand{\isdef}{\stackrel{\scalebox{0.5}{\text{$\mathsf{def}$}}}{=}}
\renewcommand{\Asterisk}{\scalebox{2}{\text{$*$}}}

\newcommand{\map}{\rightharpoonup}
\newcommand{\finmap}{\rightharpoonup_{\mathit{fin}}}
\newcommand{\terms}{\mathbb{T}}
\newcommand{\rank}{\mathfrak{K}}
\newcommand{\seq}{\bumpeq}
\newcommand{\sneq}{\not\seq}

\newcommand{\loc}{\mathrm{loc}}
\newcommand{\front}{\mathrm{Fr}}

\newcommand{\asys}{\mathcal{S}}
\newcommand{\arule}{\rho}
\newcommand{\aprob}{\mathcal{P}}
\newcommand{\seqset}{\Sigma}
\newcommand{\unfold}[1]{\Rightarrow_{#1}}

\newcommand{\tw}{\mathrm{tw}}

\newcommand{\mso}{$\mathsf{MSO}$}

\newcommand{\aastore}{\overline{\astore}}

\newcommand{\amorph}{\gamma}

\newcommand{\inheap}{\mathsf{loc}}
\newcommand{\nofvex}{\dot\exists}
\newcommand{\heapex}{
        \exists_{\mathsf{h}}
}
\newcommand{\nheapex}{
        \exists_{\neg\mathsf{h}}
}
\newcommand{\nheapall}{
        \forall_{\neg\mathsf{h}}
}
\newcommand{\context}[1]{\Gamma_{[#1]}}
\newcommand{\swand}{
\mathrel{\mbox{$\hspace*{-0.03em}\mathord{-}\hspace*{-0.4em}
 \mathord{-}\hspace*{-0.36em}\scalebox{0.9}{$\mathord{\bullet}$}$\hspace*{-0.005em}}}
}
\newcommand{\aistore}{\dot{\astore}}
\newcommand{\imodels}[1]{~\models^{\scalebox{.7}{$\bullet$}}_{#1}~}

\newcommand{\sunfold}[1]{\rightsquigarrow_{#1}}
\newcommand{\aaistore}{\dot{\overline{\astore}}}
\newcommand{\aroot}{\mathsf{root}}
\newcommand{\lroots}[1]{\mathrm{roots}_{\mathsf{lhs}}({#1})}
\newcommand{\rroots}[1]{\mathrm{roots}_{\mathsf{rhs}}({#1})}
\newcommand{\roots}[1]{\mathrm{roots}({#1})}
\newcommand{\core}[1]{\mathsf{Core}(#1)}
\newcommand{\istoreq}[1]{\approx_{#1}}
\newcommand{\aistorep}{\dot{\astore}'}
\newcommand{\aaistorep}{\dot{\overline{\astore}'}}
\newcommand{\allvars}[1]{\mathcal{V}_{#1}}
\newcommand{\coreset}[1]{\mathfrak{C}_{#1}}
\newcommand{\coreabs}[2]{\mathcal{C}_{#2}({#1})}
\newcommand{\coretrans}[1]{\mathcal{T}\left({#1}\right)}
\newcommand{\coresep}[1]{\circledast_{#1}}
\newcommand{\witset}[4]{\mathcal{W}_{#1}({#2},{#3},{#4})}
\newcommand{\profile}{\mathcal{F}}

\newcommand{\rem}[2]{\mathsf{rem}({#1},{#2})}
\newcommand{\add}[2]{\mathsf{add}({#1},{#2})}
\newcommand{\allocates}[2]{\mathcal{A}({#1},{#2})}
\newcommand{\conseq}{\Vdash}
\newcommand{\deriv}[1]{\Vvdash_{#1}}

\newcommand{\arestricted}{an e-restricted\xspace}
\newcommand{\restricted}{e-restricted\xspace}
\newcommand{\restrictedness}{e-restrictedness\xspace}
\newcommand{\Restricted}{E-restricted\xspace}
\newcommand{\Restrictedness}{E-restrictedness\xspace}

\newcommand{\shortVersionOnly}[1]{}
\newcommand{\longVersionOnly}[1]{#1}

\newcommand{\newAppendixSection}[1]{}

\newcommand{\putInAppendix}[1]{#1}

\newcommand{\additionalMaterial}[4]{#4}

\newcommand{\optionalProof}[3]{\proof{#3}}

\begin{document}


\title{Decidable Entailments in Separation
  Logic with Inductive Definitions: Beyond  Established Systems}

\author{Mnacho Echenim\inst{1}, Radu Iosif\inst{2} and Nicolas Peltier\inst{1}}


\institute{Univ. Grenoble Alpes, CNRS, LIG, F-38000 Grenoble France
\and
Univ. Grenoble Alpes, CNRS, VERIMAG, F-38000 Grenoble France}

\maketitle

\mycomment[np]{new title}

\begin{abstract}
We define a class of Separation Logic
\cite{IshtiaqOHearn01,Reynolds02} formul{\ae}, whose entailment
problem \emph{given formul{\ae} $\phi, \psi_1, \ldots, \psi_n$, is
  every model of $\phi$ a model of some $\psi_i$?} is
\twoexptime-complete. The formul{\ae} in this class are existentially
quantified separating conjunctions involving predicate atoms,
interpreted by the least sets of store-heap structures that satisfy a
set of inductive rules, which is also part of the input to the
entailment problem. Previous work
\cite{IosifRogalewiczSimacek13,KatelaanMathejaZuleger19,PZ20} consider
\emph{established} sets of rules, meaning that every existentially
quantified variable in a rule must eventually be bound to an
\emph{allocated} location, i.e.\ from the domain of the heap. In
particular, this guarantees that each structure has treewidth bounded
by the size of the largest rule in the set. In contrast, here we show
that establishment, although sufficient for decidability (alongside
two other natural conditions), is not necessary, by providing a
condition, called \emph{equational restrictedness}, which applies
syntactically to (dis-)equalities. The entailment problem is more
general in this case, because equationally restricted rules define
richer classes of structures, of unbounded treewidth. In this paper we
show that\ \begin{inparaenum}[(1)]
\item every established set of rules can be converted into an
  equationally restricted one and
\item the entailment problem is \twoexptime-complete in the latter
  case, thus matching the complexity of entailments for established
  sets of rules \cite{KatelaanMathejaZuleger19,PZ20}.
\end{inparaenum}
\end{abstract}

\section{Introduction}
  
Separation Logic ($\seplog$) \cite{IshtiaqOHearn01,Reynolds02} is
widely used to reason about programs manipulating recursively linked
data structures, being at the core of several industrial-scale static
program analysis techniques
\cite{DBLP:conf/nfm/CalcagnoDDGHLOP15,DBLP:conf/cav/BerdineCI11,DBLP:conf/cav/DudkaPV11}.
Given an integer $\rank \geq 1$, denoting the number of
fields in a record datatype, and an 
infinite set $\locs$ of
memory locations (addresses), the assertions in this logic describe
{\em heaps}, that are finite partial functions mapping locations to
records, i.e., $\rank$-tuples of locations. A location $\ell$
in the domain of the heap is said to be {\em allocated} and the
\emph{points-to} atom $x \mapsto (y_1,\dots,y_\rank)$ states that the
location associated with $x$ refers to the tuple of locations
associated with $(y_1,\dots,y_\rank)$.  The {\em separating
  conjunction} $\phi * \psi$ states that the formul{\ae} $\phi$ and
$\psi$ hold in non-overlapping parts of the heap, that have disjoint
domains. This connective allows for modular program analyses, because
the formul{\ae} specifying the behaviour of a program statement refer
only to the small (local) set of locations that are manipulated by
that statement, with no concern for the rest of the program's
state.

Formul{\ae} consisting of points-to atoms connected with
separating conjunctions describe heaps of bounded size only. To reason
about recursive data structures of unbounded sizes (lists, trees,
etc.), the base logic is enriched by predicate symbols, with a
semantics specified by user-defined inductive rules. For instance, the
rules: $\mathsf{excls}(x,y) \Leftarrow \exists z~.~ x\mapsto (z,y) * z
\not \seq \mathsf{c}$ and $\mathsf{excls}(x,y) \Leftarrow \exists z
\exists v~.~ x\mapsto (z,v) * \mathsf{excls}(v,y) * z \not \seq
\mathsf{c}$ describe a non-empty list segment, whose elements are
records with two fields: the first is a data field, that keeps a list
of locations, which excludes the location assigned to the global
constant $\mathsf{c}$, and the second is used to link the records in a
list whose head and tail are pointed to by $x$ and $y$, respectively.

An important problem in program verification, arising during
construction of Hoare-style correctness proofs, is the discharge of
verification conditions, that are entailments of the form $\phi \vdash
\psi_1, \ldots, \psi_n$, where $\phi$ and $\psi_1, \ldots, \psi_n$ are
separating conjunctions of points-to, predicates and (dis-)equalities,
also known as {\em symbolic heaps}.  The \emph{entailment problem}
then asks if \emph{every model of $\phi$ is a model of some $\psi_i$?}
In general, the entailment problem is undecidable and becomes
decidable when the inductive rules used to interpret the predicates
satisfy three restrictions
\cite{IosifRogalewiczSimacek13}: \begin{inparaenum}[(1)]
\item {\em progress}, stating that each rule allocates {\em
  exactly} one memory cell,
\item {\em connectivity}, ensuring that the allocated memory cells
  form a tree-shaped structure, and
\item {\em establishment}, stating that all existentially quantified
  variables introduced by an inductive rule must be assigned to some
  allocated memory cell, in every structure defined by that rule.
\end{inparaenum}
For instance, the above rules are progressing and connected but not
established, because the $\exists z$ variables are not explicitly
assigned an allocated location, unlike the $\exists v$ variables,
passed as first parameter of the $\mathsf{excls}(x,y)$ predicate, and
thus always allocated by the points-to atoms $x \mapsto (z,y)$ or $x
\mapsto (z,v)$, from the first and second rule defining
$\mathsf{excls}(x,y)$, respectively.


The argument behind the decidability of a progressing, connected and
established entailment problem is that every model of the left-hand
side is encoded by a graph whose treewidth\footnote{The treewidth of a
  graph is a parameter measuring how close the graph is to a tree, see
  \cite[Ch. 11]{flum-grohe-book} for a definition.} is bounded by the
size of the largest symbolic heap that occurs in the problem
\cite{IosifRogalewiczSimacek13}. Moreover, the progress and
connectivity conditions ensure that the set of models of a symbolic
heap can be represented by a Monadic Second Order (\mso) logic formula
interpreted over graphs, that can be effectively built from the
symbolic heap and the set of rules of the problem. The decidability of
entailments follows then from the decidability of the satisfiability
problem for \mso\ over graphs of bounded treewidth (Courcelle's
Theorem) \cite{Courcelle90}. Initially, no upper bound better than
elementary recursive was known to exist. Recently, a
\twoexptime\ algorithm was proposed
\cite{KatelaanMathejaZuleger19,PMZ20} for sets of rules satisfying
these three conditions, and, moreover, this bound was shown to be
tight \cite{DBLP:conf/lpar/EchenimIP20}.

Several natural questions arise: are the progress, connectivity and
establishment conditions really necessary for the decidability of
entailments? How much can these restriction be relaxed, without
jeopardizing the complexity of the problem? Can one decide entailments
that involve sets of heaps of unbounded treewidth? In this paper, we
answer these questions by showing that entailments are still
\twoexptime-complete when the establishment condition is replaced by a
condition on the (dis-)equations occurring in the symbolic heaps of
the problem. Informally, such (dis-)equations must be of the form $x
\seq \mathsf{c}$ ($x \sneq \mathsf{c}$), where $\mathsf{c}$ ranges
over some finite and fixed set of globally visible constants
(including special symbols such as $\nil$, that denotes a
non-allocated address, but also any free variable occurring on the
left-hand side of the entailment). We also relax slightly the progress
and connectivity conditions, by allowing forest-like heap structures
(instead of just trees), provided that every root is mapped to a
constant symbol. These entailment problems are called
\emph{equationally restricted} (\emph{\restricted}, for short). For
instance, the entailment problem $\mathsf{excls}(x,y) *
\mathsf{excls}(y,z) \vdash \mathsf{excls}(x,z)$, with the above rules,
falls in this category.


We prove that the \restricted condition loses no generality compared
to establishment, because any established entailment problem can be
transformed into an equivalent \restricted entailment problem.
\Restricted problems allow reasoning about structures that contain
dangling pointers, which frequently occur in practice, especially in
the context of modular program analysis. Moreover, the set of
structures considered in \arestricted entailment problem may contain
infinite sequences of heaps of strictly increasing treewidths, that
are out of the scope of established problems
\cite{IosifRogalewiczSimacek13}.
 
The decision procedure for \restricted problems proposed in this paper
is based on a similar idea as the one given, for established problems,
in \cite{PMZ20,PZ20}. We build a suitable abstraction of the set of
structures satisfying the left-hand side of the entailment bottom-up,
starting from points-to and predicate atoms, using abstract operators
to compose disjoint structures, to add and remove variables, and to
unfold the inductive rules associated with the predicates. The
abstraction is precise enough to allow checking that all the models of
the left-hand side fulfill the right-hand side of the entailment and
also general enough to ensure termination of the entailment checking
algorithm.

Although both procedures are similar, there are essential differences
between our work and \cite{PMZ20,PZ20}.  First, we show that instead
of using a specific language for describing those abstractions, the
considered set of structures can themselves be defined in $\seplog$,
by means of formul{\ae} of some specific pattern called {\em core
  formul{\ae}}. Second, the fact that the systems are not established
makes the definition of the procedure much more difficult, due to the
fact that the considered structures can have an unbounded
treewidth. This is problematic because, informally, this boundedness
property is essential to ensure that the abstractions can be described
using a finite set of variables, denoting the {\em frontier} of the
considered structures, namely the locations that can be shared with
other structures. In particular, the fact that disjoint heaps may
share unallocated (or ``unnamed'') locations complexifies the
definition of the composition operator. This problem is overcome by
considering a specific class of structures, called {\em normal
  structures}, of bounded treewidth, and proving that the validity of
an entailment can be decided by considering only normal structures.

In terms of complexity, we show that the running time of our algorithm
is doubly exponential w.r.t.\ the maximal size among the symbolic
heaps occurring in the input entailment problem (including those in
the rules) and simply exponential w.r.t.\ the number of such symbolic
heaps (hence w.r.t.\ the number of rules). This means that the
\twoexptime\ upper bound is preserved by any reduction increasing
exponentially the number of rules, but increasing only polynomially
the size of the rules. On the other hand, the \twoexptime-hard lower
bound is proved by a reduction from the membership problem for
exponential-space bounded Alternating Turing Machines
\cite{DBLP:conf/lpar/EchenimIP20}.  \shortVersionOnly{ Due to space
  restrictions, most proofs are shifted to an appendix.  }

\longVersionOnly{
\mycomment[np]{minor changes in the structure (maybe omitted from conference version)}

The remainder of the paper is structured as follows.  In Section
\ref{sec:prel}, all the necessary notions concerning Separation logic
are recalled, and in Section \ref{sec:restricted}, we define the
fragment of entailment problems that we are considering.  In
particular, we formally define the notions of connected, established
and \restricted problems.  In Section \ref{sec:normalize} we introduce
a preprocessing step, which transforms any problem into an equivalent
{\em normalized} one, satisfying many properties that will be
essential in the following.  In Section \ref{sec:established} we show
that \restricted problems are, in a sense to be specified formally,
strictly more general than established ones.  In Section
\ref{sec:injective-normal} we show that the considered entailments can
be tested by focusing on a specific class of structures, called {\em
  normal structures}.  In Section \ref{sec:core} we define {\em core
  formul{\ae}}, which are SL formul{\ae} of specific patterns used to
describe suitable abstractions of structures, and we define an
algorithm to test entailment based on these abstractions. Each
structure is represented by its {\em profile}, defined as the set of
core formul{\ae} it satisfies, along with some additional conditions.
In Section \ref{sec:coreabs}, we show how such profiles can be
effectively constructed and in Section \ref{sec:complexity} the
complexity of the procedure is analyzed and the main result of the
paper is stated.  }

\section{Separation Logic with Inductive Definitions}
 \label{sec:prel}
 
Let $\nat$ denote the set of natural numbers. For a countable set $S$,
we denote by $\card{S} \in \nat \cup \set{\infty}$ its
cardinality. For a partial mapping $f : A \rightharpoonup B$, let
$\dom(f) \isdef \set{x \in A \mid f(x) \in B}$ and $\img(f) \isdef
\set{f(x) \mid x \in \dom(f)}$ be its domain and range, respectively.
We say that $f$ is \emph{total} if $\dom(f) = A$, written $f : A
\rightarrow B$ and \emph{finite}, written $f : A \finmap B$ if
$\card{\dom(f)} < \infty$. Given integers $n$ and $m$, we denote by
$\interv{n}{m}$ the set $\set{n,n+1, \ldots, m}$, so that
$\interv{n}{m} = \emptyset$ if $n > m$. For a relation $\lhd \subseteq
A \times A$, we denote by $\lhd^*$ its reflexive and transitive
closure.

For an integer $n\geq0$, let $A^n$ be the set of $n$-tuples with
elements from $A$. Given a tuple $\vec{a} = (a_1, \ldots, a_n)$ and $i
\in \interv{1}{n}$, we denote by $\vec{a}_i$ the $i$-th element of
$\vec{a}$ and by $\len{\vec{a}} \isdef n$ its length. By $f(\vec{a})$
we denote the tuple obtained by the pointwise application of $f$ to
the elements of $\vec{a}$. \shortVersionOnly{If multiplicity and order
  of the elements are not important, we blur the distinction between
  tuples and sets, using the set-theoretic notations $x \in \vec{a}$,
  $\vec{a} \cup \vec{b}$, $\vec{a} \cap \vec{b}$ and $\vec{a}
  \setminus \vec{b}$.}  \longVersionOnly{By a slight abuse of
  notation, we write $a \in \vec{a}$ if $a=\vec{a}_i$, for some $i \in
  \interv{1}{n}$. Given tuples $\vec{a}$ and $\vec{b}$, we slightly
  abuse notations by defining the sets $\vec{a} \cup \vec{b} \isdef
  \set{x \mid x \in \vec{a} \text{ or } x \in \vec{b}}$, $\vec{a} \cap
  \vec{b} \isdef \set{x \mid x \in \vec{a} \text{ and } x \in
    \vec{b}}$ and $\vec{a} \setminus \vec{b} \isdef \set{x \mid x \in
    \vec{a} \text{ and } x \not\in \vec{b}}$.}

Let $\vars = \set{x,y,\ldots}$ be an infinite countable set of logical
first-order variables and $\preds = \set{p,q,\ldots}$ be an infinite
countable set (disjoint from $\vars$) of relation symbols, called
\emph{predicates}, where each predicate $p$ has arity $\#p \geq 0$. We
also consider a finite set $\const$ of \emph{constants}, of known
bounded cardinality, disjoint from both $\vars$ and
$\preds$. Constants will play a special r\^ole in the upcoming
developments and the fact that $\const$ is bounded is of a particular
importance. A \emph{term} is either a variable or a constant and we
denote by $\terms \isdef \vars \cup \const$ the set of terms.

Throughout this paper we consider an integer $\rank\geq1$ that,
intuitively, denotes the number of fields in a record
datatype. Although we do not assume $\rank$ to be a constant in any of
the algorithms presented in the following, considering that every
datatype has exactly $\rank$ records simplifies the definition. The
logic $\seplogk{\rank}$ is the set of formul{\ae} generated
inductively by the syntax:
\vspace*{-.5\baselineskip}
\[\begin{array}{rcl}
\phi & := & \emp \mid t_0 \mapsto (t_1, \ldots, t_\rank) \mid p(t_1,
\ldots, t_{\#p}) \mid t_1 \teq t_2 \mid \phi_1 * \phi_2 \mid \phi_1
\wedge \phi_2 \mid \neg\phi_1 \mid \exists x ~.~ \phi_1 \\[-4mm]
\end{array}\]
where $p \in \preds$, $t_i \in \terms$ and $x \in \vars$. Atomic
propositions of the form $t_0 \mapsto (t_1, \ldots, t_\rank)$ are
called \emph{points-to atoms} and those of the form $p(t_1, \ldots,
t_{\#p})$ are \emph{predicate atoms}. If $\rank=1$, we write $t_0
\mapsto t_1$ for $t_0 \mapsto (t_1)$. 

The connective $*$ is called \emph{separating conjunction}, in
contrast with the classical conjunction $\wedge$. The \emph{size} of a
formula $\phi$, denoted by $\size{\phi}$, is the number of occurrences
of symbols in it.  We write $\fv{\phi}$ for the set of \emph{free}
variables in $\phi$ and $\term{\phi} \isdef \fv{\phi} \cup \const$. A
formula is \emph{predicate-free} if it has no predicate atoms. As
usual, $\phi_1 \vee \phi_2 \isdef \neg(\neg\phi_1 \wedge \neg\phi_2)$
and $\forall x ~.~ \phi \isdef \neg\exists x ~.~ \neg\phi$. For
a set of variables $\vec{x} = \set{x_1,\ldots, x_n}$ and a quantifier $Q \in
\set{\exists, \forall}$, we write $Q \vec{x} ~.~ \phi \isdef Q x_1
\ldots Q x_n ~.~ \phi$. By writing $t_1 = t_2$ ($\phi_1 = \phi_2$) we
mean that the terms (formul{\ae}) $t_1$ and $t_2$ ($\phi_1$ and
$\phi_2$) are syntactically the same.

A \emph{substitution} is a partial mapping $\sigma : \vars \map
\terms$ that maps variables to terms. We denote by $[t_1/x_1, \ldots,
  t_n/x_n]$ the substitution that maps the variable $x_i$ to $t_i$,
for each $i \in \interv{1}{n}$ and is undefined elsewhere. By
$\phi\sigma$ we denote the formula obtained from $\phi$ by
substituting each variable $x \in \fv{\phi}$ by $\sigma(x)$ (we assume
that bound variables are renamed to avoid collisions if needed). By
abuse of notation, we sometimes write $\sigma(x)$ for $x$, when $x
\not\in \dom(\sigma)$.  

To interpret $\seplogk{\rank}$ formul{\ae}, we consider an infinite
countable set $\locs$ of \emph{locations}. The semantics of
$\seplogk{\rank}$ formul{\ae} is defined in terms of \emph{structures}
$(\astore, \aheap)$, where:\begin{compactitem}
\item $\astore : \terms \map \locs$ is a partial mapping of terms into
  locations, called a \emph{store}, that interprets at least all the
  constants, i.e.\ $\const \subseteq \dom(\astore)$ for every store
  $\astore$, and
\item $\aheap : \locs \finmap \locs^\rank$ is a finite partial mapping
  of locations into $\rank$-tuples of locations, called a \emph{heap}.
\end{compactitem}
Given a heap $\aheap$, let $\loc(\aheap) \isdef \set{\ell_0, \ldots,
  \ell_\rank \mid \ell_0 \in \dom(\aheap),~ \aheap(\ell_0) = (\ell_1,
  \ldots, \ell_\rank)}$ be the set of locations that occur in the heap
$\aheap$. Two heaps $\aheap_1$ and $\aheap_2$ are \emph{disjoint} iff
$\dom(\aheap_1) \cap \dom(\aheap_2) = \emptyset$, in which case their
\emph{disjoint union} is denoted by $\aheap_1 \uplus \aheap_2$,
otherwise undefined.  The \emph{frontier between $\aheap_1$ and
  $\aheap_2$} is the set of common locations
$\front(\aheap_1,\aheap_2) \isdef \loc(\aheap_1) \cap
\loc(\aheap_2)$. Note that disjoint heaps may have nonempty
frontier. The \emph{satisfaction relation} $\models$ between
structures $(\astore,\aheap)$ and predicate-free $\seplogk{\rank}$
formul{\ae} $\phi$ is defined recursively on the structure of
formul{\ae}:
\vspace*{-.5\baselineskip}
\[\begin{array}{rclcl}
(\astore, \aheap) & \models & t_1 \teq t_2 & \iff &
t_1, t_2 \in \dom(\astore) \text{ and } \astore(t_1) = \astore(t_2) \\
(\astore, \aheap) & \models & \emp & \iff & \aheap = \emptyset \\
(\astore, \aheap) & \models & t_0 \mapsto (t_1, \ldots, t_\rank) & \iff &
t_0, \ldots, t_\rank \in \dom(\astore),\ \dom(\aheap) = \set{\astore(t_0)} \text{ and }
\aheap(\astore(t_0)) = (\astore(t_1), \ldots, \astore(t_\rank)) \\
(\astore, \aheap) & \models & \phi_1 \wedge \phi_2 & \iff &
(\astore, \aheap) \models \phi_i,~ i = 1,2 \\
(\astore, \aheap) & \models & \neg\phi_1 & \iff & \fv{\phi_1} \subseteq \dom(\astore) 
\text{ and } (\astore, \aheap) \not\models \phi_1 \\
(\astore, \aheap) & \models & \phi_1 * \phi_2 & \iff & \text{there exist heaps
  $\aheap_1$, $\aheap_2$ such that $\aheap = \aheap_1 \uplus \aheap_2$} 
\text{ and $(\astore, \aheap_i) \models \phi_i$, $i=1,2$} \\
(\astore, \aheap) & \models & \exists x ~.~ \phi & \iff &
(\astore[x \leftarrow \ell], \aheap) \models \phi, \text{ for some location $\ell \in \locs$} \\[-2mm]
\end{array}\]
where $\astore[x \leftarrow \ell]$ is the store, with domain
$\dom(\astore) \cup \set{x}$, that maps $x$ to $\ell$ and behaves like
$\astore$ over $\dom(\astore) \setminus \set{x}$. For a tuple of
variables $\vec{x} = (x_1,\ldots,x_n)$ and locations $\overline{\ell}
= (\ell_1,\ldots,\ell_n)$, we call the store $\astore[\vec{x}
  \leftarrow \overline{\ell}] \isdef \astore[x_1 \leftarrow \ell_1]
\ldots [x_n \leftarrow \ell_n]$ an \emph{$\vec{x}$-associate} of
$\astore$.  A structure $(\astore,\aheap)$ such that $(\astore,\aheap)
\models \phi$, is called a \emph{model} of $\phi$.  Note that
$(\astore,\aheap) \models \phi$ only if $\fv{\phi} \subseteq
\dom(\astore)$.

The fragment of \emph{symbolic heaps} is obtained by confining the
negation and conjunction to the formul{\ae} $t_1 \seq t_2 \isdef t_1
\teq t_2 \wedge \emp$ and $t_1 \sneq t_2 \isdef \neg t_1 \teq t_2
\wedge \emp$, called \emph{equational atoms}, by abuse of language. We
denote by $\shk{\rank}$ the set of symbolic heaps, formally defined
below:
\vspace*{-.5\baselineskip}
\[\begin{array}{rcl}
\phi & := & \emp \mid t_0 \mapsto (t_1, \ldots, t_\rank) \mid p(t_1,
\ldots, t_{\#p}) \mid t_1 \seq t_2 \mid t_1 \sneq t_2 \mid \phi_1 *
\phi_2 \mid \exists x ~.~ \phi_1 \\[-4mm]
\end{array}\]
Given quantifier-free symbolic heaps $\phi_1, \phi_2 \in \shk{\rank}$,
it is not hard to check that $\exists x ~.~ \phi_1 * \exists y ~.~
\phi_2$ and $\exists x \exists y ~.~ \phi_1 * \phi_2$ have the same
models. Consequently, each symbolic heap can be written in prenex
form, as $\phi = \exists x_1 \ldots \exists x_n ~.~ \psi$,
\shortVersionOnly{where $\psi$ is a quantifier-free separating
  conjunction of points-to atoms and (dis-)equalities. A variable $x
  \in \fv{\psi}$ is \emph{allocated} in $\phi$ iff there exists a
  (possibly empty) sequence of equalities $x \seq \ldots \seq t_0$ and
  a points-to atom $t_0 \mapsto (t_1, \ldots, t_\rank)$ in $\psi$. }
\longVersionOnly{where:
  \begin{equation}\label{eq:symbolic-heap}
    \psi = \Asterisk_{i=1}^\alpha t_0^i \mapsto (t_1^i, \ldots, t_\rank^i) ~*~
    \Asterisk_{j=\alpha+1}^\beta p_j(t_1^j, \ldots, t_{\#p}^j) ~*~
    \Asterisk_{k=\beta+1}^{\gamma} t_1^k \seq t_2^k ~*~
    \Asterisk_{\ell=\gamma+1}^{\delta} t_1^\ell \sneq t_2^\ell
  \end{equation}
  for some integers $0 \leq \alpha \leq \beta \leq \gamma \leq
  \delta$.  A variable $x \in \fv{\phi}$ is \emph{allocated} in a
  symbolic heap $\phi$ if, using the notations from
  (\ref{eq:symbolic-heap}), either $x \in \{t_0^1,\ldots,t_0^\alpha\}$
  or there exists a sequence of terms $(t_1, \ldots, t_j)$ such that
  $j\geq 2$, $x = t_1$, $\set{t_i, t_{i+1}} = \{t_1^k, t_2^k\}$ for
  some $k \in \interv{\beta+1}{\gamma}$ and $t_j \in
  \{t_0^1,\ldots,t_0^\alpha\}$. Clearly, if $\phi$ is satisfiable
and predicate-free then $x$ is allocated in $\phi$ if and only if
$\astore'(x) \in \dom(\aheap)$ holds for every
$(x_1,\ldots,x_n)$-associate $\astore'$ of $\astore$ such that
$(\astore',\aheap) \models \psi$. }

The predicates from $\preds$ are intepreted by a given set $\asys$ of
\emph{rules} $p(x_1, \ldots, x_{\#p}) \Leftarrow \arule$, where
$\arule$ is a symbolic heap, such that $\fv{\arule} \subseteq \{x_1,
\ldots, x_{\#_p}\}$. We say that $p(x_1, \ldots, x_{\#p})$ is the
\emph{head} and $\arule$ is the \emph{body} of the rule. For
conciseness, we write $p(x_1, \ldots, x_{\#p}) \Leftarrow_\asys
\arule$ instead of $p(x_1, \ldots, x_{\#p}) \Leftarrow \arule \in
\asys$. In the following, we shall often refer to a given set of rules
$\asys$.

\begin{definition}[Unfolding]\label{def:unfolding}
  A formula $\psi$ is a \emph{step-unfolding} of a formula $\phi \in
  \seplogk{\rank}$, written $\phi \unfold{\asys} \psi$, if $\psi$ is
  obtained by replacing an occurrence of an atom $p(t_1, \ldots,
  t_{\#p})$ in $\phi$ with $\arule[t_1/x_1, \ldots, t_{\#p}/x_{\#p}]$,
  for a rule $p(x_1, \ldots, x_{\#p}) \Leftarrow_\asys \arule$. An
  \emph{unfolding} of $\phi$ is a formula $\psi$ such that $\phi
  \unfold{\asys}^* \psi$.
\end{definition}
It is easily seen that any unfolding of a symbolic heap is again a
symbolic heap. We implicitly assume that all bound variables are
$\alpha$-renamed throughout an unfolding, to avoid name clashes.
Unfolding extends the semantics from predicate-free to arbitrary
$\seplogk{\rank}$ formul{\ae}:

\begin{definition}\label{def:unfolding-semantics}
  Given a structure $(\astore,\aheap)$ and a formula $\phi \in
  \seplogk{\rank}$, we write $(\astore,\aheap) \models_\asys \phi$ iff
  there exists a predicate-free unfolding $\phi \unfold{\asys}^* \psi$
  such that $(\astore,\aheap) \models \psi$. In this case,
  $(\astore,\aheap)$ is an \emph{$\asys$-model of $\phi$}. For two
  formul{\ae} $\phi, \psi \in \seplogk{\rank}$, we write $\phi
  \models_\asys \psi$ iff every $\asys$-model of $\phi$ is an
  $\asys$-model of $\psi$.
\end{definition}
Note that, if $(\astore,\aheap) \models_\asys \phi$, then
$\dom(\astore)$ might have to contain constants that do not occur in
$\phi$. For instance if $p(x) \Leftarrow_\asys x \mapsto \mathsf{a}$
is the only rule with head $p(x)$, then any $\asys$-model
$(\astore,\aheap)$ must map $\mathsf{a}$ to some location, which is
taken care of by the assumption $\const \subseteq \dom(\astore)$, that
applies to any store.
%

\begin{definition}[Entailment]\label{def:entailment}
  Given symbolic heaps $\phi, \psi_1, \ldots, \psi_n$, such that
  $\phi$ is quantifier-free and $\fv{\phi} = \fv{\psi_1} = \ldots =
  \fv{\psi_n} = \emptyset$, the \emph{sequent} $\phi \vdash \psi_1,
  \ldots, \psi_n$ is \emph{valid for $\asys$} iff $\phi \models_\asys
  \bigvee_{i=1}^n \phi_i$. An \emph{entailment problem} $\aprob =
  (\asys,\seqset)$ consists of a set of rules $\asys$ and a set
  $\seqset$ of sequents, asking whether each sequent in $\seqset$ is
  valid for $\asys$.
\end{definition}
Note that we consider entailments between formul{\ae} without free
variables. This is not restrictive, since any free variable can be
replaced by a constant from $\const$, with no impact on the validity
status or the computational complexity of the problem.  We silently
assume that $\const$ contains enough constants to allow this
replacement. For conciseness, we write $\phi \vdash_\aprob
\psi_1,\ldots,\psi_n$ for $\phi \vdash \psi_1,\ldots,\psi_n \in
\seqset$, where $\seqset$ is the set of sequents of $\aprob$. The
following example shows an entailment problem asking whether the
concatenation of two acyclic lists is again an acyclic list:

\begin{example}\label{ex:acyclic-lists}
The entailment problem below consists of four rules, defining the
predicates $\mathsf{ls}(x,y)$ and $\mathsf{sls}(x,y,z)$,
respectively, and two sequents: 
\vspace*{-.5\baselineskip}
\[\begin{array}{rcl}
\mathsf{ls}(x,y) & \Leftarrow & x \mapsto y * x \sneq y \mid 
\exists v ~.~ x \mapsto v * \mathsf{ls}(v,y) * x \sneq y 
\\
\mathsf{sls}(x,y,z) & \Leftarrow & x \mapsto y * x \sneq y * x \sneq z \mid 
\exists v ~.~ x \mapsto v * \mathsf{sls}(v,y,z) * x \sneq y * x \sneq z 
\\
\mathsf{ls}(a,b) * \mathsf{ls}(b,c) & \vdash & 
\exists x ~.~ a \mapsto x * \mathsf{ls}(x,c) * a \sneq c \hspace*{5mm}
\mathsf{sls}(a,b,c) * \mathsf{ls}(b,c) \vdash \exists x ~.~ a \mapsto x * \mathsf{ls}(x,c) * a \sneq c \\[-2mm]
\end{array}\]
Here $\mathsf{ls}(x,y)$ describes non-empty acyclic list segments with
head and tail pointed to by $x$ and $y$, respectively. The first
sequent is invalid, because $c$ can be allocated within the list
segment defined by $\mathsf{ls}(a,b)$, in which case the entire list
has a cycle starting and ending with the location associated with
$c$. To avoid the cycle, the left-hand side of the second sequent uses
the predicate $\mathsf{sls}(x,y,z)$ describing an acyclic list segment
from $x$ to $y$ that skips the location pointed to by $z$. The second
sequent is valid. \hfill$\blacksquare$
\end{example}

The complexity analysis of the decision procedure described in this
paper relies on two parameters. First, the \emph{width} of an
entailment problem $\aprob=(\asys,\seqset)$ is (roughly) the maximum
among the sizes of the symbolic heaps occurring in $\aprob$ and the
number of constants in $\const$. Second, the \emph{size} of the
entailment problem is (roughly) the number of symbols needed to
represent it, namely:
\vspace*{-0.5\baselineskip}
\[\begin{array}{rcl}
\probwidth{\aprob} & \isdef & \max
\big(\{\size{\arule} + \#p \mid p(x_1,\ldots,x_{\#p}) \Leftarrow_\asys \arule\} \cup 
\{\size{\psi_i} \mid \psi_0 \vdash_\aprob \psi_1,\ldots,\psi_n \} \cup 
\set{\card{\const}} \big) \\
\size{\aprob} & \isdef & \sum_{p(x_1,\ldots,x_{\#p}) ~\Leftarrow_\asys ~\arule} (\size{\arule}+\#p) + 
\sum_{\psi_0 ~\vdash_\aprob~ \psi_1,\ldots,\psi_n} \sum_{i=1}^n \size{\psi_i} 
\end{array}\]
In the next section we give a transformation of an entailment problems
with a time complexity that is bounded by the product of the size and
a simple exponential of the width of the input, such that, moreover,
the width of the problem increases by a polynomial factor only. The
latter is instrumental in proving the final \twoexptime\ upper bound
on the complexity of the entailment problem.

To alleviate the upcoming technical details, we make the following
assumption:

\begin{assumption}\label{ass:dist-const}
 Distinct constants are always associated with distinct locations: for
 all stores $\astore$, and for all $c,d \in \const$, we have $c \not =
 d$ only if $\astore(c) \neq \astore(d)$.
\end{assumption}
This assumption loses no generality, because one can enumerate all the
equivalence relations on $\const$ and test the entailments separately
for each of these relations, by replacing all the constants in the
same class by a unique representative\footnote{The replacement must be
  performed also within the inductive rules, not only in the
  considered formul{\ae}.}, while assuming that constants in distinct
classes are mapped to distinct locations. The overall complexity of
the procedure is still doubly exponential, since the number of such
equivalence relations is bounded by the number of partitions of
$\const$, that is $2^{\bigO(\card{\const}\cdot\log\card{\const})} =
2^{\bigO(\card{\probwidth{\aprob}}\cdot\log\card{\probwidth{\aprob}})}$,
for any entailment problem $\aprob$. Thanks to Assumption
\ref{ass:dist-const}, the considered symbolic heaps can be, moreover,
safely assumed not to contain atoms $c \bowtie d$, with $\bowtie \in
\{ \seq,\sneq \}$ and $c,d \in \const$, since these atoms are either
unsatisfiable or equivalent to $\emp$.

\section{Decidable Classes of Entailments}
\label{sec:dec-entailments}

In general, the entailment problem (Definition \ref{def:entailment})
is undecidable and we refer the reader to
\cite{DBLP:conf/atva/IosifRV14,AntonopoulosGorogiannisHaaseKanovichOuaknine14}
for two different proofs. A first attempt to define a naturally
expressive class of formul{\ae} with a decidable entailment problem
was reported in \cite{IosifRogalewiczSimacek13}. The entailments
considered in \cite{IosifRogalewiczSimacek13} involve sets of rules
restricted by three conditions, recalled below, in a slightly
generalized form.

First, the \emph{progress} condition requires that each rule adds to
the heap exactly one location, associated either to a constant or to a
designated parameter. Formally, we consider a mapping $\aroot : \preds
\rightarrow \nat \cup \const$, such that $\aroot(p) \in
\interv{1}{\#p} \cup \const$, for each $p \in \preds$. The term
$\aroot(p(t_1,\ldots,t_{\#p}))$ denotes either $t_i$ if $\aroot(p) = i
\in \interv{1}{\#p}$, or the constant $\aroot(p)$ itself if $\aroot(p)
\in \const$. The notation $\aroot(\alpha)$ is extended to points-to
atoms $\alpha$ as $\aroot(t_0 \mapsto (t_1,\dots,t_\rank)) \isdef
t_0$. Second, the \emph{connectivity} condition requires that all
locations added during an unfolding of a predicate atom form a set of
connected trees (a forest) rooted in locations associated either with
a parameter of the predicate or with a constant.

\begin{definition}[Progress \& Connectivity]\label{def:progress-connectivity}
  A set of rules $\asys$ is \emph{progressing} if each rule in $\asys$
  is of the form $p(x_1, \ldots, x_{\#p}) \Leftarrow \exists z_1
  \ldots \exists z_m ~.~ \aroot(p(x_1,\ldots,x_{\#p})) \mapsto (t_1,
  \ldots, t_\rank) * \psi$ and $\psi$ contains no occurrences of
  points-to atoms. Moreover, $\asys$ is \emph{connected} if
  $\aroot(q(u_1,\ldots,u_{\#q})) \in \{t_1, \ldots, t_\rank\} \cup
  \const$, for each predicate atom $q(u_1,\ldots,u_{\#q})$ occurring
  in $\psi$. An entailment problem $\aprob = (\asys,\seqset)$ is
  progressing (connected) if $\asys$ is progressing (connected).
\end{definition}
The progress and connectivity conditions can be checked in polynomial
time by a syntactic inspection of the rules in $\asys$, even if the
$\aroot(.)$ function is not known \`{a}~priori. Note that this
definition of connectivity is less restrictive that the definition
from \cite{IosifRogalewiczSimacek13}, that asked for
$\aroot(q(u_1,\ldots,u_{\#q})) \in \{t_1, \ldots, t_\rank\}$.  For
instance, the set of rules $\{ \mathsf{p}(x) \Leftarrow \exists y~.~ x
\mapsto y * \mathsf{p}(y) * \mathsf{p}(\mathsf{c}), \mathsf{p}(x)
\Leftarrow x \mapsto \nil \}$, where $\mathsf{c} \in \const$ is
progressing and connected (with $\aroot(\mathsf{p}) = 1$) in the sense
of Definition \ref{def:progress-connectivity}, but not connected in
the sense of \cite{IosifRogalewiczSimacek13}, because $\mathsf{c}
\not\in (y)$. Note also that nullary predicate symbols are allowed,
for instance $\mathsf{q}() \Leftarrow \mathsf{c} \mapsto \nil$ is
progressing and connected (with $\aroot(\mathsf{q}) =
\mathsf{c}$). Further, the entailment problem from Example
\ref{ex:acyclic-lists} is both progressing and connected.

Third, the \emph{establishment} condition is defined, slightly
extended from its original statement \cite{IosifRogalewiczSimacek13}:

\begin{definition}[Establishment]\label{def:establishment}
  Given a set of rules $\asys$, a symbolic heap $\exists x_1 \ldots
  \exists x_n ~.~ \phi$, where $\phi$ is quantifier-free, is
  $\asys$-\emph{established} iff every $x_i$ for $i \in \interv{1}{n}$
  is allocated in each predicate-free unfolding $\phi \unfold{\asys}^*
  \varphi$. A set of rules $\asys$ is \emph{established} if the body
  $\arule$ of each rule $p(x_1,\ldots,x_{\#p}) \Leftarrow_\asys
  \arule$ is $\asys$-established. An entailment problem $\aprob =
  (\asys,\seqset)$ is established if $\asys$ is established, and
  \emph{strongly established} if, moreover, $\phi_i$ is
  $\asys$-established, for each sequent $\phi_0 \vdash_\aprob
  \phi_1,\ldots,\phi_n$ and each $i \in \interv{0}{n}$.
\end{definition}
For example, the entailment problem from Example
\ref{ex:acyclic-lists} is strongly established.

\longVersionOnly{
  \subsection*{\Restricted Entailments }
  \label{sec:restricted}
}

In this paper, we replace establishment with a new condition that, as
we show, preserves the decidability and computational complexity of
progressing, connected and established entailment problems. The new
condition can be checked in time linear in the size of the
problem. This condition, called \emph{equational restrictedness}
(\emph{\restrictedness}, for short), requires that each equational
atom occurring in a formula involves at least one constant. We will
show that the \restrictedness condition is more general than
establishment, in the sense that every established problem can be
reduced to an equivalent \restricted problem (Theorem
\ref{thm:established-restricted}). Moreover, the class of structures
defined using \restricted symbolic heaps is a strict superset of the
one defined by established symbolic heaps.

\begin{definition}[\Restrictedness]\label{def:restricted}
  A symbolic heap $\phi$ is \emph{\restricted} if, for every
  equational atom $t \bowtie u$ from $\phi$, where $\bowtie \in
  \set{\seq,\sneq}$, we have $\set{t,u} \cap \const \neq \emptyset$. A
  set of rules $\asys$ is \emph{\restricted} if the body $\arule$ of
  each rule $p(x_1,\ldots,x_{\#p}) \Leftarrow_\asys \arule$ is
  \restricted. An entailment problem $\aprob=(\asys,\seqset)$ is
  \emph{\restricted} if $\asys$ is \restricted and $\phi_i$ is
  \restricted, for each sequent $\phi_0 \vdash_\aprob
  \phi_1,\ldots,\phi_n$ and each $i \in \interv{0}{n}$.
\end{definition}
For instance, the entailment problem from Example
\ref{ex:acyclic-lists} is not \restricted, because several rule bodies
have disequalities between parameters, e.g.\ $\mathsf{ls}(x,y)
\Leftarrow x \mapsto y * x \sneq y$. However, the set of rules $\{
\mathsf{ls_c}(x) \Leftarrow x \mapsto c * x \sneq \mathsf{c},
\mathsf{ls_c}(x) \Leftarrow \exists y ~.~ x \mapsto y *
\mathsf{ls_c}(y) * x \sneq \mathsf{c} \}$, where $\mathsf{c} \in
\const$ and $\mathsf{ls_c}$ is a new predicate symbol, denoting an
acyclic list ending with $\mathsf{c}$, is \restricted.  Note that any
atom $\mathsf{ls}(x,y)$ can be replaced by $\mathsf{ls_y}(x)$,
provided that $y$ occurs free in a sequent and can be viewed as a
constant.

We show next that every established entailment problem (Definition
\ref{def:establishment}) can be reduced to \arestricted entailment
problem (Definition \ref{def:restricted}).  The transformation incurs
an exponential blowup, however, as we show, the blowup is exponential
only in the width and polynomial in the size of the input
problem. This is to be expected, because checking \restrictedness of a
problem can be done in linear time, in contrast with checking
establishment, which is at least co-\np-hard
\cite{JansenKatelaanMathejaNollZuleger17}.

\longVersionOnly{
\section{Pre-Processing Step: Normalizing Entailements}
\label{sec:normalize}
\mycomment[np]{added section}
}

We begin by showing that each problem can be translated into an
equivalent \emph{normalized} problem:
\begin{definition}[Normalization]\label{def:normalized}\hfill
  \begin{compactenum}[(1)]
  \item A symbolic heap $\exists \vec{x} ~.~ \psi \in \shk{\rank}$,
    where $\psi$ is quantifier-free, is \emph{normalized} iff for
    every atom $\alpha$ in $\psi$:\begin{compactenum}[a.]
    \item\label{it1:normalized} if $\alpha$ is an equational atom,
      then it is of the form $x \sneq t$ ($t \sneq x$), where $x \in
      \vec{x}$,
    \item\label{it1bis:normalized} every variable $x \in \fv{\psi}$
      occurs in a points-to or predicate atom of $\psi$,
    \item\label{it2:normalized} if $\alpha$ is a predicate atom
      $q(t_1,\ldots,t_{\#q})$, then $\{t_1, \ldots, t_{\#q}\} \cap
      \const = \emptyset$ and $t_i \neq t_j$, for all $i \neq j \in
      \interv{1}{\#q}$.
    \end{compactenum}
  \item A set of rules $\asys$ is \emph{normalized} iff for each rule
    $p(x_1,\ldots,x_{\#p}) \Leftarrow_\asys \rho$, the symbolic heap
    $\rho$ is normalized and, moreover: \begin{compactenum}[a.]
    \item\label{it3:normalized} For every $i \in \interv{1}{\#p}$ and
      every predicate-free unfolding $p(x_1,\ldots,x_{\#p})
      \unfold{\asys}^* \varphi$, $\varphi$ contains a points-to atom
      $t_0 \mapsto (t_1,\ldots,t_\rank)$, such that $x_i \in
      \{t_0,\ldots,t_\rank\}$.
    \item\label{it4:normalized} There exist sets $\allocpar{p}{\asys}
      \subseteq \interv{1}{\#p}$ and $\alloconst{p}{\asys} \subseteq
      \const$ such that, for each predicate-free unfolding
      $p(x_1,\ldots,x_{\#p}) \unfold{\asys}^* \varphi$: \begin{compactitem}
      \item $i \in \allocpar{p}{\asys}$ iff $\varphi$ contains an atom
        $x_i \mapsto (t_1,\ldots,t_\rank)$, for every $i\in
        \interv{1}{\#p}$,
      \item $c \in \alloconst{p}{\asys}$ iff $\varphi$ contains an
        atom $c \mapsto (t_1,\ldots,t_\rank)$, for every $c\in
        \const$.         
      \end{compactitem}
      \item\label{it5:normalized} For every predicate-free unfolding
        $p(x_1,\ldots,x_{\#p}) \unfold{\asys}^* \varphi$, if $\varphi$
        contains an atom $t_0 \mapsto (t_1,\dots,t_\rank)$ such that
        $t_0\in \vars \setminus \{ x_1,\dots,x_{\#p} \}$, then
        $\varphi$ also contains atoms $t_0 \not \seq c$, for every $c
        \in \const$. 
    \end{compactenum}
  \item An entailment problem $\aprob = (\asys,\seqset)$ is
    \emph{normalized} if $\asys$ is normalized and, for each sequent
    $\phi_0 \vdash_\aprob \phi_1,\ldots,\phi_n$ the symbolic heap
    $\phi_i$ is normalized, for each $i \in \interv{0}{n}$.
  \end{compactenum}
\end{definition}
The intuition behind Condition (\ref{it3:normalized}) is that no term
can ``disappear'' while unfolding an inductive definition.  Condition
(\ref{it4:normalized}) states that the set of terms eventually
allocated by a predicate atom is the same in all unfoldings. This
allows to define the set of symbols that occur freely in a symbolic
heap $\phi$ and are necessarily allocated in every unfolding of
$\phi$, provided that the set of rules is normalized:

\begin{definition}\label{def:alloc}
Given a normalized set of rules $\asys$ and a symbolic heap $\phi \in
\shk{\rank}$, the set $\alloc{\phi}{\asys}$ is defined recursively on
the structure of $\phi$:
\vspace*{-.5\baselineskip}
\[\begin{array}{rclcrcl}
\alloc{t_0 \mapsto (t_1,\ldots,t_\rank)}{\asys} & \isdef & \set{t_0} &&
\alloc{p(t_1,\ldots,t_{\#p})}{\asys} & \isdef & \set{t_i \mid i \in \allocpar{p}{\asys}} \\
\alloc{t_1 \bowtie t_2}{\asys} & \isdef & \emptyset,~ \bowtie \in \{\seq,\sneq\} &&&& \cup~ \alloconst{p}{\asys} \\
\alloc{\phi_1 * \phi_2}{\asys} & \isdef & \alloc{\phi_1}{\asys} \cup \alloc{\phi_2}{\asys} && 
\alloc{\exists x ~.~ \phi_1}{\asys} & \isdef & \alloc{\phi_1}{\asys} \setminus \set{x}
\end{array}\]
\end{definition}

\begin{example}
The rules $p(x,y) \Leftarrow \exists z ~.~ x \mapsto z * p(z,y) * x
\not \seq y$ and $p(x,y) \Leftarrow \exists z~.~ x \mapsto z$ are not
normalized, because they contradict Conditions (\ref{it1:normalized})
and (\ref{it3:normalized}) of Definition \ref{def:normalized},
respectively.  A set $\asys$ containing the rules $q(x,y) \Leftarrow
\exists z~.~ x \mapsto y * q(y,z)$ and $q(x,y) \Leftarrow x \mapsto y$
is not normalized, because it is not possible to find a set
$\allocpar{q}{\asys}$ satisfying Condition (\ref{it4:normalized}).
Indeed, if $2 \in \allocpar{q}{\asys}$ then the required equivalence
does not hold for the second rule (because it does not allocate $y$),
and if $2 \not \in \allocpar{q}{\asys}$ then it fails for the first
one (since the predicate $q(y,z)$ allocates $y$). On the other hand,
$\asys' = \{ p(x,y) \Leftarrow \exists z~.~ x \mapsto z * p(z,y) * z
\not \seq x * z \not \seq \nil, p(x,y) \Leftarrow x \mapsto y, q(x,y)
\Leftarrow \exists z~.~ x \mapsto y * q(y,z) * z \not \seq \nil$ ,
$q(x,y) \Leftarrow x \mapsto y * r(y), r(x) \Leftarrow x \mapsto
\nil\}$ is normalized (assuming $\const = \{ \nil \}$), with
$\allocpar{p}{\asys'} = \allocpar{r}{\asys'} =\{ 1 \}$,
$\allocpar{q}{\asys'} = \{ 1,2 \}$ and $\alloconst{\pi}{\asys'} =
\emptyset$, for all $\pi \in \set{p,q,r}$. Then $\alloc{p(x_1,x_2) *
  q(x_3,x_4) * r(x_5)}{\asys'} = \{ x_1,x_3,x_4,x_5
\}$. \hfill$\blacksquare$
\end{example}

The following lemma states that every entailment problem can be
transformed into a normalized entailment problem, by a transformation
that preserves \restricted-ness and (strong) establishment.
\begin{lemma}\label{lemma:normalized}
  An entailment problem $\aprob$ can be translated to an equivalent
  normalized problem $\aprob_n$, such that $\probwidth{\aprob_n} =
  \bigO(\probwidth{\aprob}^2)$ in time $\size{\aprob} \cdot
  2^{\bigO(\probwidth{\aprob}^2)}$. Further, $\aprob_n$ is \restricted
  and (strongly) established if $\aprob$ is \restricted and (strongly)
  established.
\end{lemma}
\optionalProof{Lemma \ref{lemma:normalized}}{sec:dec-entailments}{ Let
  $\aprob = (\asys,\seqset)$ be an input entailment problem. We
  transform $\aprob$ in order to meet points (\ref{it1:normalized}),
  (\ref{it2:normalized}), (\ref{it3:normalized}),
  (\ref{it4:normalized}) and (\ref{it5:normalized}) of Definition
  \ref{def:normalized}, as follows.

  \noindent(\ref{it1:normalized}) First, we apply exhaustively, to
  each symbolic heap occurring in $\aprob$, the following
  transformations, for each term $t \in \terms$:
  \begin{eqnarray}
    \exists x ~.~ x \seq t * \phi & \leadsto & \phi[t/x] \label{eq:exists-subst} \\
    t \seq t * \phi & \leadsto & \phi \label{eq:equal-remove}
  \end{eqnarray}
  Note that, at this point, there are no equality atoms involving an
  existentially quantified variable (recall that equalities between
  constants can be dismissed since they are either trivially false or
  equivalent to $\emp$). We apply the following transformations, that
  introduce disequalities between the remaining existential variables
  and the rest of the terms.
  \begin{eqnarray}
    p(\vec{x}) & \Leftarrow & \exists x ~.~ \arule \leadsto \set{\begin{array}{rcl} p(\vec{x}) & \Leftarrow & \arule[t/x] \\
        p(\vec{x}) & \Leftarrow & \exists x ~.~ \arule * x \sneq t \end{array}} \label{eq:rule-split} \\[-1mm]
    && \begin{array}{l} \text{for all $t \in (\fv{\arule} \setminus \set{x}) \cup \const$,
           where $x \sneq t$ does not occur in $\arule$} \end{array}\nonumber \\[2mm]
    \phi \vdash \psi_1,\ldots, \exists x ~.~ \psi_i, \ldots, \psi_n 
    & \leadsto & 
    \phi \vdash \psi_1,\ldots,\psi_{i-1},\psi_i[t/x],\exists x ~.~ x \sneq t * \psi_i,\ldots,\psi_n\label{eq:rhs-split} \\[-1mm]
    && \begin{array}{l} \text{for all $t \in \const$, such that $x \sneq t$ does not occur in $\psi_i$} \end{array}\nonumber
  \end{eqnarray}  
  Let $\aprob_1 = (\asys_1,\seqset_1)$ be the result of applying the
  transformations (\ref{eq:exists-subst}-\ref{eq:rhs-split})
  exhaustively. Because every transformation preserves the equivalence
  of rules and sequents, $\aprob_1$ is valid iff $\aprob$ is
  valid. Note that, by Definition \ref{def:entailment}, there are no
  free variables occurring in a sequent from $\seqset$. Then the only
  remaining equality atoms $t \seq u$ occurring in $\aprob_1$ must
  occur in a rule $p(x_1,\ldots,x_{\#p}) \Leftarrow_{\asys_1} \arule$
  and neither $t$ nor $u$ can be an existentially quantified variable,
  hence $t,u \in \{x_1,\ldots,x_{\#p}\} \cup \const$. Before
  proceeding further with Condition (\ref{it1:normalized}), we make
  sure that Condition (\ref{it2:normalized}) is satisfied.

  \noindent(\ref{it2:normalized}) Let $q(t_1,\ldots,t_{\#q})$ be a
  predicate atom occurring in a rule or a sequent from $\aprob_1$,
  where $t_1, \ldots, t_{\#q} \in \terms$, and let $(t_{i_1}, \ldots,
  t_{i_m})$ be the subsequence obtained by removing the terms from the
  set $\{t_i \mid i \in \interv{1}{\#q},~ \exists j < i ~.~ t_i=t_j\}
  \cup \const$ from $(t_1, \ldots, t_{\#q})$. We consider a fresh
  predicate symbol $q_{i_1,\ldots,i_m}$, of arity $m$, with the new
  rules $q_{i_1,\ldots,i_m}(x_1,\ldots,x_m) \Leftarrow \arule\sigma$,
  for each rule $q(x_1,\ldots,x_{\#q}) \Leftarrow_\asys \arule$, where
  the substitution $\sigma$ is defined such that, for all $j \in
  \interv{1}{\#q}$: \begin{compactitem}
  \item $\sigma(x_j) \isdef x_{i_\ell}$ if $t_j = t_{i_\ell}$, for some
    $\ell \in \interv{1}{m}$,
  \item $\sigma(x_j) \isdef t_j$ if $t_j \in \const$, and
  \item $\sigma(x_j) \isdef x_j$, otherwise. 
  \end{compactitem}
  Note that the definition of the sequence $(t_{i_1}, \ldots,
  t_{i_m})$ guarantees that such a substitution exists and it is
  unique. If the rule body obtained by applying the substitution
  $\sigma$ contains a disequality $t \sneq t$, for some $t \in
  \terms$, we eliminate the rule. Otherwise, we apply transformation
  (\ref{eq:equal-remove}) to the newly obtained rule to eliminate
  trivial equalities. Finally, we replace each occurrence of
  $q(t_1,\ldots,t_{\#q})$ in $\aprob_1$ with
  $q_{i_1,\ldots,i_m}(t_{i_1}, \ldots, t_{i_m})$. Because
  $q(t_1,\ldots,t_m)$ and $q_{i_1,\ldots,i_m}(t_{i_1},\ldots,t_{i_m})$
  have the same step unfoldings, they have the same predicate-free
  unfoldings and this transformation preserves equivalence, yielding a
  problem that satisfies condition (\ref{it2:normalized}). Let
  $\aprob_2 = (\asys_2, \seqset_2)$ be the outcome of this
  transformation, where $\asys_2$ is the set of newly introduced rules
  and $\seqset_2$ is obtained from $\seqset_1$ by the replacement of
  each predicate atom $q(t_1,\ldots,t_{\#q})$ with
  $q_{i_1,\ldots,i_m}(t_{i_1}, \ldots, t_{i_m})$. It is easy to check
  that $\aprob_2$ and $\aprob_1$ have the same validity status, which
  is that of $\aprob$.

  \noindent(\ref{it1:normalized}) We will now finish the proof of
  Condition (\ref{it1:normalized}).  Since the transformation
  (\ref{eq:exists-subst}) removes equalities involving an
  existentially quantified variable and the equalities between
  constants can be eliminated as explained above, the only equalities
  that occur in the body of a rule $p(x_1, \ldots, x_{\#p})
  \Leftarrow_{\asys_2} \arule$ are of the form $x_i \seq t$, where $i
  \in \interv{1}{\#p}$ and $t \in \{x_j \mid j \in \interv{1}{\#p}, j
  \neq i\} \cup \const$. We show that if such an equality occurs in
  the body of a rule, then this rule can safely be removed because any
  unfolding involving it generates an unsatisfiable symbolic heap.
  Let $p(u_1, \ldots, u_{\#p})$ be a predicate atom that occurs in a
  some unfolding of a symbolic heap from $\aprob$ and assume a
  step-unfolding that substitutes $p(u_1, \ldots, u_{\#p})$ with
  $\arule[u_1/x_1, \ldots, u_{\#p}/x_{\#p}]$. We distinguish two
  cases: \begin{compactenum}[(i)]
  \item\label{it1:1:normalized} $t = x_j$, for some $j \in
    \interv{1}{\#p} \setminus \set{i}$: by point
    (\ref{it2:normalized}), $u_i$ and $u_j$ must be distinct terms. If
    $u_i, u_j \in \const$, then $u_i \sneq u_j$ necessarily holds, by
    Assumption \ref{ass:dist-const}, thus the equality $x_i \seq t$ is
    false when $x_i,x_j$ are instantiated by $u_i,u_j$. Otherwise, if
    $u_i \in \vars$ (the case $u_j \in \vars$ is symmetric) then $u_i$
    and $u_j$ were necessarily introduced by existential quantifiers,
    in which case the disequality $u_i \sneq u_j$ has been asserted by
    transformations (\ref{eq:rule-split}) or (\ref{eq:rhs-split}),
    thus $x_i \seq t$ is false when $x_i$ is replaced by $u_i$.
  \item\label{it1:2:normalized} $t \in \const$: by a similar argument
    we show that that all the relevant instances of the equality $x_i
    \seq t$ are unsatisfiable.
  \end{compactenum}
  Consequently, if an equality occurs in a rule, then this the rule
  can safely be removed. 

  \noindent(\ref{it1bis:normalized}) To ensure that all variables occur within a
  points-to or predicate atom, we apply exhaustively the following
  transformation to each symbolic heap in the problem:
  \begin{equation}\label{eq:remove-diseq}
    \exists x ~.~ \Asterisk_{i=1}^n x \sneq t_i * \psi \leadsto \psi \text{, if $x \not\in \fv{\psi}$}
  \end{equation}
  Let $\aprob_3=(\asys_3,\seqset_2)$ be the outcome of this
  transformation. Because $\locs$ is infinite, any formula $\exists x
  ~.~ \Asterisk_{i=1}^n x \sneq t_i$ is equivalent to
  $\emp$. Consequently, $\aprob_3$ and $\aprob_2$ have the same
  validity status as $\aprob$ and $\aprob_3$ satisfies conditions
  (\ref{it1:normalized}), (\ref{it1bis:normalized}) and
  (\ref{it2:normalized}).

  \noindent(\ref{it3:normalized}+\ref{it4:normalized}) For each
  predicate symbol $p$ that occurs in $\asys_3$, we consider the
  predicate symbols $p_{X,Y,Z,A,B,C}$, of arities $\#p$ each, where
  $(X,Y,Z)$ is a partition of $\interv{1}{\#p}$ and $(A,B,C)$ is a
  partition of $\const$, along with the following rules:
  $p_{X,Y,Z,A,B,C}(x_1,\ldots,x_{\#p}) \Leftarrow \arule'$ if and only
  if $p(x_1, \ldots, x_{\#p}) \Leftarrow_{\asys_3} \arule$ and
  $\arule'$ is obtained from $\arule$ by replacing each predicate atom
  $q(t_1,\ldots,t_{\#q})$ by a predicate atom
  $q_{X',Y',Z',A',B',C'}(t_1,\ldots,t_{\#q})$, for some partition
  $(X',Y',Z')$ of $\interv{1}{\#q}$ and some partition $(A',B',C')$ of
  $\const$, such that the following holds. For each $i \in
  \interv{1}{\#p}$: \begin{compactitem}
  \item $i \in X$ iff either a points-to atom $x_i \mapsto
    (t_1,\ldots,t_\rank)$ occurs in $\arule$, or $\arule$ contains a
    predicate atom $r_{X'', Y'', Z'', A'', B'', C''}(t_1, \ldots,
    t_{\#r})$ such that $x_i=t_j$ and $j \in X''$,
  \item $i \in Y$ iff either $x_i \in
    \{t_1,\ldots,t_\rank\}$ for a points-to atom $t_0 \mapsto
    (t_1,\ldots,t_\rank)$ occurring in $\arule$, or $\arule$ contains
    a predicate atom $r_{X'', Y'', Z'', A'', B'', C''}(t_1, \ldots,
    t_{\#r})$ such that $x_i=t_j$ and $j \in Y''$.
  \end{compactitem}
  Further, for each constant $c \in
  \constset$: \begin{compactitem}
  \item $c \in A$ iff a points-to atom $c \mapsto (t_1,
    \ldots, t_\rank)$ occurs in $\arule$ or $\arule$ contains a
    predicate atom $r_{X'', Y'', Z'', A'', B'', C''}(t_1, \ldots,
    t_{\#r})$ such that $c \in A''$,
  \item $c \in B$ iff either $c \in \{t_1, \ldots,
    t_\rank\}$, for a points-to atom $t_0 \mapsto (t_1, \ldots,
    t_\rank)$ occurring in $\arule$ or $\arule$ contains a predicate
    atom $r_{X'', Y'', Z'', A'', B'', C''}(t_1, \ldots, t_{\#r})$ such
    that $c \in B''$, 
  \end{compactitem}
  Let $\seqset_4$ (resp.\ $\asys_4$) be the set of sequents
  (resp.\ rules) obtained by replacing each predicate atom
  $p(t_1,\ldots,t_{\#p})$ with $p_{X,Y,Z,A,B,C}(t_1,\ldots,t_{\#p})$,
  for some partition $(X,Y,Z)$ of $\interv{1}{\#p}$ and some partition
  $(A,B,C)$ of $\constset$. For each predicate symbol
  $p_{X,Y,Z,A,B,C}$ we consider a fresh predicate symbol
  $\overline{p}_{X,Y,A,B}$, of arity $\#\overline{p} \isdef \#p -
  \card{Z}$, and each predicate atom
  $p_{X,Y,Z,A,B,C}(t_1,\ldots,t_{\#p})$ occurring in either $\asys_4$
  or $\seqset_4$ is replaced by $\overline{p}_{X,Y,A,B}(t_{i_1},
  \ldots, t_{i_m})$, where $t_{i_1}, \ldots, t_{i_m}$ is the
  subsequence of $t_1,\ldots,t_{\#p}$ obtained by removing the terms
  from $\set{t_i \mid i \in Z}$ and each atom involving these terms is
  removed from $\asys_4$ and $\seqset_4$. Let the result of this
  transformation be denoted by $\aprob_n = (\asys_n,\seqset_n)$, with
  $\allocpar{\overline{p}_{X,Y,A,B}}{\asys_n} \isdef X$ and
  $\alloconst{\overline{p}_{X,Y,A,B}}{\asys_n} \isdef A$. Properties
  \ref{it3:normalized} and \ref{it4:normalized} follow from the
  definition of the rules of $\overline{p}_{X,Y,A,B}$ by an easy
  induction on the length of the unfolding. The equivalence between
  the validity of $\aprob_n$ and the validity of $\aprob_4$ is based
  on the following:

  \begin{fact}\label{fact:exist-redundant}
    Let $\phi$ be a symbolic heap occurring in a sequent from
    $\seqset_4$, $\phi \unfold{\asys_4}^* \psi$ be a predicate-free
    unfolding of $\phi$ and $p_{X,Y,Z,A,B,C}(t_1, \ldots, t_{\#p})$ be
    a predicate atom that occurs at some intermediate step of this
    predicate-free unfolding. Then each variable $t_i \in \fv{\psi}$,
    such that $i \in Z$, occurs existentially quantified in a
    subformula $\exists t_i ~.~ \Asterisk_{j=1}^n t_i \sneq u$ of
    $\psi$ and nowhere else.
  \end{fact}
  \proof{ Since $\fv{\phi} = \emptyset$, it must be the case that
    $x_i$ has been introduced as an existentially quantified variable
    by an intermediate unfolding step. We show, by induction on the
    length of the unfolding from the point where the variable was
    introduced that $t_i$ cannot occur in a points-to atom. \qed}

  Since $\locs$ is infinite, any formula $\exists x ~.~
  \Asterisk_{j=1}^n x \sneq u_j$ is trivially satisfied in any
  structure $(\astore,\aheap)$, such that $\{u_1,\ldots,u_n\} \in
  \dom(\astore)$. By Fact \ref{fact:exist-redundant}, it follows that
  eliminating the terms $\set{t_i \mid i \in Z}$ from each predicate
  atom $p_{X,Y,Z,A,B,C}(t_1, \ldots, t_{\#p})$ preserves equivalence. 
  
  \noindent(\ref{it5:normalized}) The exhaustive application of rules
  (\ref{eq:rule-split}) and (\ref{eq:rhs-split}), that add all possibe
  disequalities between existentially quantified variables and
  constants, ensures that Condition (\ref{it5:normalized}) is
  satisfied. Consequently, $\aprob_n$ is normalized.

  Assume now that $\aprob$ is \restricted, namely that each equational
  atom $t \bowtie u$ occurring in $\aprob$ is such that $\set{t,u}
  \cap \const \neq \emptyset$. Note that the transformations
  (\ref{eq:rule-split}) and (\ref{eq:rhs-split}) may introduce
  disequalities $x \sneq t'$, where $x$ is an existentially quantified
  variable. In the case where $\aprob$ is \restricted, we apply these
  rules only for $t \in \const$. Suppose that, after applying rules
  (\ref{eq:exists-subst}-\ref{eq:equal-remove}) exhaustively, there
  exist some equality $t \seq u$ in a rule, such that neither $t$ nor
  $u$ is an existentially quantified variable. But since $\aprob$ is
  \restricted, $\set{t,u} \cap \const \neq \emptyset$ and this rule
  will be eliminated by the disequalities introduced by the modified
  versions of the transformations (\ref{eq:rule-split}) and
  (\ref{eq:rhs-split}). Finally, if $\aprob$ is (strongly) established
  then $\aprob_n$ is (strongly) established, because the
  transformation does not introduce new existential quantifiers and
  preserves equivalence.

  Let us now compute the time complexity of the normalization
  procedure and the width of the output entailment problem. Observe
  that transformations (\ref{eq:exists-subst}--\ref{eq:rhs-split})
  either instantiate existentially quantified variables, add or remove
  equalities, thus they can be applied $\bigO(\size{\aprob})$ times,
  increasing the width of the problem by at most
  $\bigO(\size{\aprob})$.
  After the exhaustive application of transformations
  (\ref{eq:exists-subst}-\ref{eq:rhs-split}), the number of rules in
  $\asys$ and the number of sequents in $\seqset$ has increased by a
  factor of $2^{\probwidth{\aprob}}$ and the width of the problem by a
  linear factor. Then $\size{\aprob_1} = \bigO(\size{\aprob}\cdot
  2^{\probwidth{\aprob}})$ and $\probwidth{\aprob_1} =
  \bigO(\probwidth{\aprob})$.
  The transformation of step (\ref{it2:normalized}) increases the
  number of rules in $\asys_1$ by a factor of $2^\alpha =
  2^{\bigO(\probwidth{\aprob_1})} = 2^{\bigO(\probwidth{\aprob}^2)}$,
  where $\alpha = \max\{\#p \mid p(x_1,\ldots,x_{\#p})
  \Leftarrow_{\asys_1} \arule\} \leq \probwidth{\aprob}$ and does not
  change the width of the problem, i.e.\ $\size{\aprob_2} =
  \size{\aprob} \cdot 2^{\bigO(\probwidth{\aprob}^2)}$ and
  $\probwidth{\aprob_2} = \bigO(\probwidth{\aprob}^2)$. Next, going
  from $\aprob_2$ to $\aprob_3$ does not increase the bounds on the
  size or width of the problem and we trivially obtain
  $\size{\aprob_3} = \size{\aprob} \cdot
  2^{\bigO(\probwidth{\aprob}^2)}$ and $\probwidth{\aprob_3} =
  \bigO(\probwidth{\aprob}^2)$. Finally, going from $\aprob_3$ to
  $\aprob_4$ increases the size of the problem by a factor of
  $2^{3\alpha}\cdot 2^{3\card{\const}}$ and, because $\card{\const}
  \leq \probwidth{\aprob}$, by the definition of $\probwidth{\aprob}$,
  we obtain $\size{\aprob_n} = \size{\aprob} \cdot
  2^{\bigO(\probwidth{\aprob}^2)}$ and $\probwidth{\aprob_n} =
  \bigO(\probwidth{\aprob}^2)$. Finally, the entire procedure has to
  be repeated for each partition $\constpart$ of the set of constants
  $\const$. Since the number of partitions is $2^{\bigO(\card{\const}
    \cdot \log_2\card{\const})} = 2^{\bigO(\probwidth{\aprob} \cdot
    \log_2\probwidth{\aprob})}$, we obtain that the size of the result
  is $\size{\aprob} \cdot 2^{\bigO(\probwidth{\aprob}^2)}$. Since the
  increase in the size of the output problem is mirrored by the time
  required to obtain it, the execution of the procedure takes time
  $\size{\aprob} \cdot 2^{\bigO(\probwidth{\aprob}^2)}$. \qed}

\begin{example}
The entailment problem $\aprob = (\asys,~ \set{p(\mathsf{a},\mathsf{b}) \vdash \exists x,y~.~ q(x,y)})$ with:
\vspace*{-.5\baselineskip}
\[
\asys \isdef \left\{\begin{tabular}{rclcrcl}
$p(x,y)$ & $\Leftarrow$ &  $\exists z ~.~ x \mapsto z * p(z,y) * x \not \seq y$ & \quad &
$q(x,y)$ & $\Leftarrow$ & $\exists z~.~ x \mapsto y * q(y,z) * z \not \seq \mathsf{a} * z\not \seq \mathsf{b}$ \\
$p(x,y)$ & $\Leftarrow$  & $\exists z~.~ x \mapsto z$ & \quad & 
$q(x,y)$ & $\Leftarrow$ & $x \mapsto y$ \\[-2mm]
\end{tabular}\right\}
\]
may be transformed into $(\asys',~ \set{p_1() \vdash \exists x,y~.~ q_1(x,y), \exists x,y~.~ q_2(x,y)})$, where:
\vspace*{-.5\baselineskip}
\[
\asys' \isdef \left\{\begin{tabular}{rclcrcl}
$p_1()$ & $\Leftarrow$ &  $\exists z ~.~ \mathsf{a} \mapsto z * p_2(z) * z \not \seq \mathsf{a} * z\not \seq \mathsf{b}$ &  
$p_1()$ & $\Leftarrow$ &  $\mathsf{a} \mapsto \mathsf{b} * p_3()$ \\ 
$p_1()$ & $\Leftarrow$  & $\exists z~.~ \mathsf{a} \mapsto z$ &
$p_2(x)$ & $\Leftarrow$ &  $x \mapsto \mathsf{b} * p_3()$ \\ 
$p_2(x)$ & $\Leftarrow$ &  $\exists z ~.~ x \mapsto z * p_2(z) * z \not \seq \mathsf{a} * z \not \seq \mathsf{b}$ & 
$p_2(x)$ & $\Leftarrow$  & $\exists z~.~ x \mapsto z$ \\
$p_3()$ & $\Leftarrow$  & $\exists z~.~\mathsf{b} \mapsto z$ &
$q_1(x,y)$ & $\Leftarrow$ & $\exists z~.~ x \mapsto y * q_1(y,z) * z \not \seq \mathsf{a} * z\not \seq \mathsf{b}$ \\ 
$q_1(x,y)$ & $\Leftarrow$ & $\exists z~.~ x \mapsto y * q_2(y,z) * z \not \seq \mathsf{a} * z\not \seq \mathsf{b}$ & 
$q_2(x,y)$ & $\Leftarrow$ & $x \mapsto y$ \\[-2mm]
\end{tabular}\right\}
\]
The predicate atoms $p_1(), p_2(x)$ and $p_3()$ are equivalent to
$p(\mathsf{a},\mathsf{b})$, $p(x,\mathsf{b})$ and
$p(\mathsf{b},\mathsf{b})$, respectively. $q(x,y)$ is equivalent to
$q_1(x,y) \vee q_2(x,y)$.  Note that $p_2(x)$ is only used in a
context where $x\not \seq b$ holds, thus this atom may be omitted from
the rules of $p_2()$. Recall that $\mathsf{a}$ and $\mathsf{b}$ are
mapped to distinct locations, by Assumption
\ref{ass:dist-const}. \hfill$\blacksquare$
\end{example}

\longVersionOnly{

\section{Comparing \Restricted and Established Problems}
\label{sec:established}

}

We show that every established problem $\aprob$ can be reduced to
\arestricted problem in time linear in the size and exponential in the
width of the input, at the cost of a polynomial increase of its width: 

\additionalMaterial{Proof of Theorem
  \ref{thm:established-restricted}}{sec:dec-entailments}{app:established}{
  \longVersionOnly{ First, we show that every established entailment
    problem can be reduced to a strongly established entailment
    problem (Definition \ref{def:establishment}) that is, moreover,
    normalized: }
\begin{lemma}\label{lemma:strongly-established}
  Every established entailment problem $\aprob = (\asys,\seqset)$ can
  be reduced in time $\size{\aprob} \cdot
  2^{\bigO(\probwidth{\aprob}^2)}$ to a normalized and strongly
  established entailment problem $\aprob_e$, such that
  $\probwidth{\aprob_e} = \bigO(\probwidth{\aprob}^2)$.
\end{lemma}
\proof{ First, we use Lemma \ref{lemma:normalized} to reduce $\aprob$
  to an established normalized problem $\aprob_n =
  (\asys_n,\seqset_n)$ in time $\size{\aprob} \cdot
  2^{\bigO(\probwidth{\aprob}^2)}$, such that
  $\size{\aprob_n}=\size{\aprob} \cdot
  2^{\bigO(\probwidth{\aprob}^2)}$ and
  $\probwidth{\aprob_n}=\bigO(\probwidth{\aprob}^2)$. Second, given a
  symbolic heap $\phi$ and a variable $x$, we define the set of
  symbolic heaps   $\allocates{\phi}{x}$ recursively on the structure of
  $\phi$, as follows:
  \[\begin{array}{rcll}
  \allocates{t_1 \bowtie t_2}{x} & \ \isdef\  & \emptyset & \\
  \allocates{t_0 \mapsto (t_1,\ldots,t_\rank)}{x} & \isdef & \set{t_0 \mapsto (t_1,\ldots,t_\rank) * x \seq t_0} \\
  \allocates{p(t_1,\ldots,t_{\#p})}{x} & \isdef & \set{\overline{p}(x,t_1,\ldots,t_{\#p})} \\
  \allocates{\phi_1 * \phi_2}{x} & \isdef & \bigcup_{i=1,2} \set{\phi_i * \psi \mid \psi \in \allocates{\phi_{3-i}}{x}}
  \end{array}\]
  where $\overline{p}$ is a fresh predicate symbol not occurring in
  $\aprob$, of arity $\#\overline{p} \isdef \#p+1$ and the set of inductive
  rules is updated by replacing each rule $p(x_1,\ldots,x_{\#p})
  \Leftarrow_\asys \arule$ by the set of rules
  $\{\overline{p}(x_0,x_1,\ldots,x_{\#p}) \Leftarrow \psi \mid \psi
  \in \allocates{\arule}{x_0}\}$. It is straightforward to show by induction that
  if $(\astore,\aheap)$ is a structure such that $(\astore,\aheap)
  \models_\asys \psi$ for some $\psi \in \allocates{\phi}{x}$, then we
  have $\astore(x) \in \dom(\aheap)$. Observe that
  $\card{\allocates{\phi}{x}} \leq 2^{\size{\phi}}$ and $\size{\psi} =
  \bigO(\size{\phi})$, for each $\psi \in \allocates{\phi}{x}$.

  Let $\phi_0 \vdash_{\aprob_n} \phi_1, \ldots, \phi_n$ be a sequent
  from $\aprob_n$ and $(\astore,\aheap)$ be a structure such that
  $(\astore,\aheap) \models_{\asys_n} \phi_0$. By Definition
  \ref{def:entailment}, $\phi_0$ is quantifier-free. Assume that
  $\phi_1 = \exists x ~.~ \psi_1$ (the argument is repeated for all
  existential quantifiers occurring in $\phi_1, \ldots, \phi_n$).
  Note that, since $\aprob_n$ is normalized, $x$ occurs in a points-to
  or a predicate atom in $\phi_1$. This implies that $x$ necessarily
  occurs in a points-to atom in each symbolic heap $\varphi_1$
  obtained by a predicate-free unfolding $\phi_1 \unfold{\asys_n}^*
  \varphi_1$, by point (\ref{it3:normalized}) of Definition
  \ref{def:normalized}. Thus, $\astore'(x) \in \loc(\aheap)$, for each
  $x$-associate $\astore'$ of $\astore$ such that $(\astore',\aheap)
  \models \psi_1$. Since $\asys_n$ is established, each location from
  $\loc(\aheap)$ belongs to $\astore(\const) \cup \dom(\aheap)$, thus
  $\astore'(x) \in \astore(\const) \cup \dom(\aheap)$. Hence $\phi_1$
  can safely be replaced by the set of symbolic heaps
  $\set{\psi_1[t/x] \mid t \in \const} \cup \set{\exists x ~.~ \varphi
    \mid \varphi \in \allocates{\psi_1}{x}}$. Applying this
  transformation to each existentially quantified variable occurring
  in a sequent from $\aprob_n$ yields a strongly established problem
  $\aprob'$. Moreover, the reduction of $\aprob_n$ to $\aprob'$
  requires $\size{\aprob_n} \cdot 2^{\bigO(\probwidth{\aprob_n})} =
  \size{\aprob} \cdot 2^{\bigO(\probwidth{\aprob}^2)}$ time and the
  width of the outcome is $\probwidth{\aprob'} =
  \bigO(\probwidth{\aprob_n}) = \bigO(\probwidth{\aprob}^2)$.  \qed}}

\begin{theorem}\label{thm:established-restricted}
  Every established entailment problem $\aprob = (\asys, \seqset)$ can
  be reduced in time $\size{\aprob} \cdot
  2^{\bigO(\probwidth{\aprob}^2)}$ to normalized \arestricted problem
  $\aprob_r$, such that $\probwidth{\aprob_r} =
  \bigO(\probwidth{\aprob})$.
\end{theorem}
\optionalProof{Theorem
  \ref{thm:established-restricted}}{sec:dec-entailments}{ Lemma
  \ref{lemma:strongly-established}, we can reduce $\aprob$ to a
  normalized strongly established entailment problem $\aprob_e =
  (\asys_e, \seqset_e)$ in time $\size{\aprob} \cdot
  2^{\bigO(\probwidth{\aprob}^2)}$, such that $\probwidth{\aprob_e} =
  \bigO(\probwidth{\aprob}^2)$. Let $\phi \unfold{\asys_e}^* \varphi$
  be an arbitrary predicate-free unfolding of a symbolic heap $\phi$
  on the right-hand side of a sequent in $\seqset_e$, where $\varphi =
  \exists x_1 \ldots \exists x_n ~.~ \psi$ and $\psi$ is
  quantifier-free. Because $\aprob_e$ is normalized, there are no
  equalities in $\psi$. Let $x \sneq y$ be a disequality from $\psi$,
  where $\set{x,y} \cap \const = \emptyset$. By Definition
  \ref{def:entailment}, all variables from $\aprob_e$ are
  existentially quantified, thus it must be the case that $x,y \in
  \set{x_1,\ldots,x_n}$. Because $\aprob_e$ is strongly established,
  $\phi$ is $\asys_e$-established, thus both $x$ and $y$ are allocated
  in $\psi$. Moreover, since there are no equalities in $\psi$, there
  must exist two distinct points-to atoms $x \mapsto
  (t_1,\ldots,t_\rank)$ and $y \mapsto (u_1,\ldots,u_\rank)$ in $\psi$
  such that, $(\astore,\aheap) \models_{\asys_e} \phi$ implies
  $(\astore',\aheap') \models_{\asys_e} x \mapsto (t_1,\ldots,t_\rank)
  * y \mapsto (u_1,\ldots,u_\rank)$, for any structure
  $(\astore,\aheap)$, for some heap $\aheap' \subseteq \aheap$ and
  $\astore'$ is a $(x_1,\ldots,x_n)$-associate of $\astore$. But then
  $(\astore',\emptyset) \models_{\asys_e} x \sneq y$ and, since the
  choice of the structure $(\astore,\aheap)$ was arbitrary, we can
  remove any disequality $x \sneq y$ such that $\set{x,y} \cap
  \constset = \emptyset$ from $\aprob_e$. This transformation takes
  time $\bigO(\size{\aprob_e}) = \size{\aprob} \cdot
  2^{\bigO(\probwidth{\aprob}^2)}$ and does not increase the width of
  the problem. The outcome of is \arestricted entailment
  problem. \qed}
 
 \shortVersionOnly{ \noindent The class of \restricted problems is
   more general than the class of established problems, in the
   following sense: for each established problem
   $\aprob=(\asys,\seqset)$, the treewidth of each $\asys$-model of a
   $\asys$-established symbolic heap $\phi$ is bounded by
   $\probwidth{\aprob}$ \cite{IosifRogalewiczSimacek13}, while
   \restricted symbolic heaps may have infinite sequences of models
   with strictly increasing treewidth:
   \begin{example}\label{ex:grids}
     Consider the set of rules $\{\mathsf{lls}(x,y) \Leftarrow x
     \mapsto (y,\nil),~ \mathsf{lls}(x,y) \Leftarrow \exists z \exists
     v ~.~ x \mapsto (z,v) * \mathsf{lls}(z,y)\}$. The existentially
     quantified variable $v$ in the second rule in never allocated in
     any predicate-free unfolding of $\mathsf{lls(a,b)}$, thus the set
     of rules is not established. However, it is trivially
     \restricted, because no equational atoms occur within the
     rules. Among the models of $\mathsf{lls(a,b)}$, there are all
     $n\times n$-square grid structures, known to have treewidth $n$,
     for $n > 1$ \cite{RobertsonSeymour84}. \hfill$\blacksquare$
   \end{example}
} 
 \longVersionOnly{ 

   As a concluding remark, we show that the class of \restricted is
   more general than the class of established entailment problems, in
   the following sense. Let $\aprob = (\asys,\aprob)$ be an
   established entailment problem. Each structure $(\astore,\aheap)$
   can be associated with a unique integer $\tw(\astore,\aheap) \geq
   0$, called its \emph{treewidth}. The formal definition of the
   treewidth is given below, for reasons of self-containment, however
   the argument can be followed without it.

   A \emph{labeled tree} is a graph $(N,E,\lambda)$, where $N$ is a
   finite set of nodes, $E \subseteq N \times N$ is an undirected edge
   relation and $\lambda : N \rightarrow 2^\locs$ is a labeling
   function. Moreover, there is a unique node $r \in N$, such that for
   each node $n \in N \setminus \set{r}$ there exists a unique path
   from $r$ to $n$. A set of nodes $M \subseteq N$ is said to be
   \emph{connected} if there is a path between any two nodes in the
   set.
   \begin{definition}\label{def:tw} 
     Given a structure $(\astore,\aheap)$, a tree decomposition of
     $(\astore,\aheap)$ is a labeled tree $T=(N,E,\lambda)$, such
     that: \begin{compactenum}
     \item for each $\ell \in \loc(\aheap)$, the set $\set{n \in N \mid
       \ell \in \lambda(n)}$ is nonempty and connected,
     \item for each $\ell_1 \in \dom(\aheap)$ and $\ell_2 \in
       \aheap(\ell_1)$, we have $\ell_1,\ell_2 \in \lambda(n)$, for some
       $n \in N$.
     \end{compactenum}
     The \emph{treewidth} of $T$ is $\tw(T) \isdef
     \max\set{\card{\lambda(n)} \mid n \in N} - 1$ and the
     \emph{treewidth} of $\aheap$ is $\tw(\astore,\aheap) \isdef
     \min\{\tw(T) \mid T \text{ is a tree decomposition of }
     (\astore,\aheap)\}$.
   \end{definition}
   As shown in \cite{IosifRogalewiczSimacek13}, the treewidth of each
   $\asys$-model of a $\asys$-established symbolic heap $\phi$ is
   bounded by $\probwidth{\aprob}$. However, if $\aprob$ is
   \restricted but not established, there can be infinitely many
   $\asys$-models $(\astore_1, \aheap_1), (\astore_2, \aheap_2),
   \ldots$ of \arestricted symbolic heap, such that $\tw(\astore_1,
   \aheap_1) < \tw(\astore_2, \aheap_2) < \ldots$, as shown by the
   example below:
   \begin{example}\label{ex:grids}
     Consider the following set of rules:
     \[\begin{array}{rcl}
     \mathsf{lls}(x,y) & \Leftarrow & x \mapsto (y,\nil) \\
     \mathsf{lls}(x,y) & \Leftarrow & \exists z \exists v ~.~ x \mapsto (z,v) * \mathsf{lls}(z,y)
     \end{array}\]
     \begin{figure}[htb]
       \centerline{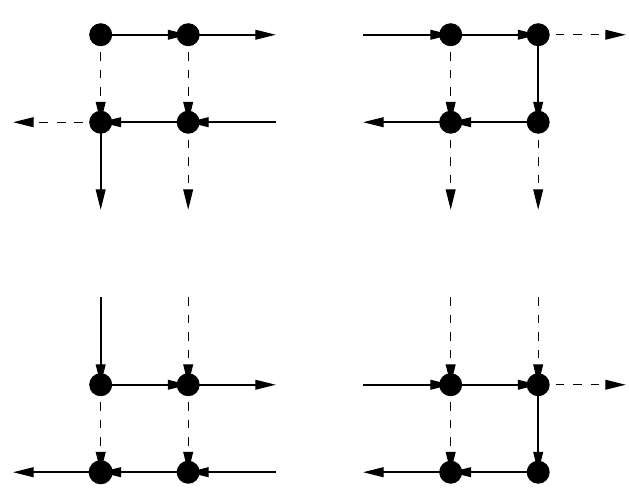}
       \caption{\label{fig:grids}}
     \end{figure}
     The existentially quantified variable $v$ in the second rule in
     never allocated in any predicate-free unfolding of
     $\mathsf{lls}(a,b)$, thus the set of rules is not
     established. However, it is trivially \restricted, because no
     equational atoms occur within the rules. Among the models of
     $\mathsf{lls}(a,b)$, there is an infinite set of $n\times
     n$-square grid structures depicted in Figure \ref{fig:grids}, for
     all $n > 1$. It is known that the treewdith of an $n \times
     n$-square grid is $n$
     \cite{RobertsonSeymour84}. \hfill$\blacksquare$
   \end{example}
 }

\section{Normal Structures}
\label{sec:injective-normal}

The decidability of \restricted entailment problems relies on the fact
that, to prove the validity of a sequent, it is sufficient to consider
only a certain class of structures, called \emph{normal}, that require
the variables not mapped to the same location as a constant to be
mapped to pairwise distinct locations:
\begin{definition}\label{def:normal}
  A structure $(\astore,\aheap)$ is a \emph{normal $\asys$-model} of a
  symbolic heap $\phi$ iff there exists:\begin{compactenum}
  \item\label{it1:normal} a predicate-free unfolding $\phi
    \unfold{\asys} \exists \vec{x} ~.~ \psi$, where $\psi$ is
    quantifier-free, and
  \item\label{it2:normal} an $\vec{x}$-associate $\aastore$ of
    $\astore$, such that $(\aastore,\aheap) \models_\asys \psi$ and
    $\aastore(x)=\aastore(y) \wedge x \neq y \Rightarrow \aastore(x)
    \in \astore(\const)$, for all $x,y \in \fv{\psi}$.
  \end{compactenum}
\end{definition}

\begin{example}
\label{ex:normal}
Consider the formula $\varphi = p(x_1) * p(x_2)$, with $p(x)
\Leftarrow_{\asys} \exists z~.~ x \mapsto z$ and $\const = \{
\mathsf{a} \}$.  Then the structures: $(\astore,\aheap)$ and
$(\astore,\aheap')$ with $\astore = \{ (x_1,\ell_1),(x_2,\ell_2),
(\mathsf{a},\ell_3) \}$, $\aheap = \{ (\ell_1,\ell_3), (\ell_2,\ell_3)
\}$ and $\aheap' = \{ (\ell_1,\ell_4), (\ell_2,\ell_5) \}$ are normal
models of $\varphi$.  On the other hand, if $\aheap'' = \{
(\ell_1,\ell_4), (\ell_2,\ell_4) \}$ (with $\ell_4 \not = \ell_3$)
then $(\astore,\aheap'')$ is a model of $\varphi$ but it is not
normal, because any associate of $\astore$ will map the existentials
from the predicate-free unfolding of $p(x_1) * p(x_2)$ into the same
location, different from $\astore(\mathsf{a})$. \hfill$\blacksquare$
\end{example}

Since the left-hand side symbolic heap $\phi$ of each sequent $\phi
\vdash \psi_1, \ldots, \psi_n$ is quantifier-free and has no free
variables (Definition \ref{def:entailment}) and moreover, by
Assumption \ref{ass:dist-const}, every constant is associated a
distinct location, to check the validity of a sequent it is enough to
consider only structures with injective stores. We say that a
structure $(\aistore,\aheap)$ is \emph{injective} if the store
$\aistore$ is injective. As a syntactic convention, by stacking a dot
on the symbol denoting the store, we mean that the store is
injective. \longVersionOnly{Moreover, we denote by $\phi
  \imodels{\asys} \psi$ the fact that each injective $\asys$-model of
  $\phi$ is a model of $\psi$.}

\additionalMaterial{Additional Material for Normal Structures}{sec:injective-normal}{app:injective-normal}{

\longVersionOnly{
A natural question is: are normal models closed under the composition
induced by the separating conjunction? If $(\astore,\aheap)$ is a
normal $\asys$-model of $\phi_1 * \phi_2$, there exist normal
$\asys$-models $(\astore, \aheap_i)$ of $\phi_i$, for $i=1,2$, such
that $\aheap = \aheap_1 \uplus \aheap_2$. The converse is not true, in
general, and requires further conditions:
}

\begin{definition}\label{def:companions}
  Given symbolic heaps $\phi_1,\phi_2 \in \shk{\rank}$, a pair of
  structures $\tuple{(\astore_1, \aheap_1),(\astore_2, \aheap_2)}$ is
  a \emph{normal $\asys$-companion for $(\phi_1,\phi_2)$} iff
  $(\astore_i, \aheap_i)$ is a normal $\asys$-model of $\phi_i$, for
  $i=1,2$ and: \begin{compactenum}
  \item\label{it1:def:companions} $\aastore_1(t) = \aastore_2(t)$, for
    each term $t \in \fv{\psi_1} \cap \fv{\psi_2} \cup \const$,
  \item\label{it2:def:companions} $\aastore_i(\vec{x}_i) \cap
    \aastore_{3-i}(\fv{\psi_{3-i}}) \subseteq \aastore_i(\const)$, for $i=1,2$,
  \end{compactenum}
  where $\phi_i \unfold{\asys}^* \exists \vec{x}_i ~.~ \psi_i$ are the
  predicate-free unfoldings and $\aastore_i$ is the
  $\vec{x}_i$-associate of $\astore_i$ satisfying conditions
  (\ref{it1:normal}) and (\ref{it2:normal}) of Definition
  \ref{def:normal}, for $i=1,2$, respectively. The normal
  $\asys$-companion $\tuple{(\astore_1, \aheap_1),(\astore_2,
    \aheap_2)}$ is, moreover, \emph{injective} iff $\astore_1$ and
  $\astore_2$ are injective and $\astore_1(\fv{\phi_1} \setminus
  \fv{\phi_2}) \cap \astore_2(\fv{\phi_2} \setminus \fv{\phi_1}) =
  \emptyset$.
\end{definition}

\begin{lemma}\label{lemma:companions}
  Given symbolic heaps $\phi_1,\phi_2 \in \shk{\rank}$, a structure
  $(\astore,\aheap)$ is a (injective) normal $\asys$-model of $\phi_1
  * \phi_2$ iff there exists a (injective) normal $\asys$-companion
  $\tuple{(\astore_1, \aheap_1), (\astore_2,\aheap_2)}$ for $(\phi_1,
  \phi_2)$, such that $\aheap=\aheap_1\uplus\aheap_2$.
\end{lemma}
\proof{``$\Rightarrow$'' Let
  $(\astore,\aheap)$ be a normal $\asys$-model of $\phi_1 *
  \phi_2$. Then there exists a predicate-free unfolding $\phi_1 *
  \phi_2 \unfold{\asys}^* \exists \vec{x}_1 ~.~ \psi_1 * \exists
  \vec{x}_2 ~.~ \psi_2$ such that $\psi_1$ and $\psi_2$ are
  quantifier-free and $(\astore,\aheap) \models \exists \vec{x}_1 ~.~
  \psi_1 * \exists \vec{x}_2 ~.~ \psi_2$. By $\alpha$-renaming if
  necessary, we can assume that $\vec{x}_i \cap \fv{\psi_{3-i}} =
  \emptyset$, for $i=1,2$, thus $(\astore,\aheap) \models \exists
  \vec{x}_1 \exists \vec{x}_2 ~.~ \psi_1 * \psi_2$. Hence there exist
  an $(\vec{x}_1 \cup \vec{x}_2)$-associate $\aastore$ of $\astore$
  and two disjoint heaps $\aheap_1$ and $\aheap_2$, such that $\aheap
  = \aheap_1 \uplus \aheap_2$ and $(\aastore,\aheap_i) \models
  \psi_i$, for $i=1,2$.  Let $\astore_i \isdef \astore$, for $i=1,2$,
  so that $\astore=\astore_1 \cup \astore_2$. By considering the
  $\vec{x}_i$-associate of $\astore$ defined as the restriction of
  $\aastore$ to $\vec{x}_i \cup \dom(\astore)$ and using the fact that
  $(\astore,\aheap)$ is a normal $\asys$-model of $\phi_1 * \phi_2$,
  it is easy to check that $(\astore_i,\aheap_i)$ is a normal
  $\asys$-model of $\phi_i$.
 Further, points (\ref{it1:def:companions}) and
 (\ref{it2:def:companions}) of Definition \ref{def:companions} are
 easy checks. Finally, if $\astore$ is injective then trivially
 $\astore_1$ and $\astore_2$ are injective and
 $\astore_1(\term{\phi_1} \setminus \term{\phi_2}) \cap
 \astore_2(\term{\phi_2} \setminus \term{\phi_1}) =
 \astore(\term{\phi_1} \setminus \term{\phi_2}) \cap
 \astore(\term{\phi_2} \setminus \term{\phi_1}) = \emptyset$.

  \noindent''$\Leftarrow$'' If $(\astore_i, \aheap_i)$ is a normal
  $\asys$-model of $\phi_i$, then there exist predicate-free
  unfoldings $\phi_i \unfold{\asys}^* \exists \vec{x}_i ~.~ \psi_i$
  and $\vec{x}_i$-associates $\aastore_i$ of $\astore_i$, that satisfy
  the points (\ref{it1:normal}) and (\ref{it2:normal}) of Definition
  \ref{def:normal}. By an $\alpha$-renaming if necessary, we assume
  that $\vec{x}_1 \cap \vec{x}_2 = \emptyset$. Then $\phi_1 * \phi_2
  \unfold{\asys}^* \exists \vec{x}_1 ~.~ \psi_1 * \exists \vec{x}_2
  ~.~ \psi_2$ is a predicate-free unfolding. Let $\astore'_i$ and
  $\aastore'_i$ be the restrictions of $\astore_i$ and $\aastore_i$ to
  $\term{\phi_i}$ and $\term{\psi_i}$ for $i=1,2$, respectively. By
  point (\ref{it1:def:companions}) of Definition \ref{def:companions},
  $\astore \isdef \astore'_1 \cup \astore'_2$ is a well-defined store
  and, since $\vec{x}_1 \cap \vec{x}_2 = \emptyset$, we obtain that
  $\aastore \isdef \aastore'_1 \cup \aastore'_2$ is a well-defined
  $(\vec{x}_1 \cup \vec{x}_2)$-associate of $\astore$. To show that
  $(\astore,\aheap_1 \uplus \aheap_2)$ is a normal $\asys$-model of
  $\phi_1 * \phi_2$, let $t_1, t_2 \in \term{\psi_1} \cup
  \term{\psi_2}$ be distinct terms such that $\aastore(t_1) =
  \aastore(t_2)$ and suppose, for a contradiction, that $\aastore(t_1)
  \not \in \aastore(\const)$. Since $(\astore_i, \aheap_i)$ is a
  normal $\asys$-model of $\phi_i$, for $i=1,2$, the only interesting
  cases are $t_i \in \term{\psi_i} \setminus \term{\psi_{3-i}}$ and
  $t_i \in \term{\psi_{3-i}} \setminus \term{\psi_{i}}$. Assume $t_i
  \in \term{\psi_i} \setminus \term{\psi_{3-i}}$ for $i=1,2$, the
  other case is symmetric. Since $t_i \not\in \cst{\psi_1 * \psi_2}$,
  it must be the case that $t_i \in \vec{x}_i$, for $i=1,2$. Then
  $\aastore_1(t_1)=\aastore(t_1)=\aastore(t_2)=\aastore_2(t_2)$, which
  contradicts point (\ref{it2:def:companions}) of Definition
  \ref{def:companions}. Finally, it is easy to check that $\astore =
  \astore'_1 \cup \astore'_2$ is injective, provided that $\astore_1$
  and $\astore_2$ are injective and that $\astore_1(\term{\phi_1}
  \setminus \term{\phi_2}) \cap \astore_2(\term{\phi_2} \setminus
  \term{\phi_1}) = \astore'_1(\term{\phi_1} \setminus \term{\phi_2})
  \cap \astore'_2(\term{\phi_2} \setminus \term{\phi_1}) =
  \emptyset$. \qed}

The following lemma states an important property of normal
$\asys$-models, that will be used to build abstract composition
operators, needed to define a finite-range abstraction of an infinite
set normal structures\longVersionOnly{ (see \S\ref{sec:composition})}. 
\begin{lemma}\label{lemma:frontier}
  Given symbolic heaps $\phi_1, \phi_2 \in \shk{\rank}$ and
  $\tuple{(\aistore,\aheap_1), (\aistore,\aheap_2)}$ an injective
  normal $\asys$-companion for $(\phi_1,\phi_2)$, we have
  $\front(\aheap_1,\aheap_2) \subseteq \aistore\left(\fv{\phi_1} \cap
  \fv{\phi_2} \cup \const\right)$.
\end{lemma}
\proof{Let $\ell \in
  \front(\aheap_1, \aheap_2) = \loc(\aheap_1) \cap \loc(\aheap_2)$ be
  a location, $\phi_i \unfold{\asys}^* \exists \vec{x}_i ~.~ \psi_i$
  be predicate-free unfoldings and $\aastore_i$ be the
  $\vec{x}_i$-associates of $\aistore$ that satisfy points
  (\ref{it1:def:companions}) and (\ref{it2:def:companions}) of
  Definition \ref{def:companions}, such that $(\aastore_i, \aheap_i)
  \models \psi_i$, for $i=1,2$. By $\alpha$-renaming, if necessary, we
  assume w.l.o.g. that $\vec{x}_i \cap \fv{\psi_{3-i}} = \emptyset$,
  for $i = 1,2$. Because $\ell \in \loc(\aheap_i)$, there exist
  points-to atoms $t^i_0 \mapsto (t^i_1,\ldots,t^i_\rank)$ in
  $\psi_i$, such that $\ell = \aastore_1(t^1_{i_1}) =
  \aastore_2(t^2_{i_2})$, for some $i_1, i_2 \in \interv{0}{\rank}$
  and all $i=1,2$. We distinguish two cases: \begin{compactitem}
  \item if $t^1_{i_1} \in \term{\phi_1}$ and $t^2_{i_2} \in
    \term{\phi_2}$, since $\aastore_i$ is a $\vec{x}_i$-associate of
    $\aistore$, $\aastore_i$ and $\aistore$ agree over
    $\term{\phi_i}$, for $i=1,2$, we obtain $\aistore(t^1_{i_1}) =
    \aastore_1(t^1_{i_1}) = \aastore_2(t^2_{i_2}) =
    \aistore(t^2_{i_2})$, thus $t^1_{i_1} = t^2_{i_2}$, because
    $\aistore$ is injective, hence $\ell \in \aistore(\term{\phi_1}
    \cap \term{\phi_2}) \subseteq \aistore(\fv{\phi_1} \cap
    \fv{\phi_2} \cup \const)$.
  \item else $t^1_{i_1} \in \term{\psi_1} \setminus \term{\phi_1} =
    \vec{x}_1 \cup \const$ (the case $t^2_{i_2} \in \term{\psi_2}
    \setminus \term{\phi_2}$ is symmetric). If $t^1_{i_1} \in \const$,
    we obtain $\ell = \aastore_1(t^1_{i_1}) = \aistore(t^1_{i_1}) \in
    \aistore(\const)$, because $\const \subseteq \dom(\aistore)$ and
    $\aastore$ agrees with $\aistore$ over $\const$. Else $t^1_{i_1}
    \in \vec{x}_1$ and we distinguish two cases: \begin{compactitem}
    \item if $t^2_{i_2} \in \cst{\psi_2}$, we obtain
      $\ell=\aaistore_2(t^2_{i_2})=\aistore(t^2_{i_2}) \in
      \aistore(\const)$, by the above argument.
    \item else $t^2_{i_2} \in \fv{\psi_2}$ and $\aastore_1(t^1_{i_1})
      = \aastore_2(t^2_{i_2}) \in \aastore(\const)$ by point
      (\ref{it2:def:companions})  of Definition
      \ref{def:companions}. \qed
    \end{compactitem}
\end{compactitem}}


\begin{example}
Consider the structures defined in Example \ref{ex:normal}.
The structure $(\astore,\aheap)$ is a normal model of $p(x_1) * p(x_2)$:
we have $(\astore,\aheap_i) \models p(x_i)$ with $\aheap_i = (\ell_i \mapsto \ell_3)$ (for $i = 1,2$), 
$\aheap = \aheap_1 \uplus \aheap_2$
and $\front(\aheap_1,\aheap_2) = \{ \ell_3 \} \subseteq \aistore(\const)$.
Similarly,
$(\astore,\aheap')$ is a normal model of $p(x_1) * p(x_2)$, 
$(\astore,\aheap_i') \models p(x_i)$ with $\aheap_i' = (\ell_i \mapsto \ell_{3+i})$ (for $i = 1,2$), 
$\aheap' = \aheap_1' \uplus \aheap_2'$
and $\front(\aheap_1',\aheap_2') = \emptyset$.
On the other hand,  $(\aistore,\aheap'')$  is not normal: we have
$(\astore,\aheap_i'') \models p(x_i)$ with $\aheap_i'' = (\ell_i \mapsto \ell_{4})$ (for $i = 1,2$), 
$\aheap'' = \aheap_1'' \uplus \aheap_2''$
and $\front(\aheap''_1,\aheap_2'') = \{ \ell_4 \} \not \subseteq \astore\left(\fv{p(x_1)} \cap
  \fv{p(x_2)} \cup \const\right)= \{ \ell_3\}$.
\end{example}
}

The key property of normal structures is that validity of \restricted
entailment problems can be checked considering only (injective) normal
structures. The intuition is that, since the (dis-)equalities
occurring in the considered formula involve a constant, it is
sufficient to assume that all the existential variables not equal to a
constant are mapped to pairwise distinct locations, as all other
structures can be obtained from such structures by applying a morphism
that preserves the truth value of the considered
formul{\ae}\shortVersionOnly{\footnote{See Appendices
    \ref{app:injective-normal} and \ref{app:normalproof} for more
    details.}}.

\putInAppendix{
The proof of this result (Lemma
  \ref{lemma:normal-entailment}) relies on the following definition
  and lemmas.

\begin{definition}\label{def:compatible}
  A total function $\amorph : \locs \rightarrow \locs$ is
  \emph{compatible} with a structure $(\astore,\aheap)$ if and only
  if, for all $\ell_1, \ell_2 \in \locs$ such that either
  $\ell_1,\ell_2 \in \dom(\aheap)$ or $\ell_1 \in \astore(\const)$, if
  $\amorph(\ell_1) = \amorph(\ell_2)$ then $\ell_1=\ell_2$. We define
  $\amorph(\aheap) \isdef \set{\langle\amorph(\ell), (\amorph(\ell_1),
    \ldots, \amorph(\ell_\rank))\rangle \mid \aheap(\ell) = (\ell_1, \ldots,
    \ell_\rank)}$, whenever $\gamma$ is compatible with
  $(\astore,\aheap)$.
\end{definition}

\begin{lemma}\label{lemma:restricted-unfolding}
  Let $\asys$ be \arestricted (resp.\ normalized) set of rules and
  $\phi$ be \arestricted formula. Then, each unfolding $\psi$ of
  $\phi$ is \restricted (resp.\ normalized).
\end{lemma}
\proof{The proof is by induction on the length of the unfolding
  sequence $\phi \unfold{\asys}^* \psi$. \qed}

\begin{lemma}\label{lemma:compatible}
  If $\asys$ is \arestricted set of rules, $\phi$ is \arestricted
  formula and $(\astore,\aheap)$ is an $\asys$-model of $\phi$, then
  for any total function $\amorph$ compatible with $(\astore,\aheap)$,
  the following hold: \begin{inparaenum}[(1)]
  \item\label{it2:compatible} $\amorph(\aheap)$ is a heap,
  \item\label{it3:compatible} $(\amorph\circ\astore,\amorph(\aheap))
    \models_\asys \phi$.
  \end{inparaenum}
\end{lemma}
\proof{ (\ref{it2:compatible}) The set $\set{\amorph(\ell) \mid \ell
    \in \dom(\aheap)}$ is finite, because $\dom(\aheap)$ is
  finite. Consider two tuples $\langle\amorph(\ell), (\amorph(\ell_1),
  \ldots, \amorph(\ell_\rank))\rangle$ and $\langle\amorph(\ell'),
  (\amorph(\ell_1'), \ldots, \amorph(\ell_\rank'))\rangle \in
  \amorph(\aheap)$ and assume that $\amorph(\ell) =
  \amorph(\ell')$. Then since $\amorph$ is compatible with
  $(\astore,\aheap)$, necessarily $\ell = \ell'$. Since $\aheap$ is a
  partial function, we have $(\ell_1, \ldots, \ell_\rank) = (\ell_1',
  \ldots, \ell_\rank')$, so that $\amorph(\aheap)$ is also a finite
  partial function.
  
  \noindent(\ref{it3:compatible}) If $(\astore,\aheap) \models_\asys
  \phi$ then there exists a predicate-free unfolding $\phi
  \unfold{\asys} \psi = \exists \vec{x} ~.~ \Asterisk_{i=1}^n t_i \seq
  u_i * \Asterisk_{i=1}^m t'_i \sneq u'_i * \Asterisk_{i=1}^k x_i
  \mapsto (t^i_1, \ldots, t^i_\rank)$, such that $(\aastore,\aheap)
  \models \psi$, for an $\vec{x}$-associate $\aastore$ of
  $\astore$. Note that $\amorph\circ\aastore$ is an
  $\vec{x}$-associate of $\amorph\circ\astore$, because $\amorph$ is
  total. Moreover, because $\phi$ and $\asys$ are both \restricted, by
  Lemma \ref{lemma:restricted-unfolding}, $\psi$ is \restricted, thus
  we can assume that $t_i \in \const$, for all $i \in \interv{1}{n}$
  and that $t'_i \in \const$, for all $i \in \interv{1}{m}$. We
  consider the three types of atoms from $\psi$ below:
  \begin{compactitem}
  \item For any $i \in \interv{1}{n}$, since $(\aastore,\emptyset)
    \models t_i \seq u_i$, we have $\aastore(t_i)=\aastore(u_i)$, thus
    $\amorph(\aastore(t_i))=\amorph(\aastore(s_i))$, leading to
    $(\amorph\circ\aastore,\emptyset) \models t_i \seq u_i$.
  \item For any $i \in \interv{1}{m}$, since $(\aastore,\emptyset)
    \models t'_i \sneq u'_i$, we have $\aastore(t'_i) \neq
    \aastore(u'_i)$. Because $t'_i \in \const$ and $(\astore,\aheap)
    \models \phi$, we have $t'_i \in \dom(\astore)$ and
    $\aastore(t'_i) = \astore(t'_i) \in \astore(\const)$. By
    Definition \ref{def:compatible}, we obtain
    $\amorph(\aastore(t'_i)) \neq \amorph(\aastore(u'_i))$, thus
    $(\amorph\circ\aastore,\emptyset) \models t'_i \sneq u'_i$.
  \item If $(\aastore,\aheap) \models \Asterisk_{i=1}^k x_i \mapsto
    (t^i_1, \ldots, t^i_\rank)$ then $\aastore(x_1), \ldots,
    \aastore(x_k)$ are pairwise distinct and $\dom(\aheap) =
    \{\aastore(x_1), \ldots, \aastore(x_k)\}$. Since $\aastore(x_1),
    \ldots, \aastore(x_k) \in \dom(\aheap)$, by Definition
    \ref{def:compatible}, we obtain that $\amorph(\aastore(x_1)),
    \ldots, \amorph(\aastore(x_k))$ are pairwise distinct and
    $\dom(\amorph(\aheap)) = \set{\amorph(\aastore(x_1)), \ldots,
      \amorph(\aastore(x_k))}$. We have $\aheap(\aastore(x_i)) =
    (\aastore(t^i_1), \ldots, \aastore(t^i_\rank))$, thus
    $\amorph(\aheap)(\aastore(x_i)) = (\amorph(\aastore(t^i_1)),
    \ldots, \amorph(\aastore(t^i_\rank)))$, for each $i \in
    \interv{1}{k}$, by Definition \ref{def:compatible} and
    $(\amorph\circ\aastore,\amorph(\aheap)) \models \Asterisk_{i=1}^k
    x_i \mapsto (t^i_1, \ldots, t^i_\rank)$. \qed
  \end{compactitem}}
}

\begin{lemma}\label{lemma:normal-entailment}
  Let $\aprob = (\asys, \seqset)$ be a normalized and \restricted
  entailment problem and let $\phi \vdash_\aprob \psi_1, \ldots,
  \psi_n$ be a sequent. Then $\phi \vdash_\aprob \psi_1, \ldots,
  \psi_n$ is valid for $\asys$ iff $(\aistore, \aheap) \models_\asys
  \bigvee_{i=1}^n \psi_i$, for each normal injective $\asys$-model
  $(\aistore,\aheap)$ of $\phi$.
\end{lemma}
\optionalProof{Lemma \ref{lemma:normal-entailment}}{sec:injective-normal}{
\label{app:normalproof} This
    direction is trivial. ``$\Leftarrow$'' Let $(\aistore,\aheap)$ be
    an injective $\asys$-model of $\phi$.  Then by Lemma
    \ref{lemma:restricted-unfolding}, there exists a predicate-free
    unfolding $\phi \unfold{\asys}^* \exists \vec{x} ~.~ \varphi$,
    where $\varphi = \Asterisk_{i=1}^m t_i \sneq u_i *
    \Asterisk_{i=1}^k x_i \mapsto (t^i_1, \ldots, t^i_\rank)$ is
    \restricted and normalized, and an $\vec{x}$-associate $\aastore$
    of $\aistore$ such that $(\aastore,\aheap) \models \varphi$.  Note
    that $\varphi$ contains no equalities since it is normalized and,
    since it is \restricted, we can assume that $t_i \in \const$, for
    all $i \in \interv{1}{m}$. We consider a store $\astore' :
    \dom(\aastore) \rightarrow \locs$ that satisfies the following
    hypothesis: \begin{compactenum}[(a)]
    \item\label{it1:normal-entailment} $\astore'(t) = \aastore(t)$,
      for each $t \in \dom(\aastore)$ such that $\aastore(t) \in
      \aastore(\const)$,
    \item\label{it2:normal-entailment} $\astore'(t) \neq \astore'(u)$,
      for all terms $t \neq u \in \dom(\aastore)$ such that
      $\aastore(t) \not\in \aastore(\const)$ or $\aastore(u) \not\in
      \aastore(\const)$.
    \end{compactenum}
    Note that such a store exists because $\locs$ is infinite, thus
    all terms that are not already mapped by $\aastore$ into locations
    from $\aastore(\const)$ can be mapped to pairwise distinct
    locations, not occurring in $\aastore(\const)$. Then we define the
    heap $\aheap' \isdef \{\langle\astore'(x_i), (\astore'(t^i_1),
    \ldots, \astore'(t^i_\rank))\rangle \mid i \in
    \interv{1}{k}\}$. To prove that $\aheap'$ is a well-defined heap,
    first note that the set $\set{\astore'(x_i) \mid i \in
      \interv{1}{k}}$ is finite and suppose, for a contradiction that
    $\astore'(x_i)=\astore'(x_j)$, for some $i \neq j \in
    \interv{1}{k}$. By point (\ref{it2:normal-entailment}), it must be
    the case that $\aastore(x_i), \aastore(x_j) \in \aastore(\const)$,
    in which case we obtain
    $\aastore(x_i)=\astore'(x_i)=\astore'(x_j)=\aastore(x_j)$, by
    point (\ref{it1:normal-entailment}), thus contradicting the fact
    that $(\aastore,\aheap) \models \Asterisk_{i=1}^k x_i \mapsto
    (t^i_1, \ldots, t^i_\rank)$. Hence the locations
    $\set{\astore'(x_i) \mid i \in \interv{1}{k}}$ are pairwise
    distinct and $\aheap'$ is a finite partial function. We prove next
    that $(\astore',\aheap') \models \varphi$, considering each type
    of atom in $\varphi$: \begin{compactitem}
    \item for any $i \in \interv{1}{m}$, since $(\aastore,\emptyset)
      \models t_i \sneq u_i$, we have $\aastore(t_i) \neq
      \aastore(u_i)$ and, since $t_i \in \const$, we obtain
      $\astore'(t_i)=\aastore(t_i) \in \aastore(\const)$. We
      distinguish the following cases: \begin{compactitem}
      \item if $\aastore(u_i) \in \aastore(\const)$ then
        $\astore'(u_i)=\aastore(u_i) \neq
        \aastore(t_i)=\astore'(t_i)$, by point
        (\ref{it1:normal-entailment}),
      \item otherwise, $\aastore(u_i) \not\in \aastore(\const)$ and
        $\astore'(t_i) \neq \astore'(u_i)$, by point
        (\ref{it2:normal-entailment}).
      \end{compactitem}
      In both cases, we have $(\astore',\emptyset) \models t_i \sneq
      u_i$.
  \item $(\astore',\aheap') \models \Asterisk_{i=1}^k x_i \mapsto
    (t^i_1, \ldots, t^i_\rank)$, by the definition of $\aheap'$.
  \end{compactitem}
  Let $\astore''$ be the restriction of $\astore'$ to
  $\dom(\aistore)$. By point (\ref{it2:normal-entailment}),
  $(\astore'',\aheap')$ is an injective normal $\asys$-model of
  $\phi$, according to Definition \ref{def:normal} (simply let
  $\astore'$ be its $\vec{x}$-associate).  Because $\astore''$ is
  injective, by the assumption of the Lemma, we obtain
  $(\astore'',\aheap') \models_\asys \psi_i$, for some $i \in
  \interv{1}{n}$, and we are left with proving the sufficient
  condition $(\aistore,\aheap)\models_\asys\psi_i$. To this end,
  consider the function $\amorph : \locs\rightarrow\locs$, defined
  as: \begin{compactitem}
  \item $\amorph(\astore''(x))=\aastore(x)$, for all $x \in
    \dom(\astore'')$,
  \item $\amorph(\ell)=\ell$, for all $\ell \in
    \locs\setminus\img(\astore'')$. 
  \end{compactitem} 
  Observe that $\amorph$ is well-defined, since by definition of
  $\astore'$, $\astore'(x)=\astore'(x') \Rightarrow
  \aastore(x)=\aastore(x')$. Below we check that $\amorph$ is
  compatible with $(\astore'',\aheap')$. Let $\ell_1, \ell_2 \in
  \locs$ be two locations such that $\amorph(\ell_1) =
  \amorph(\ell_2)$: \begin{compactitem}
  \item if $\ell_1, \ell_2 \in \dom(\aheap')$ then $\ell_1 =
    \astore''(x_i)$ and $\ell_2 = \astore''(x_j)$, for some $i, j \in
    \interv{1}{k}$, by  definition of $\aheap'$. Suppose, for a
    contradiction, that $i \neq j$. Then $\aastore(x_i) =
    \amorph(\astore''(x_i)) = \amorph(\astore''(x_j)) =
    \aastore(x_j)$, which contradicts the fact that $(\aastore,\aheap)
    \models \Asterisk_{i=1}^k x_i \mapsto (t^i_1, \ldots,
    t^i_\rank)$. Hence $i = j$, leading to $\ell_1 = \ell_2$.
  \item if $\ell_1 \in \astore''(\const)$, then let $c \in \const$ be
    a constant such that $\ell_1 = \astore''(c)$, so that
    $\amorph(\ell_1) = \aastore(c)$. Suppose, for a contradiction,
    that $\ell_2 \not\in \img(\astore'')$. Then $\amorph(\ell_2) =
    \ell_2 = \aastore(c)$, hence $\ell_2 \in \aastore(\const)$. But
    since $\aastore$ and $\astore''$ agree over $\const$, we have
    $\aastore(c) \in \astore''(\const)$. Hence $\ell_2 = \aastore(c) =
    \astore''(c)$, which contradicts with $\ell_2 \not\in
    \img(\astore'')$. Thus $\ell_2 \in \img(\astore'')$ and let
    $\ell_2 = \astore''(t)$, for some term $t$. We have
    $\amorph(\astore''(t)) = \aastore(t)$, thus $\aastore(c) =
    \amorph(\ell_2) = \amorph(\ell_1) = \aastore(t)$. By point
    (\ref{it1:normal-entailment}), we obtain $\ell_2 = \astore'(t) =
    \aastore(t) = \aastore(c) = \astore''(c) = \ell_1$.
  \end{compactitem}
  Moreover, it is easy to check that $(\aastore, \aheap) =
  (\amorph\circ\astore'', \amorph(\aheap'))$. Since $\aistore$ is the
  restriction of $\aastore$ to $\term{\phi}$, by Lemma
  \ref{lemma:compatible}, we obtain $(\aistore,\aheap) \models
  \psi_i$. \qed}

\shortVersionOnly{ Another important property of injective normal
  structures is that the frontier $\front(\aheap_1, \aheap_2)$ of a
  heap decomposition $\aheap = \aheap_1 \uplus \aheap_2$ such that
  $(\aistore,\aheap) \models_\asys \phi_1 * \phi_2$ and
  $(\aistore,\aheap_i) \models_\asys \phi_i$, for each $i = 1,2$ is
  contained in the image of the common free variables and constants
  via $\astore$, i.e.\ $\front(\aheap_1, \aheap_2) \subseteq
  \aistore(\fv{\phi_1} \cap \fv{\phi_2} \cup \const)$\footnote{See
    Lemma \ref{lemma:frontier} in Appendix \ref{app:injective-normal}
    for details.}.  }

\section{Core Formul{\ae}}
\label{sec:core}

Given \arestricted entailment problem $\aprob = (\asys, \seqset)$, the
idea of the entailment checking algorithm is to compute, for each
symbolic heap $\phi$ that occurs as the left-hand side of a sequent
$\phi \vdash_\aprob \psi_1, \ldots, \psi_n$, a finite set of sets of
formul{\ae} $\profile(\phi) = \set{F_1, \ldots, F_m}$, of some
specific pattern, called \emph{core formul{\ae}}. The set
$\profile(\phi)$ defines an equivalence relation, of finite index, on
the set of injective normal $\asys$-models of $\phi$, such that each
set $F \in \profile(\phi)$ encodes an equivalence class. Because the
validity of each sequent can be checked by testing whether every
(injective) normal model of its left-hand side is a model of some
symbolic heap on the right-hand side (Lemma
\ref{lemma:normal-entailment}), an equivalent check is that each set
$F \in \profile(\phi)$ contains a core formula entailing some formula
$\psi_i$, for $i = 1,\dots,n$. To improve the presentation, we first
formalize the notions of core formul{\ae} and abstractions by sets of
core formul{\ae}, while deferring the effective construction of
$\profile(\phi)$, for a symbolic heap $\phi$, to the next section
(\S\ref{sec:coreabs}). In the following, we refer to a given
entailment problem $\aprob = (\asys, \seqset)$.

First, we define core formul{\ae} as a fragment of
$\seplogk{\rank}$. Consider the formula $\inheap(x) \isdef \exists y_0
\ldots \exists y_\rank ~.~ y_0 \mapsto (y_1,\ldots,y_\rank) *
\bigvee_{i=0}^\rank x \teq y_i$. Note that a structure is a model of
$\inheap(x)$ iff the variable $x$ is assigned to a location from the
domain or the range of the heap. We define also the following bounded
quantifiers:
\vspace*{-.5\baselineskip}
\[\begin{array}{rclcrcl}
\nofvex x ~.~ \phi & \isdef & \exists x ~.~ 
\bigwedge_{t \in (\fv{\phi} \setminus \set{x}) \cup \const} \neg x \teq t \wedge \phi
& \hspace*{1cm} &  
\heapex x ~.~ \phi & \isdef & \nofvex x ~.~ \inheap(x) \wedge \phi 
\\
\nheapex x ~.~ \phi & \isdef & \nofvex x ~.~ \neg\inheap(x) \wedge \phi 
&&
\nheapall x ~.~ \phi & \isdef & \neg\nheapex x ~.~ \neg \phi \\[-2mm]
\end{array}\]
In the following, we shall be extensively using the $\heapex x ~.~
\phi$ and $\nheapall x ~.~ \phi$ quantifiers. The formula $\heapex x
~.~ \phi$ states that there exists a location $\ell$ which occurs in
the domain or range of the heap and is distinct from the locations
associated with the constants and free variables, such that $\phi$
holds when $x$ is associated with $\ell$. Similarly, $\nheapall x ~.~
\phi$ states that $\phi$ holds if $x$ is associated with any location
$\ell$ that is outside of the heap and distinct from all the constants
and free variables.  The use of these special quantifiers will allow
us to restrict ourselves to injective stores (since all variables and
constants are mapped to distinct locations), which greatly simplifies
the handling of equalities.

\additionalMaterial{Additional Material on Core Formul{\ae}}{sec:core}{app:core}{

The formal semantics of the bounded quantifiers is
stated below:

\begin{lemma}\label{lemma:bound-quant}
  Given a $\seplogk{\rank}$ formula $\phi$ and $x \in \fv{\phi}$, the
  following hold, for any structure
  $(\astore,\aheap)$: \begin{compactenum}
  \item\label{it:heapex} $(\astore,\aheap) \models_\asys \heapex x ~.~
    \phi$ iff $(\astore[x\leftarrow \ell],\aheap)
    \models_\asys \phi$, for some $\ell \in \loc(\aheap) \setminus
    \astore((\fv{\phi} \setminus \set{x}) \cup \const)$, 
  \item\label{it:nheapall} $(\astore,\aheap) \models_\asys \nheapall x
    ~.~ \phi$ iff $(\astore[x\leftarrow \ell],\aheap) \models_\asys
    \phi$, for all $\ell \in \locs \setminus \left[\loc(\aheap) \cup
    \astore((\fv{\phi} \setminus \set{x}) \cup \const)\right]$.
  \end{compactenum}
\end{lemma}
\proof{
First, for any structure $(\astore,\aheap)$, we have
  $(\astore,\aheap) \models \inheap(x) \iff \astore(x) \in
  \loc(\aheap)$. 

  \noindent(\ref{it:heapex}) By definition, $\heapex x ~.~ \phi$ is
  equivalent to $\exists x ~.~ \bigwedge_{t \in (\fv{\phi} \setminus
    \set{x}) \cup \const} \neg x \teq y \wedge \inheap(x) \wedge \phi$.

  \noindent(\ref{it:nheapall}) By definition, $\nheapall x ~.~ \phi$
  is equivalent to $\forall x ~.~ (\bigwedge_{t \in (\fv{\phi}
    \setminus \set{x}) \cup \const} \neg x \teq t \wedge \neg\inheap(x))
  \rightarrow \phi$. \qed}}

\shortVersionOnly{
\begin{wrapfigure}{R}{0.5\textwidth}
  \vspace*{-0.5\baselineskip}
  \centerline{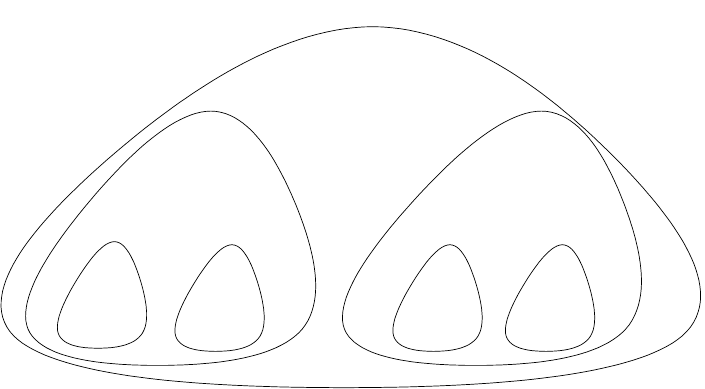}
  \caption{Inductive Definition of Context Predicates}
  \label{fig:context}
  \vspace*{-\baselineskip}
\end{wrapfigure}
}
\longVersionOnly{
  \begin{figure}[htb]
    \centerline{\input{context.pdf_t}}
    \caption{Inductive Definition of Context Predicates}
    \label{fig:context}    
  \end{figure}
}

The main ingredient used to define core formul{\ae} are \emph{context
  predicates}.  Given a tuple of predicate symbols $(p, q_1, \ldots,
q_n) \in \preds^{n+1}$, where $n\geq 0$, we consider a {context}
predicate symbol $\context{p,q_1,\ldots,q_n}$ of arity
$\#p+\sum_{i=1}^n \#q_i$. 
\longVersionOnly{ The rules defining the
  semantics of the context predicate atoms are given below:
\begin{eqnarray}
\context{p,p}(\vec{x},\vec{y}) & \Leftarrow & \vec{x} \seq \vec{y} 
\text{, where $\vec{x} \cap \vec{y} = \emptyset$ and $\len{\vec{x}} = \len{\vec{y}} = \#p$}, \label{rule:emp} \\
\context{p,q_1,\ldots,q_n}(\vec{x},\vec{y}_1,\ldots,\vec{y}_n) & \Leftarrow & \exists \vec{v} ~.~ \psi\sigma * 
\Asterisk_{j=1}^m \context{p_j,q_{i^j_1},\ldots, q_{i^j_{k_j}}}(\sigma(\vec{w}_j),\vec{y}_{i^j_1}, \ldots, \vec{y}_{i^j_{k_j}}\!\!\!\!),~
\label{rule:notemp}
\end{eqnarray}
for each rule $p(\vec{x}) \Leftarrow_\asys \exists \vec{z} ~.~ \psi *
\Asterisk_{j=1}^m p_j(\vec{w}_j)$ where
$\vec{x},\vec{z},\vec{y}_1,\ldots,\vec{y}_n$ are pairwise disjoint
tuples of variables; each substitution $\sigma : \vec{z} \map \vec{x}
\cup \bigcup_{i=1}^n \vec{y}_i$ where $\vec{v} = \vec{z} \setminus
\dom(\sigma)$ and pairwise disjoint (and possibly empty) sets
$\setof{\{i^j_1, \ldots, i^j_{k_j}\}}{j \in \interv{1}{m}}$ with a
union equal to $\interv{1}{n}$.  Let $\coreset{\asys}$ be the set of
rules for contexts (\ref{rule:emp}) and (\ref{rule:notemp}), induced
by the rules from $\asys$.  The satisfaction relation
$(\astore,\aheap) \models_{\coreset{\asys}} \varphi$ is defined as
usual, w.r.t. the set of rules $\coreset{\asys}$ (Definition
\ref{def:unfolding-semantics}).}
The informal intuition of a context predicate atom
$\context{p,q_1,\ldots,q_n}(\vec{t},\vec{u}_1,\ldots,\vec{u}_n)$ is
the following: a structure $(\astore,\aheap)$ is a model of this atom
if there exist models $(\astore,\aheap_i)$ of $q_i(\vec{u}_i)$,
$i\in\interv{1}{n}$ respectively, with mutually disjoint heaps, an
unfolding $\psi$ of $p(\vec{t})$ in which the atoms $q_i(\vec{u}_i)$
occur, and an associate $\astore'$ of $\astore$ such that
$(\astore',\aheap \uplus \biguplus_{i=1}^n \aheap_i)$ is a model of
$\psi$. 

For readability's sake, we adopt a notation close in spirit to
$\seplog$'s separating implication (known as the magic wand), and we
write $\Asterisk_{i=1}^n q_i(\vec{y}_i) \swand p(\vec{x})$ for
$\context{p,q_1,\ldots,q_n}(\vec{x},\vec{y}_1,\ldots,\vec{y}_n)$ and
$\emp \swand p(\vec{x})$, when $n=0$\footnote{Context predicates are
  similar to the {\em strong magic wand} introduced in
  \cite{NTKY2018}. A context predicate $\alpha \swand \beta$ is also
  related to the usual separating implication $\alpha \wand \beta$ of
  separation logic, but it is not equivalent. Intuitively, $\wand$
  represents a difference between two heaps, whereas $\swand$ removes
  some atoms in an unfolding. For instance, if $p$ and $q$ are defined
  by the same inductive rules, up to a renaming of predicates, then
  $p(x) \wand q(x)$ always holds in a structure with an empty heap,
  whereas $p(x) \swand q(x)$ holds if, moreover, $p(x)$ and $q(x)$ are
  the same atom.}. The set of rules defining the interpretation of
context predicates is the least set defined by the inference rules
below, denoted $\coreset{\asys}$:
\begin{equation} \tag{$\mathsf{I}$}\label{infrule:emp}
 \infer[\vec{x} \cap \vec{y} = \emptyset]{
  p(\vec{x}) \swand p(\vec{y}) \Leftarrow_{\coreset{\asys}} \vec{x} \seq \vec{y}
}{} 
\end{equation}
\vspace*{-\baselineskip}
\begin{equation}\tag{$\mathsf{II}$}\label{infrule:notemp}
  \infer[\begin{array}{l}
      \vec{x}, \vec{z}, \vec{y}_1, \ldots, \vec{y}_n \text{ pairwise disjoint} \\
      \sigma : \vec{z} \map \vec{x} \cup \bigcup_{i=1}^n \vec{y}_i \\
      \vec{v} = \vec{z} \setminus \dom(\sigma)
  \end{array}]{
    \Asterisk_{i=1}^n q_i(\vec{y}_i) \swand p(\vec{x}) \Leftarrow_{\coreset{\asys}} 
    \exists \vec{v} ~.~ \psi\sigma * \Asterisk_{j=1}^m \left(\gamma_j \swand p_j(\sigma(\vec{w}_j))\right)
  }{
    p(\vec{x}) \Leftarrow_{\asys} \exists \vec{z} ~.~ \psi * \Asterisk_{j=1}^m p_j(\vec{w}_j) \hspace*{4mm} 
    \Asterisk_{i=1}^n q_i(\vec{y}_i) = \Asterisk_{j=1}^m \gamma_j
  } 
\end{equation}

Note that $\coreset{\asys}$ is not progressing, since the rule for
$p(\vec{x}) \swand p(\vec{y})$ does not allocate any
location. However, if $\asys$ is progressing, then the set of rules
obtained by applying (\ref{infrule:notemp}) only is also progressing.
Rule (\ref{infrule:emp}) says that each predicate atom $p(\vec{t})
\swand p(\vec{u})$, such that $\vec{t}$ and $\vec{u}$ are mapped to
the same tuple of locations, is satisfied by the empty heap. To
understand rule (\ref{infrule:notemp}), let $(\astore,\aheap)$ be an
$\asys$-model of $p(\vec{t})$ and assume there are a predicate-free
unfolding $\psi$ of $p(\vec{t})$ and an associate $\astore'$ of
$\astore$, such that $q_1(\vec{u}_1) , \ldots, q_n(\vec{u}_n)$ occur
in $\psi$ and $(\astore',\aheap) \models_{\asys} \psi$
(Fig. \ref{fig:context}). If the first unfolding step is an instance
of a rule $p(\vec{x}) \Leftarrow_\asys \exists \vec{z} ~.~ \psi *
\Asterisk_{j=1}^m p_j(\vec{w}_j)$ then there exist a
$\vec{z}$-associate $\aastore$ of $\astore$ and a split of $\aheap$
into disjoint heaps $\aheap_0, \ldots, \aheap_m$ such that
$(\aastore,\aheap_0) \models \psi[\vec{t}/\vec{x}]$ and
$(\aastore,\aheap_j) \models_\asys \Asterisk_{j=1}^m
p_j(\vec{w}_j)[\vec{t}/\vec{x}]$, for all $j \in
\interv{1}{m}$. Assume, for simplicity, that $\vec{u}_1 \cup \ldots
\cup \vec{u}_n \subseteq \dom(\aastore)$ and let $\aaheap_1, \ldots,
\aaheap_n$ be disjoint heaps such that $(\aastore,\aaheap_i)
\models_\asys q_i(\vec{u}_i)$. Then there exists a partition
$\big\{\{i_{j,1}, \ldots, i_{j,k_j}\} \mid j\in \interv{1}{m}\big\}$
of $\interv{1}{n}$, such that $\aaheap_{i_{j,1}}, \ldots,
\aaheap_{i_{j,k_j}} \subseteq \aheap_j$, for all $j \in
\interv{1}{m}$. Let $\gamma_j \isdef \Asterisk_{\ell=1}^{k_j}
q_\ell(\vec{u}_\ell)$, then $(\aastore, \aheap_j \setminus
(\aaheap_{i_{j,1}} \cup \ldots \cup \aaheap_{i_{j,k_j}}))
\models_{\coreset{\asys}} \gamma_j \swand
p_j(\vec{w}_j)[\vec{t}/\vec{x}]$, for each $j \in \interv{1}{m}$.
This observation leads to the inductive definition of the semantics
for $\Asterisk_{i=1}^n q_i(\vec{u}_i) \swand p(\vec{t})$, by the rule
that occurs in the conclusion of (\ref{infrule:notemp}), where the
substitution $\sigma : \vec{z} \map \vec{x} \cup \bigcup_{i=1}^n
\vec{y}_i$ is used to instantiate\footnote{Note that this
  instantiation is, in principle, redundant (i.e.\ the same rules are
  obtained if $\dom(\sigma)=\emptyset$ by chosing appropriate
  $\vec{z}$-associates) but we keep it to simplify the related
  proofs.} some of the existentially quantified variables from the
original rule $p(\vec{x}) \Leftarrow_\asys \exists \vec{z} ~.~ \psi *
\Asterisk_{j=1}^m p_j(\vec{w}_j)$.

\putInAppendix{
Below we prove the equivalence
    between the atoms $p(\vec{t})$ and $\emp \swand p(\vec{t})$. 

\begin{lemma}\label{lemma:pred-equiv}
  A structure $(\astore,\aheap)$ is an $\asys$-model of $p(\vec{t})$ if
  and only if $(\astore,\aheap)$ is a $\coreset{\asys}$-model of
  $\emp\swand p(\vec{t})$.
\end{lemma}
\proof{ ``$\Rightarrow$'' For each rule $p(\vec{x}) \Leftarrow_\asys
  \exists \vec{z} ~.~ \psi * \Asterisk_{i=1}^n q_i(\vec{y}_i)$, there
  exists a rule $\emp \swand p(\vec{x}) \Leftarrow_{\coreset{\asys}}
  \exists \vec{z} ~.~ \psi * \Asterisk_{i=1}^n \emp \swand
  q_i(\vec{y}_i)$, corresponding to the case where the substitution
  $\sigma$ is empty. The proof follows by a simple induction on the
  length of the predicate-free unfolding of
  $p(\vec{t})$. ``$\Leftarrow$''
  We prove the other direction by induction on the length of the
  predicate-free unfolding of $\emp \swand p(\vec{t})$. Assume
  $(\astore, \aheap)$ is a $\coreset{\asys}$-model of $\emp \swand
  p(\vec{t})$.  Then there exist a rule $\emp \swand p(\vec{x})
  \Leftarrow_{\coreset{\asys}} \exists \vec{v} ~.~ \psi\sigma *
  \Asterisk_{j=1}^m \left(\emp \swand p_j(\sigma(\vec{w}_j))\right)$
  in $\coreset{\aprob}$ and a $\vec{v}$-associate $\astore'$ of
  $\astore$ such that $(\astore', \aheap) \models \psi\sigma\theta *
  \Asterisk_{j=1}^m \left(\emp \swand
  p_j(\theta\circ\sigma(\vec{w}_j))\right)$. By definition of
  $\coreset{\aprob}$, this entails that $p(\vec{t})$ can be unfolded
  into $\exists \vec{z}~.~\psi\theta * \Asterisk_{j=1}^m
  p_j(\theta(\vec{w}_j))$ using the rules in $\asys$. The heap
  $\aheap$ can be decomposed into $\aheap_0\uplus \cdots\uplus
  \aheap_m$, where $(\astore', \aheap_j) \models \emp\swand
  p_j(\theta\circ\sigma(\vec{w}_j))$, for $j\in \interv{1}{m}$. By the
  induction hypothesis, $(\astore', \aheap_j)$ is an $\asys$-model of
  $p_j(\theta\circ\sigma(\vec{w}_j))$, and we deduce that $(\astore,
  \aheap)$ is an $\asys$-model of $\exists z~.~ \psi\theta *
  \Asterisk_{j=1}^m p_j(\theta(\vec{w}_j))$. \qed}

Another property of context predicate atoms is
  stated by the lemma below: 

\begin{lemma}\label{lemma:context-emp}
  If $\asys$ is progressing, then for each store (resp. injective store)
  $\astore$, we have $(\astore,\emptyset) \models_{\coreset{\asys}}
  \Asterisk_{i=1}^n q_i(\vec{u}_i) \swand p(\vec{t})$ if and only if
  $n=1$, $p = q_1$ and $\astore(\vec{t}) = \astore(\vec{u}_1)$
  (resp. $\vec{t} = \vec{u}_1$).
\end{lemma}
\proof{ ``$\Rightarrow$'' If $(\astore,\emptyset)
  \models_{\coreset{\asys}} \Asterisk_{i=1}^n q_i(\vec{u}_i) \swand
  p(\vec{t})$ then there exists a rule $\Asterisk_{i=1}^n
  q_i(\vec{u}_i) \swand p(\vec{t}) \Leftarrow_{\coreset{\asys}} \phi$
  and a substitution $\sigma$ such that $(\astore,\emptyset)
  \models_{\coreset{\asys}} \phi\sigma$, where $\sigma =
         [\vec{t}/\vec{x}, \vec{u}_1/\vec{y}_1, \ldots,
           \vec{u}_n/\vec{y}_n]$. If the rule is an instance of
         (\ref{infrule:emp}) then $n = 1$, $p = q_1$ and
         $(\astore,\emptyset) \models \vec{t} \seq \vec{u}_1$, leading
         to $\astore(\vec{t}) = \astore(\vec{u}_1)$. If, moreover
         $\astore$ is injective, we get $\vec{t} = \vec{u}_1$. Otherwise,
         if the rule is an instance of (\ref{infrule:notemp}), then since
         $\asys$ is progressing, $\phi\sigma$ must contain exactly one
         points-to atom, hence $(\astore,\emptyset)
         \models_{\coreset{\asys}} \phi\sigma$ cannot be the
         case. ``$\Leftarrow$'' This is a simple application of rule
         (\ref{infrule:emp}). \qed}}

\begin{example}
\label{ex:contexts}
Consider the set \(\asys = \{p(x) \Leftarrow \exists z_1,z_2 ~.~ x
\mapsto (z_1,z_2) * q(z_1) * q(z_2), q(x) \Leftarrow x \mapsto
(x,x)\}\). We have $(\astore,\aheap) \models_{\asys} p(x)$ with
$\astore = \{ (x,\ell_1) \}$ and $\aheap = \{ (\ell_1,\ell_2,\ell_3),
(\ell_2,\ell_2,\ell_2), (\ell_3,\ell_3,\ell_3) \}$.  The atom $q(y)
\swand p(x)$ is defined by the following non-progressing rules:
\shortVersionOnly{
\vspace*{-.5\baselineskip}
\[\begin{tabular}{rclrcl}
$q(y) \swand p(x)$ & $\Leftarrow$ & $\exists z_1,z_2 ~.~ x \mapsto (z_1,z_2) * q(y) \swand  q(z_1) *  \emp \swand q(z_2)$  &
$q(y) \swand q(x)$ & $\Leftarrow$  &  $x \seq y$ \\
$q(y) \swand p(x)$ & $\Leftarrow$ & $\exists z_1,z_2 ~.~ x \mapsto (z_1,z_2) * \emp \swand q(z_1) * q(y) \swand q(z_2)$  &
$\emp \swand q(x)$ & $\Leftarrow$  &  $x \mapsto (x,x)$ \\[-2mm]
\end{tabular}\]
}
\longVersionOnly{\[
	\begin{tabular}{rcl}
	$q(y) \swand p(x)$ & $\Leftarrow$ & $\exists z_1,z_2 ~.~ x \mapsto (z_1,z_2) * q(y) \swand  q(z_1) *  \emp \swand q(z_2)$  \\
	$q(y) \swand p(x)$ & $\Leftarrow$ & $\exists z_1,z_2 ~.~ x \mapsto (z_1,z_2) * \emp \swand q(z_1) * q(y) \swand q(z_2)$  \\
	$q(y) \swand q(x)$ & $\Leftarrow$  &  $x \seq y$ \\
	$\emp \swand q(x)$ & $\Leftarrow$  &  $x \mapsto (x,x)$ \\
	\end{tabular}
	\]}
The two rules for $q(y) \swand p(x)$ correspond to the two ways of
distributing $q(y)$ over $q(z_1)$, $q(z_2)$.  We have $\aheap =
\aheap_1 \uplus \aheap_2$, with $\aheap_1 = \{ (\ell_1,\ell_2,\ell_3),
(\ell_2,\ell_2,\ell_2)\}$ and $\aheap_2 = \{ (\ell_3,\ell_3,\ell_3)
\}$.  It is easy to check that $(\astore[y \leftarrow
  \ell_3],\aheap_1) \models_{\coreset{\asys}} q(y) \swand p(x)$, and
$(\astore[y \leftarrow \ell_3],\aheap_2) \models_{\coreset{\asys}}
q(y)$.  Note that we also have $(\astore[y \leftarrow
  \ell_2],\aheap_1') \models_{\coreset{\asys}} q(y) \swand p(x)$, with
$\aheap_1' = \{ (\ell_1,\ell_2,\ell_3), (\ell_3,\ell_3,\ell_3)\}$.
\hfill$\blacksquare$
\end{example}

Having introduced context predicates, the pattern of core formul{\ae}
is defined below:

\begin{definition}\label{def:core-formulae}
  A \emph{core formula} $\varphi$ is an instance of the pattern: 
  \[\heapex \vec{x} \nheapall \vec{y} ~.~ \Asterisk_{i=1}^n
  \left(\Asterisk_{j=1}^{k_i} q_j^i(\vec{u}^i_j) \swand
  p_i(\vec{t}_i)\right) * \Asterisk_{i=n+1}^m t^i_0 \mapsto
  (t^i_1,\ldots,t^i_\rank) \text{\quad such that:}\]
  \vspace*{-\baselineskip}
  \begin{compactenum}[(i)]
  \item\label{core-formulae:no_useless_var} each variable occurring in
    $\vec{y}$ also occurs in an atom in $\varphi$;
  \item\label{core-formulae:exists} for every variable $x \in
    \vec{x}$, either $x \in \vec{t}_i \setminus \bigcup_{i=1}^{k_i}
    \vec{u}^i_j$ for some $i \in \interv{1}{n}$, or $x = t^i_j$, for some $i
    \in \interv{n+1}{m}$ and some $j \in \interv{0}{\rank}$;
  \item\label{core-formulae:roots_are_distinct} each term $t$ occurs
    at most once as $t = \aroot(\alpha)$, where $\alpha$ is an atom of
    $\varphi$.
  \end{compactenum}
  We define moreover the set of terms $\roots{\varphi} \isdef
  \lroots{\varphi} \cup \rroots{\varphi}$, where $\lroots{\varphi}
  \isdef \{\aroot(q_j^i(\vec{u}^i_j)) \mid i \in \interv{1}{n}, j \in
  \interv{1}{k_i}\}$ and $\rroots{\varphi} \isdef
  \{\aroot(p_i(\vec{t}_i)) \mid i \in \interv{1}{n}\} \cup \{t^i_0
  \mid i \in \interv{n+1}{m}\}$.
\end{definition}
Note that an unfolding of a core formula using the rules in
$\coreset{\asys}$ is not necessarily a core formula, because of the
unbounded existential quantifiers and equational atoms that occur in
the rules from $\coreset{\asys}$.  Note also that a core formula
cannot contain an occurrence of a predicate of the form $p(\vec{t})
\swand p(\vec{t})$ because otherwise, Condition
(\ref{core-formulae:roots_are_distinct}) of Definition
\ref{def:core-formulae} would be violated.

\putInAppendix{

The following lemma states a
    technical result about core formul{\ae}, that will be used in the proof
    of Lemma \ref{lemma:coresep}: 
 
  \begin{lemma}\label{lemma:lhs-root}
    For each quantifier-free core formula $\varphi$, each injective
    $\coreset{\asys}$-model $(\aistore,\aheap)$ of $\varphi$ such
    that $\card{\aheap} \geq 1$, and each term $t \in
    \lroots{\varphi}$, we have $\aistore(t) \in \loc(\aheap) \cup
    \aistore(\const)$.
  \end{lemma}
  \proof{ Let $\varphi$ be a quantifier-free core formula of the
    following form (cf. Definition \ref{def:core-formulae}):
    \begin{equation}\label{eq:qf-core}
      \Asterisk_{i=1}^n 
      \left(\Asterisk_{j=1}^{k_i} q_j^i(\vec{u}^i_j) \swand p_i(\vec{t}_i)\right) *
      \Asterisk_{i=n+1}^m x_i \mapsto (t^i_1,\ldots,t^i_\rank) 
    \end{equation}
    The proof goes by induction on $\card{\aheap}$. In the base case,
    $\card{\aheap}=1$, we prove first that the formula contains
    exactly one points-to or predicate atom. Suppose, for a
    contradiction, that it contains two or more atoms, i.e.\ $\varphi
    = \alpha_1 * \ldots * \alpha_m$, for $m \geq 2$. If $\alpha_1$ and
    $\alpha_2$ are points-to atoms, it cannot be the case that
    $(\aistore,\aheap)$ is a $\coreset{\asys}$-model of $\varphi$,
    thus we distinguish two cases: \begin{compactitem}
    \item If $\alpha_1=\Asterisk_{j=1}^{k} q_j(\vec{u}_j) \swand
      p(\vec{t})$ and $\alpha_2$ is a points-to atom then, since
      $\card{\aheap} = 1$, we must have $(\aistore, \emptyset)
      \models_{\coreset{\asys}} \alpha_1$ and $(\aistore,\aheap)
      \models \alpha_2$. By Lemma \ref{lemma:context-emp}, we obtain
      $k=1$ and $q_1(\vec{u}_1) = p(\vec{t})$, which violates the
      condition on the uniqueness of roots in $q_1(\vec{u}_1) \swand
      p(\vec{t})$, in Definition \ref{def:core-formulae}.
    \item Otherwise, $\alpha_1$ and $\alpha_2$ are both predicate atoms; we
      assume that $(\aistore, \emptyset) \models_{\coreset{\asys}}
      \alpha_1$ (the case $(\aistore, \emptyset)
      \models_{\coreset{\asys}} \alpha_2$ is identical). We obtain a
      contradiction by the argument used at the previous point.
    \end{compactitem}
    If $\varphi$ consists of a single points-to atom, then
    $\lroots{\varphi} = \emptyset$ and there is nothing to
    prove. Otherwise, $\varphi$ is of the form $\alpha_1 =
    \Asterisk_{i=1}^{k} q_i(\vec{u}_i) \swand p(\vec{t})$. By Lemma
    \ref{lemma:context-emp}, since $\asys$ is progressing and
    $(\aistore,\aheap) \models_{\coreset{\asys}} \Asterisk_{i=1}^{k}
    q_i(\vec{u}_i) \swand p(\vec{t})$, either $k>1$ or $k=1$ and
    $q_1(\vec{u}_1) \neq p(\vec{t})$. By Condition (\ref{infrule:notemp}), there
    exists: \begin{compactenum}[(a)]
    \item\label{it2:lhs-root} a rule $p(\vec{x}) \Leftarrow_\asys
      \exists \vec{z} ~.~ \psi * \Asterisk_{j=1}^m p_j(\vec{w}_j)$,
    \item\label{it3:lhs-root} separating conjunctions of predicate atoms
      $\gamma_1, \ldots, \gamma_m$, such that $\Asterisk_{j=1}^m
      \gamma_j = \Asterisk_{i=1}^k q_i(\vec{y}_i)$,
    \item\label{it1:lhs-root} a substitution $\tau : \vec{z} \map
      \vec{x} \cup \bigcup_{i=1}^n \vec{y}_i$, 
    \end{compactenum}
    that induce the rule: \[\Asterisk_{i=1}^{k} q_i(\vec{y}_i) \swand
    p(\vec{x}) \Leftarrow_{\coreset{\asys}} \exists \vec{v} ~.~
    \psi\tau * \Asterisk_{j=1}^m \gamma_j \swand
    p_j(\tau(\vec{w}_j)),\] 
    where $\vec{v} = \vec{z}
    \setminus \dom(\tau)$. Assume w.l.o.g. that $(\aistore,\aheap)
    \models_{\coreset{\asys}} \Asterisk_{i=1}^{k} q_i(\vec{u}_i)
    \swand p(\vec{t})$ is the consequence of the above rule, meaning
    that:
    \[(\aistore,\aheap) \models_{\coreset{\asys}} \left(\exists \vec{v} ~.~
    \psi\tau * \Asterisk_{j=1}^m \left(\gamma_j \swand
    p_j(\tau(\vec{w}_j))\right)\right)\sigma \text{, where } \sigma =
    [\vec{t}/\vec{x}, \vec{u}_1/\vec{y}_1, \ldots,
      \vec{u}_n/\vec{y}_n].\] Let $\aastore$ be the $\vec{v}$-associate
    of $\aistore$ such that $(\aastore,\aheap)
    \models_{\coreset{\asys}} \psi\tau\sigma * \Asterisk_{j=1}^m
    \left(\gamma_j\sigma \swand
    p_j(\sigma(\tau(\vec{w}_j)))\right)$. Since $\asys$ is
    progressing, $\psi$ contains a points-to atom $t_0 \mapsto (t_1,
    \ldots, t_\rank)$, such that $(\aastore,\aheap)
    \models_{\coreset{\asys}} \left(t_0 \mapsto (t_1, \ldots,
    t_\rank)\right)\tau\sigma$ and $(\aastore,\emptyset)
    \models_{\coreset{\asys}} \Asterisk_{j=1}^m \left(\gamma_j\sigma
    \swand p_j(\sigma(\tau(\vec{w}_j)))\right)$. Now consider $t \in
    \lroots{\phi}$, then $t = \aroot(q_i(\vec{u}_i))$, for
    some $i \in \interv{1}{k}$. Since $\Asterisk_{j=1}^m
    \gamma_j\sigma = \Asterisk_{i=1}^k q_i(\vec{u}_i)$ by Condition
    (\ref{it3:lhs-root}), 
    we have $t
    \in \term{\gamma_j\sigma}$, for some $j \in \interv{1}{m}$. Since
    $(\aastore,\emptyset) \models_{\coreset{\asys}} \gamma_j\sigma
    \swand p_j(\sigma(\tau(\vec{w}_j)))$, by Lemma
    \ref{lemma:context-emp}, we have $\aastore(t) =
    \aastore(\sigma(\tau(r)))$, where $r =
    \aroot(p_j(\vec{w}_j))$. Since $\asys$ is connected, either $r \in
    \set{t_1, \ldots, t_\rank}$ or $r \in \const$, by Definition
    \ref{def:progress-connectivity}. Since $t \in \lroots{\phi}$, we
    have $\aastore(t) = \aistore(t)$, and we conclude that $\aistore(t) \in
    \loc(\aheap) \cup \aistore(\const)$. 

    \noindent For the induction step $\card{\aheap} > 1$, let $t =
    \aroot(q_j^i(\vec{u}_j^i))$, for some $i \in \interv{1}{n}$ and
    some $j \in \interv{1}{k_j}$. If $n > 1$ or $m > n$ in Equation
    (\ref{eq:qf-core}), we have $(\aistore,\aheap')
    \models_{\coreset{\asys}} \Asterisk_{j=1}^{k_i} q_j^i(\vec{u}^i_j)
    \swand p_i(\vec{t}_i)$, for some heap $\aheap' \subset \aheap$,
    such that $\card{\aheap'} \geq 1$ and, by the inductive
    hypothesis, we obtain $\aistore(t) \in \loc(\aheap') \cup
    \aistore(\const) \subseteq \loc(\aheap) \cup
    \aistore(\const)$. Otherwise, $m = n = 1$ and the argument is
    similar to the one used in the base case. \qed}

  \begin{lemma}\label{lemma:varrightonly}
    Let $\phi = \Asterisk_{j=1}^{k} q_j(\vec{u}_j) \swand
    p_i(\vec{t})$ be a core formula and let $(\aistore,\aheap)$ be an
    injective structure.  If $\asys$ is progressing and normalized,
    $(\aistore,\aheap) \models_{\coreset{\asys}} \phi$ and $x \in
    \vec{t} \setminus (\bigcup_{j=1}^k \vec{u}_j)$ then $\aistore(x)
    \in \loc(\aheap)$.
\end{lemma}
  \proof{ We reason by induction on $\card{\aheap}$.  If $\aheap =
    \emptyset$ then by Lemma \ref{lemma:context-emp},
     we must have $k = 1$ and $\vec{u}_1 = \vec{t}$, thus
    $\vec{t} \setminus (\bigcup_{j=1}^k \vec{u}_j)$ is empty, which
    contradicts our hypothesis.  Otherwise, by definition of the rules
    in $\coreset{\asys}$, there exists a rule $p(\vec{x})
    \Leftarrow_\asys \exists \vec{z}~\,~ \psi * \Asterisk_{j=1}^{m}
    p_j(\vec{w}_j)$, an associate $\aaistore$ of $\aistore$ and a substitution $\sigma: \vec{z} \map \vec{x}
    \cup \bigcup_{j=1}^m \vec{y}_j$ such that  $(\aaistore,\aheap)
    \models_{\coreset{\asys}} \psi\sigma\theta * \Asterisk_{j=1}^{m}
    \left(\gamma_j \swand p_j(\sigma(\vec{w}_j))\right)\theta$, where 
    $\Asterisk_{j=1}^{m} \gamma_j = \Asterisk_{j=1}^{k}
    q_j(\vec{y}_j)$ and $\theta = [\vec{t}/\vec{x},
      \vec{u}_1/\vec{y}_1, \ldots, \vec{u}_m/\vec{y}_m]$. Since
    $\asys$ is normalized, by Condition \ref{it3:normalized} in Definition \ref{def:normalized}, $x$ occurs in all unfoldings of
    $p(\vec{t})$. Thus either $x$ occurs in $\psi\theta$ (hence also
    in $\psi\sigma\theta$), or $x$ occurs in $\vec{w}_j\theta$ (hence in
    $\vec{w}_j\sigma\theta$) for some $j \in \interv{1}{m}$.  In the
    former case, necessarily $x \in \loc(\aheap)$, because $\psi$ is a
    points-to atom, since $\asys$ is progressing.  In the latter
    case, we have $(\aistore,\aheap') \models_{\coreset{\asys}}
    \gamma_j \swand p_j(\sigma(\vec{w}_j))\theta$, for some subheap
    $\aheap'$ of $\aheap$, with $\card{\aheap'} < \card{\aheap}$.
    Since $x\not \in \bigcup_{j=1}^k \vec{u}_j$ by hypothesis and
    $\Asterisk_{j=1}^{m} \gamma_j = \Asterisk_{j=1}^{k}
    q_j(\vec{y}_j)$ we have $x \not \in \fv{\gamma_j\theta}$, thus $x
    \in \vec{w}_j \setminus \fv{\gamma_j}\theta$. By the induction
    hypothesis, we deduce that $x \in \loc(\aheap')$, hence $x \in
    \loc(\aheap)$. \qed}}

Lemma \ref{lemma:coretrans} shows that any symbolic heap is equivalent
to an effectively computable finite disjunction of core formul{\ae},
when the interpretation of formul{\ae} is restricted to injective
structures. For a symbolic heap $\phi \in \shk{\rank}$, we define the
set $\coretrans{\phi}$, recursively on the structure of $\phi$,
implicitly assuming w.l.o.g. that $\emp * \phi = \phi * \emp = \phi$:
\[\begin{array}{lcll}
\coretrans{\emp} & \isdef & \set{\emp} & 
\hspace*{-1cm}\coretrans{t_0 \mapsto (t_1,\ldots,t_\rank)} \isdef \set{t_0 \mapsto (t_1,\ldots,t_\rank)} \\
\coretrans{p(\vec{t})} & \isdef & \set{\emp \swand p(\vec{t})} &
\hspace*{-1cm}\coretrans{\Asterisk_{i=1}^n q_i(\vec{u}_i) \swand p(\vec{t})} \isdef \set{\Asterisk_{i=1}^n q_i(\vec{u}_i) \swand p(\vec{t})} \\
\coretrans{t_1 \seq t_2} & \isdef & \left\{\begin{array}{cc}
\set{\emp} & \mbox{ if $t_1 = t_2$} \\
\emptyset & \mbox{ if $t_1 \neq t_2$}
\end{array}\right. &
\hspace*{-1cm}\coretrans{t_1 \sneq t_2} \isdef \left\{\begin{array}{cc}
\emptyset & \mbox{ if $t_1 = t_2$} \\
\set{\emp} & \mbox{ if $t_1 \neq t_2$}
\end{array}\right. \\
\coretrans{\phi_1 * \phi_2} & \isdef &
\set{\psi_1 * \psi_2 \mid \psi_i \in \coretrans{\phi_i},~ i = 1,2} \\
\coretrans{\exists x ~.~ \phi_1} & \isdef & \{\heapex x ~.~ \psi \mid \psi \in \coretrans{\phi_1}\} 
\cup \{\psi \mid \psi \in \coretrans{\phi_1[t/x]}, & t \in (\fv{\phi_1} \setminus \set{x}) \cup \const\}
\end{array}\]
For instance, if $\phi = \exists x~.~ p(x,y) * x \not \seq y$ and $\const = \{
\mathsf{c} \}$, then $\coretrans{\phi} = \{ \heapex x~.~ \emp
\swand p(x,y),\ \emp \swand p(\mathsf{c},y) \}$. 

\putInAppendix{
\begin{proposition}\label{prop:coretrans-subs}
  Consider a quantifier-free symbolic heap $\varphi$ and an injective
  substitution $\sigma$. If $\phi \in \coretrans{\varphi}$ then
  $\phi\sigma \in \coretrans{\varphi\sigma}$.
\end{proposition} 

The following lemmas relate a symbolic heap $\phi$ with the core
formul{\ae} $\psi \in \coretrans{\phi}$, by considering separately the
cases where $\phi$ is quantifier-free, or existentially quantified. In
the latter case, we require moreover that the set of rules providing the
interpretation of predicates be normalized.

  \begin{lemma}\label{lemma:coretrans-qf}  	
    Given a quantifier-free symbolic heap $\phi \in \shk{\rank}$,
    containing only predicate atoms that are contexts, an injective
    structure $(\aistore, \aheap)$ is a $\coreset{\asys}$-model of
    $\phi$ iff $(\aistore,\aheap) \models_{\coreset{\asys}} \psi$, for
    some $\psi \in \coretrans{\phi}$.
\end{lemma}
\proof{ ``$\Rightarrow$'' By induction
  on the structure of $\phi$. We consider the following
  cases: \begin{compactitem}
  \item $\phi=\emp$, $\phi=t_0 \mapsto (t_1, \ldots, t_\rank)$ and
    $\phi=\Asterisk_{i=1}^n q_i(\vec{u}_i) \swand p(\vec{t})$: in these
    cases, the only element in $\coretrans{\phi}$ is $\phi$ itself and we have the result.
  \item $\phi=t_1 \seq t_2$: since $(\aistore, \aheap) \models t_1 \seq
    t_2$, we have $\aistore(t_1) = \aistore(t_2)$ and $\aheap =
    \emptyset$. Since $\aistore$ is injective, we obtain $t_1 = t_2$,
    $\coretrans{\phi} = \set{\emp}$ and $(\aistore,\aheap) \models
    \emp$, because $\aheap = \emptyset$.
  \item $\phi=t_1 \sneq t_2$: since $(\aistore, \aheap) \models t_1 \sneq
    t_2$, we have $\aistore(t_1) \neq \aistore(t_2)$ and $\aheap =
    \emptyset$, therefore $t_1 \neq t_2$, $\coretrans{t_1 \sneq t_2} =
    \set{\emp}$ and $(\aistore, \aheap) \models \emp$, because $\aheap
    = \emptyset$.
  \item $\phi=\phi_1 * \phi_2$: since $(\aistore, \aheap) \models_\asys
    \phi_1 * \phi_2$, there exist heaps $\aheap_1$ and $\aheap_2$,
    such that $\aheap = \aheap_1 \uplus \aheap_2$ and $(\aistore,
    \aheap_i) \models_\asys \phi_i$, for $i = 1,2$. By the inductive
    hypothesis, there exists $\psi_i \in \coretrans{\phi_i}$ such that
    $(\aistore, \aheap_i) \models_{\coreset{\asys}} \psi_i$, for
    $i=1,2$. Then $(\aistore, \aheap) \models_{\coreset{\asys}} \psi_1
    * \psi_2$, where $\psi_1 * \psi_2 \in \coretrans{\phi_1 *
      \phi_2}$.
  \end{compactitem}
  \noindent``$\Leftarrow$'' By induction on the structure of $\phi$,
  we consider only the equational atoms below, the proofs in the remaining cases
  are straightforward: \begin{compactitem}
  \item $\phi=t_1 \seq t_2$: since there exists $\psi \in \coretrans{\phi}$
    such that $(\aistore,\aheap) \models_\asys \psi$, necessarily $\coretrans{\phi} = \set{\emp}$, which implies that $t_1 = t_2$.
    Since $(\aistore,\aheap) \models \emp$, $\aheap = \emptyset$ and
    $(\aistore, \aheap) \models t_1 \seq t_2$.
  \item $\phi=t_1 \sneq t_2$: since there exists $\psi \in
    \coretrans{\phi}$ such that $(\aistore,\aheap) \models_\asys \psi$, necessarily $\coretrans{\phi} = \set{\emp}$, which implies that
    $t_1 \neq t_2$. Since $(\aistore, \aheap) \models \emp$, $\aheap =
    \emptyset$ and $(\aistore, \aheap) \models t_1 \sneq t_2$, by 
    injectivity of $\aistore$. \qed
  \end{compactitem}}}

\begin{lemma}\label{lemma:coretrans}
  Assume $\asys$ is normalized. Consider  \arestricted normalized symbolic
  heap $\phi \in \shk{\rank}$ with no occurrences of context
  predicate symbols, and an injective structure $(\aistore, \aheap)$,
  such that $\dom(\aistore) = \fv{\phi} \cup \const$. We have
  $(\aistore,\aheap) \models_\asys \phi$ iff
  $(\aistore,\aheap) \models_{\coreset{\asys}} \psi$, for some $\psi
  \in \coretrans{\phi}$.
\end{lemma}
\optionalProof{Lemma \ref{lemma:coretrans}}{sec:core}{ ``$\Rightarrow$'' By induction on $\size{\phi}$. We consider
  the following cases: \begin{compactitem}
  \item $\phi=\emp$, $\phi=t_0 \mapsto (t_1, \ldots, t_\rank)$, $\phi=t_1 \seq t_2$,
    $\phi=t_1 \sneq t_2$ and $\phi=\phi_1 * \phi_2$: the proof is the same as the one in
    Lemma \ref{lemma:coretrans-qf}. 
  \item $\phi=p(\vec{t})$: in this case $\coretrans{\phi} = \set{\emp \swand p(\vec{t})}$ and the conclusion follows application
    of Lemma \ref{lemma:pred-equiv}.
  \item $\phi=\exists x ~.~ \phi_1$: since $(\aistore,\aheap)
    \models_\asys \exists x ~.~ \phi_1$, there exists $\ell \in \locs$
    such that $(\aistore[x \leftarrow \ell], \aheap) \models_\asys
    \phi_1$ and we distinguish the following
    cases. \begin{compactitem}
    \item If $\ell \not \in \aistore(\fv{\phi}\cup\const)$, since
      $\phi$ is normalized, by Definition \ref{def:normalized}
      (\ref{it1:normalized}) $x$ occurs in a points-to or in a
      predicate atom of $\phi_1$. Since $\asys$ is normalized, by
      Definition \ref{def:normalized} (\ref{it3:normalized}), we have
      that $\ell \in \loc(\aheap)$. Since $\dom(\aistore) =
      \fv{\phi}\cup\const$, the store $\aistore[x \leftarrow \ell]$ is
      necessarily injective, hence $(\aistore[x \leftarrow \ell],
      \aheap) \models_{\coreset{\asys}} \psi_1$, for some $\psi_1 \in
      \coretrans{\phi_1}$, by the inductive hypothesis and
      $(\aistore,\aheap) \models_{\coreset{\asys}} \heapex x ~.~
      \psi_1$, by Lemma \ref{lemma:bound-quant}.
    \item Otherwise, $\ell \in \aistore(\fv{\phi} \cup \const)$ and
      let $t \in \fv{\phi} \cup \const$ be a term such that $\ell =
      \aistore(t)$.  Then $(\aistore, \aheap) \models_\asys
      \phi_1[t/x]$ and $(\aistore,\aheap) \models_{\coreset{\asys}}
      \psi_1$, for some $\psi_1 \in \coretrans{\phi_1[t/x]}$, by the
      inductive hypothesis.
    \end{compactitem}
  \end{compactitem}

  \noindent``$\Leftarrow$'' By induction on $\size{\phi}$, considering
  the following cases: \begin{compactitem}
  \item $\phi=\emp$, $\phi=t_0 \mapsto (t_1, \ldots, t_\rank)$,
    $\phi=t_1 \seq t_2$, $\phi=t_1 \sneq t_2$ and $\phi=\phi_1 *
    \phi_2$: the proof is the same as the one in Lemma \ref{lemma:coretrans-qf}.
  \item $\phi=p(\vec{t})$: in this case $\psi = \emp \swand p(\vec{t})$ is
    the only possibility and the conclusion follows by an application
    of Lemma \ref{lemma:pred-equiv}.
  \item $\phi=\exists x ~.~ \phi_1$: by the definition of
    $\coretrans{\phi}$, we distinguish the following
    cases: \begin{compactitem}
  \item If $(\aistore,\aheap) \models_{\coreset{\asys}} \heapex x ~.~
    \psi_1$, for some $\psi_1 \in \coretrans{\phi_1}$, then
    $(\aistore[x \leftarrow \ell],\aheap) \models_{\coreset{\asys}}
    \psi_1$, for some $\ell \in \loc(\aheap) \setminus
    \aistore((\fv{\psi_1} \setminus \set{x}) \cup \const)$. By the
    definition of $\coretrans{\phi_1}$, we have $\fv{\psi_1} \subseteq
    \fv{\phi_1}$ and suppose, for a contradiction, that there exists a
    variable $y \in \fv{\phi_1} \setminus \fv{\psi_1}$. Then $y$ can
    only occur either in an equality atom $y \seq y$ or in some
    disequality $y \sneq t$, for some term $t \neq y$, and nowhere
    else. Both cases are impossible, because $\phi$ is normalized,
    thus by Condition (\ref{it1bis:normalized}) of Definition
    \ref{def:normalized}, $y$ necessarily occurs in a points-to or
    predicate atom. Hence, $\fv{\phi_1} = \fv{\psi_1}$ and
    consequently, we obtain $\ell \in \loc(\aheap) \setminus
    \aistore((\fv{\phi_1} \setminus \set{x}) \cup \const)$. Since
    $\dom(\aistore) = (\fv{\phi_1} \setminus \set{x}) \cup \const$, by
    the hypothesis of the Lemma, $\aistore[x \leftarrow \ell]$ is
    injective and, by the induction hypothesis, we obtain $(\aistore[x
      \leftarrow \ell], \aheap) \models_\asys \phi_1$, thus
    $(\aistore,\aheap) \models_\asys \phi$.
  \item Otherwise $(\aistore,\aheap) \models_{\coreset{\asys}} \psi$, for
    some $\psi \in \coretrans{\phi_1[t/x]}$ and some $t \in \fv{\phi}
    \cup \const$. By the induction hypothesis, we have
    $(\aistore,\aheap) \models_\asys \phi_1[t/x]$, thus
    $(\aistore,\aheap) \models_\asys \exists x ~.~ \phi_1$. \qed
  \end{compactitem}
  \end{compactitem}}
Next, we give an equivalent condition for the satisfaction of a
context predicate atom\longVersionOnly{ (Lemma
  \ref{lemma:simple-unfolding-sat})}, that relies on an unfolding of a
symbolic heap into a core formula:

\begin{definition}\label{def:simple-formulae}
  A formula $\varphi$ is a \emph{core unfolding} of a predicate atom
  $\Asterisk_{i=1}^n q_i(\vec{u}_i) \swand p(\vec{t})$, written
  $\Asterisk_{i=1}^n q_i(\vec{u}_i) \swand p(\vec{t})
  \sunfold{\coreset{\asys}} \varphi$, iff there
  exists: \begin{compactenum}
    \item\label{it1:simple-formulae} a rule $\Asterisk_{i=1}^n
      q_i(\vec{y}_i) \swand p(\vec{x}) \Leftarrow_{\coreset{\asys}}
      \exists \vec{z} ~.~ \phi$, where $\phi$ is quantifier free, and
  \item\label{it2:simple-formulae} a substitution $\sigma =
    [\vec{t}/\vec{x}, \vec{u}_1/\vec{y}_1, \ldots,
      \vec{u}_n/\vec{y}_n] \cup \zeta$, $\zeta \subseteq
    \{(z,t) \mid z \in \vec{z},~ t \in \vec{t} \cup \bigcup_{i=1}^n
      \vec{u}_i\}$, such that $\varphi \in \coretrans{\phi\sigma}$.
  \end{compactenum}
\end{definition}
A core unfolding of a predicate atom is always a quantifier-free
formula, obtained from the translation (into a disjunctive set of core
formul{\ae}) of the quantifier-free matrix of the body of a rule, in
which some of the existentially quantified variables in the rule occur
instantiated by the substitution $\sigma$. For instance, the rule
$\emp \swand p(x) \Leftarrow_{\coreset{\asys}} \exists y ~.~ x \mapsto
y$ induces the core unfoldings $\emp \swand p(a) \sunfold{\asys} a
\mapsto a$ and $\emp \swand p(a) \sunfold{\asys} a \mapsto u$, via the
substitutions $[a/x, a/y]$ and $[a/x, u/y]$, respectively.

\additionalMaterial{Additional Material for Core Formul{\ae}}{sec:core}{app:core-unfoldings}{

\begin{lemma}\label{lemma:simple-unfolding-sat}
  Given an injective structure $(\aistore,\aheap)$ and a context
  predicate atom $\Asterisk_{i=1}^n q_i(\vec{u}_i) \swand p(\vec{t})$,
  we have $(\aistore,\aheap) \models_{\coreset{\asys}}
  \Asterisk_{i=1}^n q_i(\vec{u}_i) \swand p(\vec{t})$ iff
  $(\aaistore,\aheap) \models_{\coreset{\asys}} \varphi$, for some
  core unfolding $\Asterisk_{i=1}^n q_i(\vec{u}_i) \swand p(\vec{t})
  \sunfold{\coreset{\asys}} \varphi$ and some injective extension
  $\aaistore$ of $\aistore$.
\end{lemma}
\proof{We assume w.l.o.g. a total well-founded order $\preceq$ on the
  set of terms $\terms$ and, for a set $T \subseteq \terms$, we denote
  by $\min_\preceq T$ the minimal term from $T$ with respect to this
  order. In the following, let $\theta \isdef
  [\vec{t}/\vec{x},\vec{u}_1/\vec{y}_1,\ldots,\vec{u}_n/\vec{y}_n]$.
  
  \noindent
  ''$\Rightarrow$'' If $(\aistore,\aheap) \models_{\coreset{\asys}}
  \Asterisk_{i=1}^n q_i(\vec{u}_i) \swand p(\vec{t})$ then there
  exists a rule $\Asterisk_{i=1}^n q_i(\vec{y}_i) \swand p(\vec{x})
  \Leftarrow_{\coreset{\asys}} \exists \vec{z} ~.~ \phi$, where $\phi$
  is quantifier-free, such that $(\aistore,\aheap)
  \models_{\coreset{\asys}} \exists \vec{z} ~.~ \phi\theta$.  Let
  $\aastore$ be a (not necessarily injective) $\vec{z}$-associate of
  $\aistore$ such that $(\aastore,\aheap) \models_{\coreset{\asys}}
  \phi\theta$. We define a substitution $\tau$, such that
  $\dom(\tau)\isdef\term{\phi\theta} \subseteq \dom(\aastore)$ and for
  each $x \in \dom(\tau)$:
  \begin{compactitem}
  \item if $x \in \dom(\aistore)$ then $\tau(x)\isdef x$,
  \item else, if $x \not\in \dom(\aistore)$ and
    $\aastore(x)=\aistore(y)$, for some $y \in \dom(\aistore)$, then
    $\tau(x) \isdef \min_\preceq \{z \in \dom(\aistore) \mid
    \aistore(z)=\aistore(y)\}$,
  \item otherwise, if $x \not\in \dom(\aistore)$ and
    $\aastore(x)\neq\aistore(y)$, for all $y \in \dom(\aistore)$, then
    $\tau(x)\isdef \min_\preceq \{y \in \dom(\aastore) \mid
    \aastore(y)=\aastore(x)\}$.
  \end{compactitem}  
  Let \(E \isdef \setof{\set{y \in \dom(\aastore) \mid
      \aastore(y)=\aastore(x)}}{x \in \dom(\aastore)}\); by
  construction, the sets in $E$ are pairwise disjoint. Let $\aaistore$
  be the restriction of $\aastore$ to the set $\dom(\aistore) \cup
  \set{\min_\preceq K \mid K \in E,~ K \cap \dom(\aistore) =
    \emptyset}$. Because $\aistore$ is injective, $\aaistore$ is
  easily shown to also be injective, thus it is an injective extension
  of $\aistore$. Moreover, because $(\aastore, \aheap)
  \models_{\coreset{\asys}} \phi\theta$ and $\aastore$ agrees with
  $\aaistore \circ \tau$ on $\dom(\aaistore)$, we deduce that
  $(\aaistore, \aheap) \models_{\coreset{\asys}} \phi(\tau \circ
  \theta)$.  We conclude by noticing that $(\aaistore, \aheap)
  \models_{\coreset{\asys}} \varphi$, for some $\varphi \in
  \coretrans{\phi(\tau\circ\theta)}$, by an application of Lemma
  \ref{lemma:coretrans-qf}, because $\phi(\tau\circ\theta)$ is
  quantifier-free.

  \noindent
  ``$\Leftarrow$'' If $\Asterisk_{i=1}^n q_i(\vec{u}_i) \swand
  p(\vec{t}) \sunfold{\asys} \varphi$, by Definition
  \ref{def:simple-formulae}, we have $\varphi \in
  \coretrans{\phi\theta}$, for some rule $\Asterisk_{i=1}^n
  q_i(\vec{y}_i) \swand p(\vec{x}) \Leftarrow_{\coreset{\asys}}
  \exists \vec{z} ~.~ \phi$, where $\phi$ is quantifier-free, and some
  substitution $\theta = [\vec{t}/\vec{x}, \vec{u}_1/\vec{y}_1,
    \ldots, \vec{u}_n/\vec{y}_i]\cup\zeta$, where $\zeta \subseteq
  \{(z,t) \mid z \in \vec{z},~ t \in \vec{t} \cup \bigcup_{i=1}^n
  \vec{u}_i\}$. Since $(\aaistore,\aheap) \models_{\coreset{\asys}}
  \varphi$ and $\phi\theta$ is quantifier-free, by Lemma
  \ref{lemma:coretrans-qf}, we obtain $(\aaistore,\aheap)
  \models_{\coreset{\asys}} \phi\theta$, hence $(\aistore,\aheap)
  \models_{\coreset{\asys}} (\exists \vec{z} ~.~ \phi)\theta$ and
  $(\aistore,\aheap) \models_{\coreset{\asys}} \Asterisk_{i=1}^n
  q_i(\vec{u}_i) \swand p(\vec{t})$ follows. \qed}

\begin{lemma}\label{lemma:bijective-unfolding-sat}
  Given a bijective structure $(\aistore,\aheap)$ and a context
  predicate atom $\Asterisk_{i=1}^n q_i(\vec{u}_i) \swand p(\vec{t})$,
  we have $(\aistore,\aheap) \models_{\coreset{\asys}}
  \Asterisk_{i=1}^n q_i(\vec{u}_i) \swand p(\vec{t})$ if and only if
  $(\aistore,\aheap) \models_{\coreset{\asys}} \varphi$, for some core
  unfolding $\Asterisk_{i=1}^n q_i(\vec{u}_i) \swand p(\vec{t})
  \sunfold{\coreset{\asys}} \varphi$. 
\end{lemma}
\proof{``$\Rightarrow$'' Let $\aistore'$ be the restriction of
  $\aistore$ to $\vec{t} \cup \bigcup_{i=1}^n \vec{u}_i$. Clearly, we
  have $(\aistore', \aheap) \models_{\coreset{\asys}}
  \Asterisk_{i=1}^n q_i(\vec{u}_i) \swand p(\vec{t})$. By Lemma
  \ref{lemma:simple-unfolding-sat}, there exists a core unfolding
  $\Asterisk_{i=1}^n q_i(\vec{u}_i) \swand p(\vec{t})
  \sunfold{\coreset{\asys}} \phi$ and an injective extension
  $\aaistore$ of $\aistore'$, such that $(\aaistore,\aheap)
  \models_{\coreset{\asys}} \phi$. Let $\tau$ be the substitution
  defined by $\tau(t) = u$ if and only if $\aaistore(t) =
  \aistore(u)$, for all $t \in \term{\phi}$. Note that, since
  $\aistore$ is bijective, for each $t \in \dom(\aaistore)$, there
  exists a unique $u \in \terms$, such that $\aaistore(t) =
  \aistore(u)$, hence $\tau$ is well-defined. Furthermore, since
  $\aaistore$ is injective, $\tau$ is also injective. We have
  $(\aistore,\aheap) \models_{\coreset{\asys}} \phi\tau$ and we are
  left with proving that $\Asterisk_{i=1}^n q_i(\vec{u}_i) \swand
  p(\vec{t}) \sunfold{\coreset{\asys}} \phi\tau$ is a core
  unfolding. By Proposition \ref{prop:coretrans-subs} we have
  $\phi\tau \in \coretrans{\varphi\sigma\tau}$, hence the result.
  ``$\Leftarrow$'' This is a consequence of Lemma
  \ref{lemma:simple-unfolding-sat}, using the fact that $\aistore$ is
  an injective extension of itself. \qed} }

\putInAppendix{
\label{ap:witness}
The following property of core formul{\ae} leads to a necessary and
sufficient condition for their satisfiability (Lemma
\ref{lemma:witness}). 
The idea is that the particular identity of
locations outside of the heap, assigned by the $\nheapall$ quantifier,
is not important when considering a model of a core formula.

\begin{definition}
For a set of locations $L \subseteq \locs$, we define $\aistore
\istoreq{L} \aistorep$ if and only if $\dom(\aistore) =
\dom(\aistorep)$ and, for each term $t \in \dom(\aistore)$, if
$\{\aistore(t), \aistorep(t)\} \cap L \neq \emptyset$ then
$\aistore(t)=\aistorep(t)$. 
\end{definition}
It is easy to check that $\istoreq{L}$ is
an equivalence relation, for each set $L \subseteq \locs$.

\begin{lemma}\label{lemma:istore-eq}
  Let $\aistore$ and $\aistorep$ be two injective stores and $\aheap$
  be a heap, such that $\aistore \istoreq{\loc(\aheap)} \aistorep$. If
  $\asys$ is progressing, then for every core formula $\varphi$, we have
  $(\aistore,\aheap) \models_{\coreset{\asys}} \varphi$ if and only if
  $(\aistorep,\aheap) \models_{\coreset{\asys}} \varphi$.
\end{lemma}
\proof{ We assume that $(\aistore,\aheap) \models_{\coreset{\asys}}
  \varphi$ and show that $(\aistorep,\aheap) \models_{\coreset{\asys}}
  \varphi$; the proof in the other direction is identical since
  $\istoreq{\loc(\aheap)}$ is symmetric. The proof is carried out by
  nested induction on $\card{\aheap}$ and $\size{\varphi}$. We assume,
  w.l.o.g., that $\dom(\aistore) = \dom(\aistore') = \fv{\varphi} \cup
  \const$. This is without loss of generality since the truth value of
  $\varphi$ in $(\aistore,\aheap)$ and $(\aistore',\aheap)$ depends
  only on the restriction of $\aistore$ (resp.\ $\aistore'$) to
  $\fv{\varphi}\cup \const$.
	
  For the base case assume that $\card{\aheap}=0$. By hypothesis,
  $\varphi = \heapex \vec{x} \nheapall \vec{y} ~.~ \Asterisk_{i=1}^n
  \left(\Asterisk_{j=1}^{k_i} q_j^i(\vec{u}^i_j) \swand
  p_i(\vec{t}_i)\right) * \Asterisk_{i=n+1}^m x_i \mapsto
  (t^i_1,\ldots,t^i_\rank)$ and since $\aheap = \emptyset$,
  necessarily, $\vec{x} = \emptyset$ and $m = 0$. Let $\aistore_1$ be
  an injective $\vec{y}$-associate of $\aistore$, where for all $y\in
  \vec{y}$, we have $\aistore_1(y) \in \locs
  \setminus\left[\aistore((\fv{\varphi} \cup \vec{y}) \cup
    \const)\right]$. Note that such a store exists because $\locs$ is
  infinite, wherease $\dom(\aistore)$ and $\vec{y}$ are both finite.
  By Lemma \ref{lemma:bound-quant} we have $(\aistore_1,
  \emptyset)\models_{\coreset{\asys}} \Asterisk_{i=1}^n
  \left(\Asterisk_{j=1}^{k_i} q_j^i(\vec{u}^i_j) \swand
  p_i(\vec{t}_i)\right)$. Thus for $i\in \interv{1}{n}$ we have
  $(\aistore_1, \emptyset) \models_{\coreset{\asys}}
  \Asterisk_{j=1}^{k_i}q_j^i(\vec{u}^i_j) \swand p_i(\vec{t}_i)$, and
  by Lemma \ref{lemma:context-emp}, we deduce that $k_i = 1$, $q_1^i =
  p_i$ and $\aistore_1(\vec{u}^i_1) = \aistore_1(\vec{t}_i)$. Since
  $\aistore_1$ is injective, we deduce that $\vec{u}^i_1 = \vec{t}_i$,
  but this is impossible because by hypothesis, the roots of a core
  formula are unique. Hence $(\aistore,\aheap)
  \not\models_{\coreset{\asys}} \varphi$ and the implication holds.

  For the induction step assume that $\card{\aheap}>0$, we consider
  the following cases:
  \begin{compactitem}
  \item $\varphi=\emp$: since $\card{\aheap}>0$, we cannot have
    $(\aistore,\aheap) \models_{\coreset{\asys}} \emp$.
  \item $\varphi=t_0 \mapsto (t_1,\ldots,t_\rank)$: in this case
    $\aheap = \{(\aistore(t_0), (\aistore(t_1), \ldots,
    \aistore(t_\rank)))\}$ and since $\aistore(t_0), \aistore(t_1),
    \ldots, \aistore(t_\rank) \in \loc(\aheap)$ and
    $\aistore\istoreq{\loc(\aheap)}\aistorep$, we also have $\aheap =
    \{(\aistorep(t_0), (\aistorep(t_1), \ldots,
    \aistorep(t_\rank)))\}$, thus $(\aistorep,\aheap) \models t_0
    \mapsto (t_1,\ldots,t_\rank)$.
  \item $\varphi=\Asterisk_{i=1}^n q_i(\vec{u}_i) \swand p(\vec{t})$:
    since $\card{\aheap} > 0$, $\varphi$ cannot be $p(\vec{t}) \swand
    p(\vec{t})$. Thus the first unfolding step is an instance of a
    rule obtained from \ref{infrule:notemp}. By Lemma
    \ref{lemma:simple-unfolding-sat}, there exists an injective
    extension $\aaistore$ of $\aistore$ such that $(\aaistore, \aheap)
    \models_{\coreset{\asys}} \psi$ where $\varphi
    \sunfold{\coreset{\asys}} \psi$, and because $\asys$ is
    progressing, $\psi$ is of the form $t_0 \mapsto
    (t_1,\ldots,t_\rank) * \psi'$.  Since the truth value of $\psi$ in
    $(\aaistore, \aheap)$ depends only on the restriction of
    $\aaistore$ to $\fv{\varphi} \cup \const$, we assume, w.l.o.g.,
    that $\dom(\aaistore)$ is finite.  The heap $\aheap$ can thus be
    decomposed into $\aheap_0 \uplus \aheap'$, where $(\aaistore,
    \aheap_0) \models_{\coreset{\asys}} t_0 \mapsto
    (t_1,\ldots,t_\rank)$ and $(\aaistore, \aheap')
    \models_{\coreset{\asys}} \psi'$. Consider the store
    $\astore_1 \isdef \{(x, \aaistore(x)) \mid x\in \dom(\aaistore)
  \setminus \dom(\aistore)\wedge \aaistore(x) \in \loc(\aheap)\}$
  and let $\aistore_1 \isdef \aistore' \cup \astore_1$.  Since
  $\dom(\aistore) = \dom(\aistorep)$ by hypothesis, $\aistore_1$ is
  well-defined. It is also injective because $\aistore'$ and
  $\aaistore$ are both injective, and if $\aistore_1(x) =
  \aistore_1(y)$, where $x\in \dom(\aistore')$ and $y\in
  \dom(\astore_1)$, then $\aistore_1(y) = \aaistore(y) \in
  \loc(\aheap)$, hence we also have $\aistore_1(x) = \aistorep(x) \in
  \loc(\aheap)$. By hypothesis $\aistore \istoreq{\loc(\aheap)}
  \aistorep$, hence $\aistorep(x) = \aistore(x) = \aaistore(x)$, so
  that $\aaistore(x) = \aaistore(y)$. Since $\aaistore$ is injective,
  we deduce that $x=y$.  Now let $\aistore_2$ be an injection from
  $\dom(\aaistore) \setminus \dom(\aistore_1)$ onto $\locs \setminus
  \left(\img(\aaistore) \cup \img(\aistore') \cup
  \loc(\aheap)\right)$. Note that such an extension necessarily exists
  since $\dom(\aaistore)$, $\dom(\aistore')$ and $\loc(\aheap)$ are
  all finite whereas $\locs$ is infinite. Let $\aaistorep \isdef
  \aistore_1 \cup \aistore_2$, it is straightforward to verify that
  $\aaistore'$ is injective and that $\aaistore \istoreq{\loc(\aheap)}
  \aaistorep$. By the inductive hypothesis we have $(\aaistorep,
  \aheap_0) \models_{\coreset{\asys}} t_0 \mapsto
  (t_1,\ldots,t_\rank)$ and $(\aaistorep, \aheap')
  \models_{\coreset{\asys}} \psi'$, and by Lemma
  \ref{lemma:simple-unfolding-sat} we deduce that $(\aistore',\aheap)
  \models_{\coreset{\asys}} \Asterisk_{i=1}^n q_i(\vec{u}_i) \swand
  p(\vec{t})$.
  \item $\varphi=\heapex x ~.~ \psi$: by Lemma
    \ref{lemma:bound-quant}, there exists an $x$-associate $\aastore$
    of $\aistore$, such that $\aastore(x) \in \loc(\aheap) \setminus
    \aistore((\fv{\psi} \setminus \set{x}) \cup \const)$ and
    $(\aastore,\aheap) \models_{\coreset{\asys}} \psi$. We distinguish
    two cases. \begin{compactitem}
    \item If $\aastore(x)=\aistore(y)$ for some $y \in \dom(\aistore)$
      then $(\aistore,\aheap) \models_{\coreset{\asys}} \psi[y/x]$
      and, by the induction hypothesis, we have $(\aistore',\aheap)
      \models_{\coreset{\asys}} \psi[y/x]$. Since $\aistore
      \istoreq{\loc(\aheap)} \aistore'$ and $\aistore(y) \in
      \loc(\aheap)$, we have $\aistore(y) =
      \aistore'(y)$. Furthermore, since $\aastore(x) \not \in
      \aistore((\fv{\psi} \setminus \set{x}) \cup \const)$,
      necessarily $y \not\in (\fv{\psi} \setminus \set{x}) \cup
      \const$ and, because $\aistore'$ is injective, $\aistore'(y)
      \not\in \aistore'((\fv{\psi} \setminus \set{x}) \cup
      \const)$. Since $\aistore'(y) = \aastore(x) \in \loc(\aheap)$
      and $(\aistore',\aheap) \models_{\coreset{\asys}} \psi[y/x]$, we
      deduce that $(\aistore',\aheap) \models_{\coreset{\asys}}
      \heapex x ~.~ \psi$.
    \item Otherwise we have $\aastore(x) \neq \aistore(y)$ for all $y
      \in \dom(\aistore)$ and $\aastore$ is therefore injective. Let
      $\aastore' \isdef \aistore'[x \leftarrow \aastore(x)]$. Suppose
      that $\aastore(x)=\aistore'(y)$, for some $y \in
      \dom(\aistore')$. Since $\aistore \istoreq{\loc(\aheap)}
      \aistore'$ we have $\dom(\aistore')=\dom(\aistore)$, hence $y
      \in \dom(\aistore)$ and, since $\aistore(y) \in \loc(\aheap)$,
      we obtain $\aistore(y)=\aistore'(y)=\aastore(x)$, in
      contradiction with the assumption of this case. Thus $\aastore'$
      is injective and, using the fact that $\aastore
      \istoreq{\loc(\aheap)} \aastore'$, we deduce that
      $(\aastore',\aheap) \models_{\coreset{\asys}} \psi$ by the
      induction hypothesis. Since $\aastore'(x) = \aastore(x) \not\in
      \dom(\aistore) = \dom(\aistore')$, we have $\aastore'(x) \not\in
      \aistore'((\fv{\psi}\setminus\set{x}) \cup \const)$. Moreover,
      $\aastore'(x) = \aastore(x) \in \loc(\aheap)$, thus
      $(\aistore',\aheap) \models_{\coreset{\asys}} \heapex x ~.~
      \psi$ by Lemma \ref{lemma:bound-quant}.
    \end{compactitem}
  \item $\nheapall x ~.~ \psi$: By Lemma \ref{lemma:bound-quant},
    $(\aistore,\aheap) \models \nheapall x ~.~ \psi$ iff $(\aistore[x
    \leftarrow \ell] \models \psi$ holds for all locations $\ell \in
    \locs$ such that $\ell \not \in \loc(\aheap) \cup
    \aistore(\fv{\nheapall x ~.~ \psi})$.  Let $\ell \in
    \locs\setminus \left[\loc(\aheap)\cup \img(\aistore)\right]$ be an
    arbitrary location. Since $\locs$ is infinite and
    $\loc(\aheap)\cup\img(\aistore)$ is finite, such a location
    exists. By definition of $\nheapall$, we have
    $(\aaistore[x\leftarrow \ell], \aheap) \models_{\coreset{\asys}}
    \psi$. Now let $\ell' \in \locs\setminus
    \left[\loc(\aheap)\cup\img(\aistore')\right]$ be an arbitrary
    location. Clearly $\aistore[x\leftarrow\ell]$ and
    $\aistorep[x\leftarrow\ell']$ are injective stores and
    $\aistore[x\leftarrow\ell] \istoreq{\loc(\aheap)}
    \aistorep[x\leftarrow\ell']$, since $\ell,\ell' \not \in
    \loc(\aheap)$. By the induction hypothesis, we have
    $(\aistorep[x\leftarrow\ell'],\aheap) \models_{\coreset{\asys}}
    \psi$ and, since the choice of $\ell' \in \locs \setminus
    \left[\img(\aistore') \cup \loc(\aheap)\right] = \locs \setminus
    \left[\aistore'(\fv{\varphi} \cup \const) \cup
      \loc(\aheap)\right]$ was arbitrary, $(\aistorep,\aheap)
    \models_{\coreset{\asys}} \nheapall x ~.~ \psi$, by definition of
    $\nheapall$.  \qed
\end{compactitem}
}

The following lemma gives an alternative condition for the
satisfiability of core formul{\ae}. Intuitively, it is sufficient to
instantiate the bounded universal quantifiers with arbitrary locations
that are not in the image of the store, nor in the range of the heap.

\begin{lemma}\label{lemma:witness}
Given a core formula $\varphi = \heapex \vec{x} \nheapall \vec{y} ~.~
\psi$, where $\psi$ is quantifier-free, and an injective structure
$(\aistore,\aheap)$, such that
$\dom(\aistore)=\fv{\varphi}\cup\const$, we have $(\aistore,\aheap)
\models_{\coreset{\asys}} \varphi$ if and only if $(\aaistore,\aheap)
\models_{\coreset{\asys}} \psi$, for some injective
$(\vec{x}\cup\vec{y})$-associate $\aaistore$ of $\aistore$, such that
$\aaistore(\vec{x}) \subseteq \loc(\aheap)$ and $\aaistore(\vec{y})
\cap \loc(\aheap) = \emptyset$.
\end{lemma}
\proof{``$\Rightarrow$''
  Since $\locs$ is infinite and $\dom(\aistore) \cup \loc(\aheap)$ is
  finite, there exists an injective $(\vec{x}\cup\vec{y})$-associate
  $\aaistore$ of $\aistore$, such that $\aaistore(\vec{x}) \subseteq
  \loc(\aheap)$, $\aaistore(\vec{y}) \cap \loc(\aheap) = \emptyset$
  and $(\aaistore,\aheap)\models_{\coreset{\asys}} \psi$, by the
  semantics of the bounded quantifiers $\heapex$ and $\nheapall$
  (see Lemma \ref{lemma:bound-quant}).  
  
  ``$\Leftarrow$'' Let $\vec{x} =
  \set{x_1,\ldots,x_n}$, $\vec{y} = \set{y_1,\ldots,y_m}$ and let
  $\ell_1, \ldots, \ell_n \in \loc(\aheap) \setminus
  \aistore\left((\fv{\psi}\setminus(\vec{x}\cup\vec{y})) \cup
  \const\right)$ and $\ell_{n+1}, \ldots, \ell_{n+m} \in \locs
  \setminus
  \left(\loc(\aheap)\cup\aistore((\fv{\psi}\setminus\vec{y})\cup\const)\right)$
  be arbitrary locations, since $\locs$ is infinite and $\fv{\psi}\cup \const \cup \loc(\aheap)$ is finite, such locations necessarily
  exist. Let $\aaistore =
  \aistore[x_1\leftarrow\ell_1,\ldots,x_{n}\leftarrow\ell_{n}]$. Then
  $\aaistore[y_1 \leftarrow \ell_{n+1}, \ldots, y_{m} \leftarrow
    \ell_{n+m}] \istoreq{\loc(\aheap)} \aaistore$, thus
  $(\aaistore[y_1 \leftarrow \ell_{n+1}, \ldots, y_{m} \leftarrow
    \ell_{n+m}],\aheap) \models_{\coreset{\asys}} \psi$, by Lemma
  \ref{lemma:istore-eq}. Since the choice of
  $\ell_{n+1},\ldots,\ell_{n+m}$ is arbitrary, we deduce that
  $(\aaistore[y_1 \leftarrow \ell_{n+1}, \ldots, y_m \leftarrow
    \ell_{n+m}],\aheap) \models_{\coreset{\asys}} \nheapall y ~.~
  \psi$ and that $(\aistore,\aheap) \models_{\coreset{\asys}} \heapex x
  \nheapall y ~.~ \psi$.  \qed}}

We now define a equivalence relation, of finite index, on the set of
injective structures. Intuitively, an equivalence class is defined by
the set of core formul{\ae} that are satisfied by all structures in
the class (with some additional conditions).  First, we introduce the
overall set of core formul{\ae}, over which these equivalence classes
are defined:

\begin{definition}\label{def:allvars}
  Let $\allvars{\aprob} \isdef \allvars{\aprob}^1 \cup
  \allvars{\aprob}^2$, such that $\allvars{\aprob}^1 \cap
  \allvars{\aprob}^2 = \emptyset$ and $\card{\allvars{\aprob}^i} =
  \probwidth{\aprob}$, for $i = 1,2$ and denote by $\core{\aprob}$ the
  set of core formul{\ae} $\varphi$ such that $\roots{\varphi} \cap
  \fv{\varphi} \subseteq \allvars{\aprob}^1$, $\roots{\varphi}
  \setminus \fv{\varphi} \subseteq \allvars{\aprob}^2 \cup \const$ and
  no variable in $\allvars{\aprob}^1$ is bound in $\varphi$.
\end{definition}
Note that $\core{\aprob}$ is a finite set, because both
$\allvars{\aprob}$ and $\const$ are finite. Intuitively,
$\allvars{\aprob}^1$ will denote ``local'' variables introduced by
unfolding the definitions on the left-hand sides of the entailments,
whereas $\allvars{\aprob}^2$ will denote existential variables
occurring on the right-hand sides.  Second, we characterize an
injective structure by the set of core formul{\ae} it satisfies:

\begin{definition}\label{def:coreabs}
  For a core formula $\varphi = \heapex \vec{x} \nheapall \vec{y} ~.~
  \psi$, we denote by $\witset{\asys}{\aistore}{\aheap}{\varphi}$ the
  set of stores $\aaistore$ that are injective
  $(\vec{x}\cup\vec{y})$-associates of $\aistore$, and such
  that:\begin{inparaenum}[(1)]
  \item\label{it1:witset} $(\aaistore,\aheap) \models_{\coreset{\asys}} \psi$,
  \item\label{it2:witset} $\aaistore(\vec{x}) \subseteq \loc(\aheap)$, and
  \item\label{it3:witset} $\aaistore(\vec{y}) \cap \loc(\aheap) =
    \emptyset$. 
  \end{inparaenum}
  The elements of this set are called \emph{witnesses for
    $(\aistore,\aheap)$ and $\varphi$}.

  The \emph{core abstraction} of an injective structure
  $(\aistore,\aheap)$ is the set $\coreabs{\aistore,\aheap}{\aprob}$
  of core formul{\ae} $\varphi \in \core{\aprob}$ for which there
  exists a witness $\aaistore \in
  \witset{\asys}{\aistore}{\aheap}{\varphi}$ such that
  $\aaistore(\lroots{\varphi}) \cap \dom(\aheap) = \emptyset$.
\end{definition}
An injective structure $(\aistore,\aheap)$ satisfies each core formula
$\varphi \in \coreabs{\aistore,\aheap}{\aprob}$\footnote{An easy
  consequence of Lemma \ref{lemma:witness}\shortVersionOnly{ in
    Appendix \ref{ap:witness}}.}, fact that is witnessed by an
extension of the store assigning the universally quantified variables
random locations outside of the heap. Further, any core formula
$\varphi$ such that $(\aistore,\aheap)\models \varphi$ and
$\lroots{\varphi} = \emptyset$ occurs in
$\coreabs{\aistore,\aheap}{\aprob}$.

Our entailment checking algorithm relies on the definition of the
\emph{profile} of a symbolic heap. Since each symbolic heap is
equivalent to a finite disjunction of existential core formul{\ae},
when interpreted over injective normal structures, it is sufficient to
consider only profiles of core formul{\ae}:

\begin{definition}\label{def:profile}
  A \emph{profile} for an entailment problem $\aprob = (\asys,
  \seqset)$ is a relation $\profile \subseteq \core{\aprob} \times
  2^{\core{\aprob}}$ such that, for any core formula $\phi \in
  \core{\aprob}$ and any set of core formul{\ae} $F \in
  2^{\core{\aprob}}$, we have $(\phi, F) \in \profile$ iff $F =
  \coreabs{\aistore, \aheap}{\aprob}$, for some injective normal
  $\coreset{\asys}$-model $(\aistore,\aheap)$ of $\phi$, with
  $\dom(\aistore) = \fv{\phi}\cup \const$.
\end{definition}
Assuming the existence of a profile, the effective construction of
which will be given in Section \ref{sec:coreabs}, the following lemma
provides an algorithm that decides the validity of $\aprob$:

\begin{lemma}\label{lemma:entailment}
Let $\aprob = (\asys, \seqset)$ be a normalized \restricted entailment
problem and $\profile \subseteq \core{\aprob} \times
2^{\core{\aprob}}$ be a profile for $\aprob$. Then $\aprob$ is valid
iff, for each sequent $\phi \vdash_\aprob
\psi_1,\ldots,\psi_n$, each core formula $\varphi \in
\coretrans{\phi}$ and each pair $(\varphi,F) \in \profile$, we have $F
\cap \coretrans{\psi_i} \neq \emptyset$, for some $i \in
\interv{1}{n}$.
\end{lemma}
\optionalProof{Lemma
  \ref{lemma:entailment}}{sec:core}{``$\Rightarrow$'' Let $\phi
  \vdash_\aprob \psi_1, \ldots, \psi_n$ be a sequent and $\varphi \in
  \coretrans{\phi}$ be a core formula. Since $\phi$ is quantifier-free
  and $\fv{\phi} = \emptyset$ (Definition \ref{def:entailment}), we
  deduce that $\varphi$ is quantifier-free and $\roots{\varphi}
  \subseteq \term{\phi} \subseteq \const$, hence $\varphi \in
  \core{\aprob}$, by Definition \ref{def:allvars}. If there is no set
  of core formul{\ae} $F \in 2^{\core{\aprob}}$ such that $(\varphi,
  F) \in \profile$, then there is nothing to prove. Otherwise, let $F
  \in 2^{\core{\aprob}}$ be a set of core formul{\ae}, such that
  $(\varphi, F) \in \profile$. By Definition \ref{def:profile}, there
  exists an injective normal $\coreset{\asys}$-model
  $(\aistore,\aheap)$ of $\varphi$, such that $F =
  \coreabs{\aistore,\aheap}{\aprob}$.  Since $\aprob$ is valid, $\phi
  \models_\asys \bigvee_{i=1}^n \psi_i$, hence there exists $i \in
  \interv{1}{n}$, such that $(\aistore,\aheap) \models_\asys \psi_i$.
  Since $\dom(\aistore)=\const=\fv{\psi_i} \cup \const$, by Lemma
  \ref{lemma:coretrans}, we obtain $(\aistore,\aheap)
  \models_{\coreset{\asys}} \zeta$, for some $\zeta \in
  \coretrans{\psi_i}$. Since $\fv{\zeta} \subseteq \fv{\psi_i} =
  \emptyset$, we also have that $(\aistore,\aheap) \models_\asys
  \zeta$. We show that $\zeta \in \core{\aprob}$. First, all predicate
  atoms in $\zeta$ are of the form $\emp \swand p(\vec{t})$, and if
  $\zeta$ contains two distinct occurrences of atoms $\emp \swand
  p(\vec{t})$ and $\emp \swand q(\vec{s})$ with $\roots{p(\vec{t})} =
  \roots{q(\vec{s})}$ then $\zeta$ cannot be satisfiable, because the
  same location cannot be allocated in two disjoint parts of the heap.
  Second, since $\aprob$ is normalized, all existential variables must
  occur in a predicate or points-to atom.  Thus all the conditions of
  Definition \ref{def:core-formulae} are satisfied. Finally, since
  $\card{\allvars{\aprob}^2} = \probwidth{\aprob} \geq \size{\psi_i}$, we may
  assume up to an $\alpha$-renaming that all the bound variables in
  $\psi_i$ are in $\allvars{\aprob}^2$, hence the same holds for
  $\zeta$. Since any predicate atom that occurs in a core formula in
  $\coretrans{\psi_i}$ is of the form $\emp \swand p(\vec{t})$, we
  have $\lroots{\psi_i\sigma} = \emptyset$. By Definition
  \ref{def:coreabs}, we have $\zeta \in
  \coreabs{\aistore,\aheap}{\aprob} = F$, thus $F \cap
  \coretrans{\psi_i} \neq \emptyset$.

  \noindent''$\Leftarrow$'' Let $\phi \vdash_\aprob \psi_1, \ldots,
  \psi_n$ be a sequent. Let $(\aistore,\aheap)$ be an $\asys$-model of
  $\phi$. Since $\fv{\phi} = \fv{\psi_1} = \dots = \fv{\psi_n} =
  \emptyset$, we may assume, w.l.o.g., that $\dom(\astore) = \const$,
  and that $\aistore$ is injective (by Assumption \ref{ass:dist-const}
  all constants are mapped to pairwise distinct locations). It is
  sufficient to prove that $(\aistore,\aheap) \models_\asys \psi_i$,
  for some $i \in \interv{1}{n}$, because in this case, we also have
  $(\astore, \aheap) \models_\asys \psi_i$. By Lemma
  \ref{lemma:normal-entailment}, it is sufficient to show that any
  injective normal $\asys$-model of $\phi$ is an $\asys$-model of
  $\psi_i$, for some $i \in \interv{1}{n}$, so let us assume that
  $(\aistore,\aheap)$ is also a normal $\asys$-model of $\phi$. Since
  $\fv{\phi} = \emptyset$, by Lemma \ref{lemma:coretrans-qf}, we have
  $(\aistore,\aheap) \models_{\coreset{\asys}} \varphi$, for some
  $\varphi \in \coretrans{\phi}$. By Definition \ref{def:profile}, we
  have $(\varphi, \coreabs{\aistore,\aheap}{\aprob}) \in \profile$,
  hence $\coreabs{\aistore,\aheap}{\aprob} \cap \coretrans{\psi_i}
  \neq \emptyset$, for some $i \in \interv{1}{n}$. Then there exists a
  core formula $\zeta \in \coretrans{\psi_i}$, such that $(\aistore,
  \aheap) \models_{\coreset{\asys}} \zeta$, by Definition
  \ref{def:coreabs} and, since $\dom(\aistore) = \const = \fv{\psi_i}
  \cup \const$, by Lemma \ref{lemma:coretrans}, we obtain $(\aistore,
  \aheap) \models_{\asys} \psi_i$. Since the choice of
  $(\aistore,\aheap)$ is arbitrary, each injective normal
  $\asys$-model of $\phi\sigma$ is a model of $\psi_i\sigma$, for some
  $i \in \interv{1}{n}$. \qed } \shortVersionOnly{ The proof relies on
  Lemma \ref{lemma:normal-entailment}, according to which entailments
  can be tested by considering only normal models. As one expects,
  Lemma \ref{lemma:coretrans} is used in this proof to ensure that the
  translation $\coretrans{.}$ of symbolic heaps into core formul{\ae}
  preserves the injective models.  }

\section{Construction of the Profile Function}
\label{sec:coreabs}

\longVersionOnly{\subsection{Construction Rules}}

For a given normalized entailment problem $\aprob = (\asys, \seqset)$,
describe the construction of a profile $\profile_\aprob \subseteq
\core{\aprob} \times 2^{\core{\aprob}}$, recursively on the structure
of core formul{\ae}. We assume that the set of rules $\asys$ is
progressing, connected and \restricted. The relation $\profile_\aprob$
is the least set satisfying the recursive constraints
(\ref{eq:pto-core}), (\ref{eq:pred-core}), (\ref{eq:sep-core}) and
(\ref{eq:ex-core}), given in this section. Since these recursive
definitions are monotonic, the least fixed point exists and is
unique. We shall prove later (Theorem \ref{thm:main}) that the least
fixed point can, moreover, be attained in a finite number of steps by
a standard Kleene iteration.

\additionalMaterial{Additional Material for the Construction of Profiles}{sec:coreabs}{app:coreabs}{}

\paragraph{Points-to Atoms} For a points-to atom $t_0 \mapsto
(t_1,\ldots,t_\rank)$, such that $t_0, \ldots, t_\rank \in
\allvars{\aprob}^1 \cup \const$, we have:
\begin{eqnarray}\label{eq:pto-core}
  &&(t_0 \mapsto (t_1,\ldots,t_\rank),~ F) \in \profile_\aprob 
  \text{, iff $F$ is the set containing $t_0 \mapsto (t_1,\ldots,t_\rank)$ and all core formul{\ae} }\nonumber\\[-1mm]
  && \text{of the form } \nheapall \vec{z} ~.~ \Asterisk_{i=1}^n ~q_i(\vec{u}_i) \swand  p(\vec{t}) \in \core{\aprob},
  \text{ where } \vec{z}=(\vec{t} \cup \vec{u}_1 \cup \ldots \cup \vec{u}_n) \setminus \left(\set{t_0,\ldots,t_\rank} \cup \const\right)\nonumber\\[-1mm]
  &&\text{ such that } \emp \swand p(\vec{t}) \sunfold{\coreset{\asys}} t_0 \mapsto (t_1,\ldots,t_\rank) * \Asterisk_{i=1}^n ~\emp \swand q_i(\vec{u}_i) 
\end{eqnarray}
For instance, if $\asys = \{ p(x) \Leftarrow \exists y,z~.~ x \mapsto
y * q(y,z),\ q(x,y) \Leftarrow x \mapsto y \}$, with
$\allvars{\aprob}^1 = \{ u,v \}$ and $\allvars{\aprob}^2 = \{ z \}$,
then $\profile_\aprob$ contains the pair $(u \mapsto v, F)$ with $F =
\{ u \mapsto v, \emp \swand q(u,v), \nheapall z~.~ q(v,z) \swand p(u)
\}$.

\putInAppendix{ 
  \longVersionOnly{We prove that  constraint (\ref{eq:pto-core})
     indeed defines the profile of a points-to atom:}

\begin{lemma}\label{lemma:pto-core}
  If $\asys$ is progressing, then for all terms $t_0, \ldots, t_\rank \in
  \allvars{\aprob}^1 \cup \const$ and all sets of core formul{\ae}
  $F \in 2^{\core{\aprob}}$, we have $(t_0 \mapsto (t_1, \ldots,
  t_\rank), F) \in \profile_\aprob$ if and only if
  $F=\coreabs{\aistore,\aheap}{\aprob}$, for some injective
  $\asys$-model $(\aistore,\aheap)$ of $t_0 \mapsto
  (t_1,\ldots,t_\rank)$, such that  $\dom(\aistore)=\set{t_0, \ldots,
    t_\rank} \cup \const$.
\end{lemma}
\proof{ Let $(\aistore,\aheap)$ be an arbitrary injective model of
  $t_0 \mapsto (t_1, \ldots, t_\rank)$ where $\dom(\aistore) =
  \set{t_0, \ldots, t_\rank} \cup \const$ and $\aheap =
  \set{(\aistore(t_0), (\aistore(t_1), \ldots, \aistore(t_\rank)))}$.
  We show $F = \coreabs{\aistore,\aheap}{\aprob}$ below, where $F$ is
  defined by (\ref{eq:pto-core}):

  \noindent
  ``$\subseteq$'' Let $\phi \in F$ and consider the following
  cases: \begin{compactitem}
  \item If $\phi = t_0 \mapsto (t_1,\ldots,t_\rank)$ then
    $(\aistore,\aheap) \models t_0 \mapsto (t_1,\ldots,t_\rank)$ and
    $\lroots{\phi}=\emptyset$, thus $\phi \in
    \coreabs{\aistore,\aheap}{\aprob}$ (see Definition
    \ref{def:coreabs}). 
  \item Otherwise, $\phi = \nheapall \vec{z} ~.~ \Asterisk_{i=1}^n
    q_i(\vec{u}_i) \swand p(\vec{t})$, where $\vec{z} =
    \left(\bigcup_{i=1}^n \vec{u}_i \cup \vec{t}\right) \setminus
    (\{t_0,\ldots,t_\rank\} \cup \const)$ 
    and $\emp \swand p(\vec{t})
    \sunfold{\coreset{\asys}} t_0 \mapsto (t_1,\ldots,t_\rank) *
    \Asterisk_{i=1}^n \emp \swand q_i(\vec{u}_i)$. Note that by the progressivity condition, we have $t_0 =
    \aroot(p(\vec{t}))$. By Definition \ref{def:simple-formulae},
    there exists a rule: \[\emp \swand p(\vec{x})
    \Leftarrow_{\coreset{\asys}} \exists \vec{v} ~.~ \psi *
    \Asterisk_{i=1}^n (\emp \swand q_i(\vec{y}_i))~~(\dagger)\] such
    that $t_0 \mapsto (t_1,\ldots,t_\rank) \in
    \coretrans{\psi\sigma}$ and $\sigma$ is an extension of
    $[\vec{t}/\vec{x},\vec{u}_1/\vec{y}_1,\ldots,\vec{y}_n/\vec{u}_n]$
    with pairs $(z,t)$, where $z \in \vec{v}$ and $t \in \vec{t} \cup
    \bigcup_{i=1}^n \vec{u}_i$. By (\ref{infrule:notemp}), the rule
    ($\dagger$) occurs because of the existence of a
    rule \[p(\vec{x}) \Leftarrow_\asys \exists \vec{w} ~.~ \varphi *
    \Asterisk_{i=1}^n q_i(\vec{z}_i)~~(\dagger\dagger)\] and a
    substitution $\tau : \vec{w} \map \vec{x}$, such that $\psi =
    \varphi\tau$, $\vec{v} = \vec{w} \setminus \dom(\tau)$ and
    $\vec{y}_i = \tau(\vec{z}_i)$, for all $i \in
    \interv{1}{n}$. Applying $\tau$ to $(\dagger\dagger)$, by
    (\ref{infrule:notemp}), we obtain the rule:
    \[\Asterisk_{i=1}^n q_i(\vec{y}_i) \swand p(\vec{x}) \Leftarrow_{\coreset{\asys}} 
    \exists \vec{v} ~.~ \psi * \Asterisk_{i=1}^n (q_i(\vec{y}_i)
    \swand q_i(\vec{y}_i))~~(\ddagger)\] Let $\aaistore$ be an
    injective $\vec{v}$-associate of $\aistore$. Such an associate
    necessarily exists, for instance if $\aaistore$ maps $\vec{v}$
    into pairwise distinct locations, that are further distinct from
    $\img(\aistore)$; since $\locs$ is infinite and $\dom(\aistore)$
    is assumed to be finite, such locations always exist. By
    $\alpha$-renaming if necessary, we can assume that $\vec{v} \cap
    \set{t_0,\ldots,t_\rank} = \emptyset$, thus $\aistore$ and
    $\aaistore$ agree on $\{t_0,\ldots,t_\rank\}$ and we obtain
    $(\aaistore,\aheap) \models t_0 \mapsto
    (t_1,\ldots,t_\rank)$. Since $t_0 \mapsto (t_1, \ldots, t_\rank)
    \in \coretrans{\psi\sigma}$, by Lemma \ref{lemma:coretrans-qf}, we
    have $(\aaistore,\aheap) \models \psi\sigma$. By Lemma
    \ref{lemma:context-emp}, we have $(\aaistore,\aheap)
    \models_{\coreset{\asys}} \psi\sigma * \Asterisk_{i=1}^n
    (q_i(\vec{u}_i) \swand q_i(\vec{u}_i))$ and, by rule ($\ddagger$)
    we obtain $(\aaistore,\aheap) \models_{\coreset{\asys}}
    \Asterisk_{i=1}^n q_i(\vec{u}_i) \swand p(\vec{t})$. There remains
    to prove that $\aaistore \in
    \witset{\asys}{\aistore}{\aheap}{\phi}$. Since there are no
    existentially quantified variables in $\phi$, it suffices to show
    that $\aaistore(\vec{z}) \cap \loc(\aheap) = \aaistore(\vec{z})
    \cap \aaistore(\set{t_0,\ldots,t_\rank}) = \aaistore(\vec{z} \cap
    \set{t_0,\ldots,t_\rank}) = \emptyset$, because $\aaistore$ agrees
    with $\aistore$ on $\set{t_0,\ldots,t_\rank}$, $\aaistore$ is
    injective and $\vec{z} \cap \set{t_0,\ldots,t_\rank} = \emptyset$,
    by (\ref{eq:pto-core}). Finally, we prove the condition of
    Definition \ref{def:coreabs}, namely that
    $\aaistore(\lroots{\nheapall\vec{z} ~.~ \Asterisk_{i=1}^n
      q_i(\vec{u}_i) \swand p(\vec{t})}) \cap \dom(\aheap) =
    \{\aaistore(\aroot(q_i(\vec{u}_i))) \mid i \in \interv{1}{n}\}
    \cap \{\aaistore(t_0)\} = \emptyset$. Suppose, for a
    contradiction, that this set is not empty, thus
    $\aaistore(t_0)=\aaistore(\aroot(q_i(\vec{u}_i)))$, for some $i
    \in \interv{1}{n}$. Because $\aaistore$ is injective, we have
    $t_0=\aroot(q_i(\vec{u}_i))$. However, this contradicts with the
    condition $\Asterisk_{i=1}^n q_i(\vec{u}_i) \swand p(\vec{t}) \in
    \core{\aprob}$, which by Definition \ref{def:core-formulae},
    requires that $\aroot(p(\vec{t})) \not = \aroot(q_i(\vec{u}_i))$,
    i.e., $t_0 \neq \aroot(q_i(\vec{u}_i))$.
  \end{compactitem}

  \noindent''$\supseteq$'' Let $\phi = \heapex \overline{\vec{x}}
  \nheapall \overline{\vec{y}} ~.~ \psi \in
  \coreabs{\aistore,\aheap}{\aprob}$ be a core formula, where $\psi$
  is quantifier-free.  Note that, since $\phi \in \core{\aprob}$, we
  have $(\overline{\vec{x}} \cup \overline{\vec{y}}) \cap
  \allvars{\aprob}^1 =\emptyset$ because no variable in
  $\allvars{\aprob}^1$ can be bound in $\phi$; thus, since
  $\set{t_0,\ldots,t_\rank} \subseteq \allvars{\aprob}^1 \cup \const$
  by hypothesis, we have:
  \[(\overline{\vec{x}}\cup\overline{\vec{y}}) \cap
  \set{t_0,\ldots,t_\rank} = \emptyset~~ (\dagger).\]
  By Definition \ref{def:coreabs}, we
  have $(\aaistore,\aheap) \models_{\coreset{\asys}} \psi$, for some
  injective witness $\aaistore \in
  \witset{\asys}{\aistore}{\aheap}{\phi}$, such that
  $\aaistore(\overline{\vec{x}}) \subseteq \loc(\aheap)$ and
  $\aaistore(\overline{\vec{y}}) \cap \loc(\aheap) = \emptyset$. Since
  $(\aistore,\aheap) \models t_0 \mapsto (t_1,\ldots,t_\rank)$, it
  must be the case that $\card{\aheap}=1$, hence $\psi$ must be of
  either one of the forms: \begin{compactitem}
  \item $v_0 \mapsto (v_1,\ldots,v_\rank)$: in this case
    $\dom(\aheap)=\{\aaistore(v_0)\}$ and
    $\aheap(\aaistore(v_0))=(\aaistore(v_1), \ldots,
    \aaistore(v_\rank))$, thus $\loc(\aheap) = \{\aaistore(v_0),
    \ldots, \aaistore(v_\rank)\}$. By ($\dagger$),
    $\aaistore$ and $\aistore$ must agree over $t_0, \ldots, t_\rank$,
    hence we have $\aaistore(t_i) = \aistore(t_i)$, for all $i \in
    \interv{0}{\rank}$. Since $(\aistore,\aheap) \models t_0 \mapsto
    (t_1, \ldots, t_\rank)$, we obtain  $\dom(\aheap) =
    \{\aistore(t_0)\} = \{\aaistore(t_0)\}$ 
    and
    $\aheap(\aistore(t_0)) = (\aistore(t_1), \ldots,
    \aistore(t_\rank)) = (\aaistore(t_1), \ldots,
    \aaistore(t_\rank))$. 
  Since $\aaistore$ is injective,  we obtain
    $v_i = t_i$, for all $i \in \interv{0}{\rank}$. By Definition
    \ref{def:core-formulae}, we have $\overline{\vec{x}} \cup
    \overline{\vec{y}} \subseteq \set{t_0, \ldots, t_\rank}$, thus
    $\overline{\vec{x}} = \overline{\vec{y}} = \emptyset$, by ($\dagger$). 
    Then we obtain $\phi = t_0 \mapsto
    (t_1,\ldots,t_\rank)$ and $\phi \in F$ follows, by
    (\ref{eq:pto-core}).
  \item $\Asterisk_{i=1}^n q_i(\vec{u}_i) \swand p(\vec{t})$: Since
    $(\aistore,\aheap) \models t_0 \mapsto (t_1,\ldots,t_\rank)$, we
    have $\loc(\aheap) = \{\aistore(t_0), \ldots, \aistore(t_\rank)\}$
    \mycomment[me]{a couple of things can be factored for both items,
      it seems}. Since $\aistore$, $\aaistore$ agree over
    $\set{t_0,\ldots,t_\rank}$, we have $\loc(\aheap) =
    \{\aaistore(t_0), \ldots, \aaistore(t_\rank)\}$ and since
    $\aaistore(\overline{\vec{x}}) \subseteq \loc(\aheap)$ and
    $\aaistore$ is injective, the only possibility is
    $\overline{\vec{x}} = \emptyset$, so that $\phi = \nheapall
    \overline{\vec{y}} ~.~ \Asterisk_{i=1}^n q_i(\vec{u}_i) \swand
    p(\vec{t})$. Since $\nheapall \overline{\vec{y}} ~.~
    \Asterisk_{i=1}^n q_i(\vec{u}_i) \swand p(\vec{t})$ is a core
    formula, by Definition \ref{def:core-formulae}, we have
    $\overline{\vec{y}} \subseteq \vec{t} \cup \bigcup_{i=1}^n
    \vec{u}_i$ and therefore $\overline{\vec{y}} \subseteq
    \left(\vec{t} \cup \bigcup_{i=1}^n \vec{u}_i\right) \setminus
    (\set{t_0, \ldots, t_\rank} \cup \const)$. Since
    $\dom(\aistore)=\set{t_0, \ldots, t_\rank} \cup \const$ and $\phi
    \in \coreabs{\aistore,\aheap}{\aprob}$, we have that $\fv{\phi} =
    \set{t_0, \ldots, t_\rank}$ and thus $\overline{\vec{y}} =
    \left(\vec{t} \cup \bigcup_{i=1}^n \vec{u}_i\right) \setminus
    (\set{t_0, \ldots, t_\rank} \cup \const)$. Indeed, all variables
    $y \in \vec{t} \cup \bigcup_{i=1}^n \vec{u}_i$ not occurring in
    $\vec{y}$ necessarily occur in $\dom(\aaistore) \setminus \vec{y}
    = \dom(\aistore)$. By (\ref{infrule:notemp}), for each
    rule \[p(\vec{x}) \Leftarrow_\asys \exists \vec{w} ~.~ \psi *
    \Asterisk_{j=1}^m p_j(\vec{z}_j)~~(\dagger\dagger)\] and each
    substitution $\tau : \vec{w} \map \vec{x} \cup \bigcup_{i=1}^n
    \vec{y}_i$, there exists a rule
    \[\Asterisk_{i=1}^n q_i(\vec{y}_i) \swand p(\vec{x}) \Leftarrow_{\coreset{\asys}}
    \exists \vec{v} ~.~ \psi\tau * \Asterisk_{j=1}^m \gamma_j \swand
    p_j(\tau(\vec{z}_j))~~(\ddagger)\] where
    $\Asterisk_{j=1}^m\gamma_j = \Asterisk_{i=1}^n q_i(\vec{y}_i)$ and
    $\vec{v} = \vec{w} \setminus \dom(\tau)$. Assume w.l.o.g. that
    $(\aaistore,\aheap) \models_{\coreset{\asys}} \Asterisk_{i=1}^n
    q_i(\vec{u}_i) \swand p(\vec{t})$ is a consequence of the above
    rule, i.e., that there exists a $\vec{v}$-associate $\astore'$ of
    $\aaistore$ such that $(\astore',\aheap) \models_{\coreset{\asys}}
    \psi\tau\sigma * \Asterisk_{j=1}^m \gamma_j\sigma \swand
    p_j(\sigma(\tau(\vec{z}_j)))$, where
    $\sigma\isdef[\vec{t}/\vec{x}, \vec{u}_1/\vec{y}_1, \ldots,
      \vec{u}_n/\vec{y}_n]$. Since $\asys$ is progressing, there is
    exactly one points-to atom in $\psi$ and, because
    $\card{\aheap}=1$, it must be the case that $(\astore',\aheap)
    \models \psi\tau\sigma$ and $(\astore',\emptyset)
    \models_{\coreset{\asys}} \gamma_j\sigma \swand
    p_j(\sigma(\tau(\vec{z}_j)))$, for each $j \in \interv{1}{m}$. To
    prove that $\phi \in F$, it is sufficient to show the existence of a
    core unfolding $\emp \swand p(\vec{t}) \sunfold{\coreset{\asys}}
    t_0 \mapsto (t_1,\ldots,t_\rank) * \Asterisk_{i=1}^n \emp \swand
    q_i(\vec{u}_i)$. To this end, we first prove  the two points of
    Definition \ref{def:simple-formulae}:

    \noindent(\ref{it1:simple-formulae}) Since $(\astore',\emptyset)
    \models_{\coreset{\asys}} \gamma_j\sigma \swand
    p_j(\sigma(\tau(\vec{z}_j)))$, for each $j \in \interv{1}{m}$, by
    Lemma \ref{lemma:context-emp}, we obtain $\gamma_j\sigma =
    p_j(\vec{w}_j)$, for a tuple of variables $\vec{w}_j \in
    \dom(\astore')$, such that $\astore'(\vec{w}_j) =
    \astore'(\sigma(\tau(\vec{z}_j)))$. Since, moreover
    $\Asterisk_{j=1}^m \gamma_j\sigma = \Asterisk_{i=1}^n
    q_i(\vec{u}_i)$, we deduce that $n = m$ and, for each $i \in
    \interv{1}{n}$, we have $q_i = p_{j_i}$, for some $j_i \in
    \interv{1}{m}$. Then, by applying (\ref{infrule:notemp}) to the
    rule ($\dagger\dagger$), using the substitution $\tau$, we obtain
    the rule:
    \[\emp \swand p(\vec{x}) \Leftarrow_{\coreset{\asys}} 
    \exists \vec{v} ~.~ \psi\tau * \Asterisk_{i=1}^n \emp \swand
    q_i(\tau(\vec{z}_{j_i}))~~(\ddagger\ddagger)\]

    \noindent(\ref{it2:simple-formulae}) Let $\mu$ be the extension of $\sigma$ 
with the pairs $(z,u)$ such that $z \in \vec{v}$
    and one of the following holds: \begin{compactitem}
    \item if $\astore'(z) = \aaistore(t_i)$, for some $i \in
      \interv{0}{\rank}$, then $u = t_i$,
    \item if $\astore'(z) = \aaistore((\vec{u}_i)_\ell)$, for
      some $i \in \interv{1}{n}$ and $\ell \in \interv{1}{\#q_i}$,
      then $u = (\vec{u}_i)_\ell$,
    \item otherwise, $u = \min_\preceq\set{v \in \vec{v} \mid
      \astore'(v) = \astore'(z)}$, where $\preceq$ is a total order on
      $\vars$.
    \end{compactitem}
    Note that, since $\aaistore$ is injective, for each $z \in
    \vec{v}$ there exist at most one pair $(z,u) \in \mu$ 
    which is well-defined. Moreover,
    we have $\mu(\tau(\vec{z}_{j_i})) = \vec{u}_i$, because
    $\astore'(\sigma(\tau(\vec{z}_j))) = \astore'(\vec{u}_i) =
    \aaistore(\vec{u}_i)$, for all $i \in \interv{1}{n}$. We now prove 
    that 
    \[\begin{array}{rcl}
    t_0 \mapsto (t_1, \ldots, t_\rank) * \Asterisk_{i=1}^n \emp \swand q_i(\vec{u}_i) 
    & \in & \coretrans{\psi\tau\mu * \Asterisk_{i=1}^n \emp \swand q_i(\mu(\tau(\vec{z}_{j_i})))} \\
    \end{array}\]
    or, equivalently, that $t_0 \mapsto (t_1, \ldots, t_\rank) \in
    \coretrans{\psi\tau\mu}$.  By a case split on the form of the atom
    $\alpha$ in $\psi\tau$, using the fact that $(\astore',\aheap)
    \models \psi\tau\sigma$: \begin{compactitem}
    \item $\alpha = u_1 \seq u_2$: we have $\astore'(\sigma(u_1)) =
      \astore'(\sigma(u_2))$, hence $\mu(u_1) = \mu(u_2)$, by
      definition of $\mu$ and $\coretrans{\alpha} = \set{\emp}$.
    \item $\alpha = u_1 \sneq u_2$: we have $\astore'(\sigma(u_1)) \neq
      \astore'(\sigma(u_2))$, hence $\mu(u_1) \neq \mu(u_2)$, by
      definition of $\mu$ and $\coretrans{\alpha} = \set{\emp}$. 
    \item $\alpha = u_0 \mapsto (u_1, \ldots, u_\rank)$: since $\asys$
      is progressing, $\alpha$ 
      is the only points-to atom in $\psi$ and
      $\dom(\aheap) = \{\aaistore(t_0)\} = \{\astore'(\sigma(u_0))$,
      $\aheap(\aaistore) = (\aaistore(t_0), \ldots,
      \aaistore(t_\rank)) = (\astore'(\sigma(u_1)), \ldots,
      \astore'(\sigma(u_\rank)))$. Then we obtain $\aaistore(t_i) =
      \astore'(\sigma(u_i))$, hence $\mu(u_i) = t_i$, for all $i \in
      \interv{0}{\rank}$ and $\coretrans{\alpha} = \set{t_0 \mapsto
        (t_1, \ldots, t_\rank)}$.
    \end{compactitem}
  We obtain the core unfolding $\emp \swand p(\vec{t})
    \sunfold{\coreset{\asys}} t_0 \mapsto (t_1,\ldots,t_\rank) *
    \Asterisk_{i=1}^n ~\emp \swand q_i(\vec{u}_i)$ and we are left
    with proving that $t_0 \not\in \{\aroot(q_i(\vec{u}_i)) \mid i \in
    \interv{1}{n}\}$. By the definition of $\mu$, there exists a
    points-to atom $u_0 \mapsto (u_1, \ldots, u_\rank)$ in $\psi\tau$,
    such that $t_0 = \mu(u_0)$. Because $\asys$ is progressing, it
    must be the case that $u_0 = \aroot(p(\vec{x}))$, hence $t_0 =
    \aroot(p(\vec{t}))$, by the definition of $\mu$. Since $\phi$ is a
    core formula, by Definition \ref{def:core-formulae}, we obtain
    $\aroot(p(\vec{t})) \not\in \set{\aroot(q_i(\vec{u}_i)) \mid i \in
      \interv{1}{n}}$ and we conclude that $\phi = \nheapall
    \overline{\vec{y}} ~.~ \Asterisk_{i=1}^n q_i(\vec{u}_i) \swand
    p(\vec{t}) \in F$, by (\ref{eq:pto-core}).  \qed
\end{compactitem}}}

\paragraph{Predicate Atoms}
Since profiles involve only the core formul{\ae} obtained by the syntactic
translation of a symbolic heap, the only predicate atoms that occur in
the argument of a profile are of the form $\emp \swand p(\vec{t})$. We
consider the constraint:
\begin{eqnarray}\label{eq:pred-core}
  {(\emp \swand p(\vec{t}),~ F)} \in \profile_\aprob \text{ if } {(\heapex \vec{y} ~.~ \psi,~ F)} \in \profile_\aprob,  
  \emp \swand p(\vec{t}) \sunfold{\coreset{\asys}} \psi\in  \core{\aprob} \text{ and } 
  \vec{y} = \fv{\psi} \setminus \vec{t}
\end{eqnarray}
\paragraph{Separating Conjunctions}
Computing the profile of a separating conjunction is the most
technical point of the construction. To ease the presentation, we
assume the existence of a binary operation called \emph{composition}:
\begin{definition}\label{def:coresep}
  Given a set $D \subseteq \allvars{\aprob}^1 \cup \const$, a binary
  operator $\coresep{D} : 2^{\core{\aprob}} \times 2^{\core{\aprob}}
  \rightarrow 2^{\core{\aprob}}$ is a \emph{composition} if
  $\coreabs{\aistore,\aheap_1}{\aprob} \coresep{D}
  \coreabs{\aistore,\aheap_2}{\aprob} =
  \coreabs{\aistore,\aheap}{\aprob}$, for any injective structure
  $(\aistore,\aheap)$, such that \begin{inparaenum}[(i)]
  \item\label{it1:def:coresep} $\dom(\aistore) \subseteq
    \allvars{\aprob}^1$,
  \item\label{it2:def:coresep} $\aheap = \aheap_1 \uplus \aheap_2$,
  \item\label{it3:def:coresep} $\front(\aheap_1, \aheap_2) \subseteq
    \aistore(\allvars{\aprob}^1 \cup \const)$,
  \item\label{it4:def:coresep} $\front(\aheap_1, \aheap_2) \cap
    \dom(\aheap) \subseteq \aistore(D) \subseteq \dom(\aheap)$.
  \end{inparaenum}
\end{definition}
We recall that $\front(\aheap_1,\aheap_2) = \loc(\aheap_1) \cap
\loc(\aheap_2)$.  If $\asys$ is a normalized set of rules, then for
any core formula $\phi$ whose only occurrences of predicate atoms are
of the form $\emp \swand p(\vec{t})$, we define
$\alloc{\phi}{\coreset{\asys}}$ as the homomorphic extension of
$\alloc{\emp \swand p(\vec{t})}{\coreset{\asys}} \isdef
\alloc{p(\vec{t})}{\asys}$ to $\phi$ (see Definition \ref{def:alloc}).
Assuming that $\asys$ is a normalized set of rules and that a
composition operation $\coresep{D}$ (the construction of which will
be described \longVersionOnly{in
  \S\ref{sec:composition}}\shortVersionOnly{below}, see Lemma
\ref{lemma:coresep}) exists, we define the profile of a separating
conjunction:
\begin{eqnarray}\label{eq:sep-core}
&& \hspace*{-8mm} {(\phi_1 * \phi_2, \add{X_1}{F_1} 
\coresep{D} \add{X_2}{F_2})} \in \profile_\aprob \text{, if } {(\phi_i,F_i)} \in \profile_\aprob 
\text{\, } X_i \isdef \fv{\phi_{3-i}} \setminus \fv{\phi_i},~i=1,2 \nonumber\\[-1mm]
&&  \hspace*{-8mm} \alloc{\phi_1}{\coreset{\asys}}\cap\alloc{\phi_2}{\coreset{\asys}} = \emptyset, \hspace*{2mm}
D \isdef \alloc{\phi_1 * \phi_2}{\coreset{\asys}} \cap (\fv{\phi_1} \cap \fv{\phi_2} \cup \const) \\[-1mm]
&& \hspace*{-8mm} \add{x}{F} \isdef \{\heapex \vec{y} \nheapall \vec{z} ~.~ \psi \mid
\heapex \vec{y} \nheapall \vec{z} \nheapall \hat{x} ~. ~\psi[\hat{x}/x] \in F\}, 
~\add{\set{x_1,\ldots,x_n}}{F} \isdef \add{x_1}{ \ldots \add{x_n}{F}} 
\nonumber
\end{eqnarray}
The choice of the set $D$ above ensures (together with the restriction
to normal models) that $\coresep{D}$ is indeed a composition
operator. Intuitively, since the considered models are normal, every
location in the frontier between the heaps corresponding to $\phi_1$
and $\phi_2$ will be associated with a variable, thus $D$ denotes the
set of allocated locations on the frontier. Note that, because
$\aprob$ is normalized, $\alloc{\phi_1 * \phi_2}{\coreset{\asys}}$ is
well-defined.  Because the properties of the composition operation
hold when the models of its operands share the same store (Definition
\ref{def:coresep}), we use the $\add{x}{F}$ function that adds free
variables (mapped to locations outside of the heap) to each core
formula in $F$.

\putInAppendix{

\longVersionOnly{ We prove below that the definition of the $D$ set in
  Equation \ref{eq:sep-core} satisfies the condition from Definition
  \ref{def:coresep}, for any normal $\asys$-companion of
  $(\phi_1,\phi_2)$ (see Definition \ref{def:companions}): }

\begin{lemma}\label{lemma:coresep-d}
  If $\asys$ is normalized, $\phi_1,\phi_2 \in \shk{\rank}$ are
  symbolic heaps and $\tuple{(\aistore,\aheap_1),
    (\aistore,\aheap_2)}$ is an injective normal $\asys$-companion for
  $(\phi_1,\phi_2)$, then:
  \[\front(\aheap_1, \aheap_2) \cap
  \dom(\aheap_1 \uplus \aheap_2) \subseteq \aistore\left(\alloc{\phi_1
    * \phi_2}{\asys} \cap \left(\fv{\phi_1} \cap \fv{\phi_2} \cup
  \const\right)\right) \subseteq \dom(\aheap_1 \uplus
  \aheap_2).\]
\end{lemma}
\proof{ Let $\ell \in \front(\aheap_1, \aheap_2) \cap \dom(\aheap_1
  \uplus \aheap_2)$ be a location. By Lemma \ref{lemma:frontier},
  since $\aistore$ is injective and $\const \subseteq
  \dom(\aistore)$, we have $\ell \in \aistore\left(\fv{\phi_1} \cap
  \fv{\phi_2} \cup \const\right)$. For $i=1,2$, let $\phi_i \unfold{\asys}^*
  \exists \vec{x}_i ~.~ \psi_i$ be the predicate-free unfolding and
  $\aastore_i$ be the $\vec{x}_i$-associate of $\aistore$ that
  satisfy  points (\ref{it1:def:companions}) and
  (\ref{it2:def:companions}) of Definition \ref{def:companions}. 
Assume that
  $\ell \in \dom(\aheap_1)$ (the case $\ell \in \dom(\aheap_2)$ is
  symmetric). Because $(\aastore_1, \aheap_1) \models \psi_1$, there
  exists a points-to atom $t_0 \mapsto (t_1, \ldots, t_\rank)$ in
  $\psi_1$, such that $\aastore_1(t_0) = \ell$. Since $\asys$ is
  normalized, by Definition \ref{def:alloc}, the set
  $\alloc{\phi_1}{\asys}$ is well-defined and we distinguish two
  cases.
  \begin{compactitem} 
  \item If $t_0 \in \alloc{\phi_1}{\asys}$, then $\ell \in
    \aistore(\alloc{\phi_1}{\asys})$, because $\aastore_1$ and
    $\aistore$ agree over $\alloc{\phi_1}{\asys}$. 
  \item Otherwise, we must have $t_0 \in \vec{x}_1$. Since $\ell \in
    \front(\aheap_1, \aheap_2)$, we have $\ell \in \loc(\aheap_2)$,
    thus there exists a points-to atom $u_0 \mapsto (u_1, \ldots,
    u_\rank)$ in $\psi_2$ such that $\ell = \aastore_2(u_i)$, for some
    $i \in \interv{1}{\rank}$. Note that $\ell = \aastore_2(u_0)$ is
    impossible, because $\ell \in \dom(\aheap_1)$. Suppose, for a
    contradiction, that $\ell \not\in \aastore(\const)$. Then $u_i \in
    \fv{\psi_2}$ must be the case, which contradicts the condition
    $\aastore_1(\vec{x}_1) \cap \aastore_2(\fv{\psi_2}) \subseteq
    \aastore(\const)$, required at point (\ref{it2:def:companions}) of
    Definition \ref{def:companions}. Hence $\ell \in \aastore(\const)$
    must be the case. Since $\ell = \aastore_1(t_0)$ either $t_0 \in
    \const$ or $t_0$ is an existentially allocated variable. The
    second case cannot occur, because of the Condition
    (\ref{it5:normalized}) of Definition \ref{def:normalized}. Then we
    have $t_0 \in \const$ and, moreover, we have $t_0 \in
    \alloc{\phi_1}{\asys}$, by Definition \ref{def:alloc}, thus $t_0
    \in \alloc{\phi_1}{\asys} \cap \const$.
  \end{compactitem}
  In each case we obtain $\ell \in \aastore_1(\alloc{\phi_1}{\asys})
  \cup \aastore_2(\alloc{\phi_2}{\asys}) \subseteq
  \aistore(\alloc{\phi_1 * \phi_2}{\asys})$, because $\aastore_i$
  agrees with $\aistore$ over $\alloc{\phi_i}{\asys}$, for $i=1,2$. We
  obtain: \[\begin{array}{rcl}
  \ell & \in & \aistore(\alloc{\phi_1 * \phi_2}{\asys}) \cap
  \aistore\left(\fv{\phi_1} \cap \fv{\phi_2} \cup \const\right) \\
  & = & \aistore\left(\alloc{\phi_1 * \phi_2}{\asys} \cap (\fv{\phi_1} \cap
  \fv{\phi_2} \cup \const)\right) \text{, because $\aistore$ is
  injective .}
  \end{array}\]
  The second inclusion follows trivially from the fact that
  $\aistore(\alloc{\phi_i}{\asys}) \subseteq \dom(\aheap_i)$, for
  $i=1,2$, which is an easy consequence of Definition \ref{def:alloc}. \qed}

\longVersionOnly{ The following lemma is used to prove the
  correctness of the profile construction for separating conjunctions,
  by stating the effect of this operation on structures: }

\begin{lemma}\label{lemma:add}
  Given an injective structure $(\aistore,\aheap)$, a variable  $x
  \not\in \dom(\aistore)$ and a location $\ell
  \not\in\loc(\aheap)\cup\img(\aistore)$, we have
  $\coreabs{\aistore[x\leftarrow\ell],\aheap}{\aprob} =
  \add{x}{\coreabs{\aistore,\aheap}{\aprob}}$.
\end{lemma}
\proof{ ``$\subseteq$'' Let $\varphi =
  \heapex\vec{x}\nheapall\vec{y}\,.\,\phi \in
  \coreabs{\aistore[x\leftarrow\ell],\aheap}{\aprob}$ be a core
  formula, where $\phi$ is quantifier-free. By Definition
  \ref{def:coreabs}, there exists a witness $\aaistore \in
  \witset{\asys}{\aistore[x\leftarrow\ell]}{\aheap}{\varphi}$, such
  that $\aaistore(\lroots{\varphi}) \cap \dom(\aheap) = \emptyset$.
  Let $\aaistore'$ be the store identical to $\aaistore$, except that
  $x \not \in \dom(\aaistore')$ and $\aaistore'(\hat{x}) =
  \aaistore(x)$, for some variable $\hat{x} \not\in
  \allvars{\aprob}^1$. Since $\ell \not\in \loc(\aheap)$, we have
  $\aaistore' \in \witset{\asys}{\aistore}{\aheap}{\heapex\vec{x}\nheapall\vec{y}\nheapall
    \hat{x}~.~\phi[\hat{x}/x]}$, because $\hat{x}\notin
  \allvars{\aprob}^1$ we have $\heapex\vec{x}\nheapall\vec{y}\nheapall
  \hat{x}~.~\phi[\hat{x}/x] \in \core{\aprob}$, hence
  $\heapex\vec{x}\nheapall\vec{y}\nheapall \hat{x}~.~\phi[\hat{x}/x]
  \in \coreabs{\aistore,\aheap}{\aprob}$, from which we deduce that
  $\varphi \in \add{x}{\coreabs{\aistore,\aheap}{\aprob}}$.

  \noindent``$\supseteq$'' Let $\varphi =
  \heapex\vec{x}\nheapall\vec{y} ~.~\phi \in
  \add{x}{\coreabs{\aistore,\aheap}{\aprob}}$, where $\phi$ is
  quantifier-free, and let $\psi =
  \heapex\vec{x}\nheapall\vec{y}\nheapall \hat{x}
  \,.\,\phi[\hat{x}/x]$.  By (\ref{eq:sep-core}), we have $\psi \in
  \coreabs{\aistore,\aheap}{\aprob}$.  By Definition
  \ref{def:coreabs}, there exists a witness $\aaistore \in
  \witset{\asys}{\aistore}{\aheap}{\psi}$, such that
  $\aaistore(\lroots{\psi}) \cap \dom(\aheap) = \emptyset$. W.l.o.g.,
  by Lemma \ref{lemma:istore-eq}, we can assume that $\aaistore$ is
  such that $\ell \neq \aaistore(y)$, for all $y \in \dom(\aaistore)$
  such that $\aaistore(y) \not\in \loc(\aheap)$.  With this
  assumption, $\aaistore[x \leftarrow \ell]$ is injective. We prove
  that $\aaistore[x \leftarrow \ell] \in
  \witset{\asys}{\aistore[x\leftarrow\ell]}{\aheap}{\varphi}$: \begin{compactitem}
  \item Let $\aaistore'$ be the store identical to $\aaistore[x
    \leftarrow \ell]$ except that the images of $x$ and $\hat{x}$ are
    switched. Since $(\aaistore,\aheap) \models_{\coreset{\asys}}
    \phi[\hat{x}/x]$ we have $(\aaistore',\aheap)
    \models_{\coreset{\asys}} \phi$.  Since $\aaistore(x), \ell
    \not\in \loc(\aheap)$ (as $\aaistore(x) = \aistore(\hat{x})$, by
    definition, and $\aistore(\hat{x}) \not \in\loc(\aheap)$, by
    Condition (\ref{it3:witset}) of Definition \ref{def:coreabs}), we
    have $\aaistore' \istoreq{\loc(\aheap)} \aaistore[x \leftarrow
      \ell]$ thus $(\aaistore[\hat{x}\leftarrow\ell],\aheap)
    \models_{\coreset{\asys}} \phi$, by Lemma \ref{lemma:istore-eq}.
  \item Since $x\not\in\vec{x}$, we have
    $\aaistore[x\leftarrow\ell](\vec{x}) = \aaistore(\vec{x})
    \subseteq \loc(\aheap)$.
  \item Since $\ell \not\in \loc(\aheap)$ and $\aaistore(\vec{y}) \cap
    \loc(\aheap) = \emptyset$, we have
    $\aaistore[x\leftarrow\ell](\vec{y}) \cap \loc(\aheap) =
    \emptyset$.
  \end{compactitem}
  Since $\lroots{\heapex\vec{x}\nheapall\vec{y}\nheapall \hat{x} ~.~\phi[\hat{x}/x]} =
  \lroots{\varphi}$, we have $\aaistore(\lroots{\varphi}) \cap
  \dom(\aheap) = \emptyset$, thus $\aaistore[x \leftarrow \ell] \in
  \witset{\asys}{\aistore[x\leftarrow\ell]}{\aheap}{\varphi}$, which implies
  $\varphi \in \coreabs{\aistore[x\leftarrow\ell],\aheap}{\aprob}$. \qed}
}

\paragraph{Existential Quantifiers}
Since profiles involve only core formul{\ae} obtained by the syntactic
translation of a symbolic heap (Lemma \ref{lemma:entailment}), it is
sufficient to consider only existentially quantified core formul{\ae},
because the syntactic translation $\coretrans{.}$ does not produce
universal quantifiers. The profile of an existentially quantified core
formula is given by the constraint:
\begin{eqnarray}\label{eq:ex-core}
  &&(\heapex x' ~.~ \phi[x'/x],~ \rem{x}{F}) \in \profile_\aprob
  \text{, if $x \in \fv{\phi}$, $x' \in \allvars{\aprob}^2$, $x'$ not bound in $\phi$, ${(\phi,F)} \in \profile_\aprob$}, \qquad  \\[-1mm]
  && \rem{x}{F} \isdef \{\heapex \hat{x} ~.~ \psi[\hat{x}/x] \mid \psi \in F,~ x \in \fv{\psi}, \text{ $\hat{x}$ not
    in $\psi$}\} \cap \core{\aprob} \cup \{\psi \mid \psi \in F,~ x \not\in \fv{\psi}\} \nonumber \\[-1mm]
  && \rem{\set{x_1,\ldots,x_n}}{F} \isdef \rem{x_1}{ \ldots \rem{x_n}{F} \ldots}
  \nonumber
\end{eqnarray}
Note that $\hat{x}$ is a fresh variable, which is not bound or free in $\psi$. 
In particular, if $x \in \roots{\psi}$, then we must have $\hat{x}\in \allvars{\aprob}^2$, so
that $\heapex \hat{x} ~.~ \psi[\hat{x}/x] \in \core{\aprob}$. 
Similarly the variable $x$ is replaced by a fresh variable $x' \in \allvars{\aprob}^2$ in $\heapex x' ~.~ \phi[x'/x]$ to ensure that 
$\heapex x' ~.~ \phi[x'/x]$ is a core formula. 


\putInAppendix{

\longVersionOnly{ The following lemma is used to prove the
  correctness of the profile construction for bounded existential
  quantifiers, by stating the effect of the above function on
  structures: }

\begin{lemma}\label{lemma:rem}
  Given an injective structure $(\aistore,\aheap)$ and a variable $x \in
  \dom(\aistore) \cap \allvars{\aprob}^1$ such that $\aistore(x) \in \loc(\aheap)$, we
  have $\coreabs{\aistore',\aheap}{\aprob} =
  \rem{x}{\coreabs{\aistore,\aheap}{\aprob}}$, where $\aistore'$ is the
  restriction of $\aistore$ to $\dom(\aistore) \setminus \set{x}$.
\end{lemma}
\proof{ First note that because $\aistore$ is injective, $\aistore'$ is necessarily injective,
  thus $\coreabs{\aistore',\aheap}{\aprob}$ is well defined. We prove both inclusions.
  
  ``$\subseteq$'' Let $\varphi = \heapex \vec{x} \nheapall \vec{y} ~.~
  \phi \in \coreabs{\aistore',\aheap}{\aprob}$ be a core formula,
  where $\phi$ is quantifier-free. By Definition \ref{def:coreabs}, there exists a witness $\aaistorep \in
  \witset{\asys}{\aistore'}{\aheap}{\varphi}$, such that
  $\aaistorep(\lroots{\varphi}) \cap \dom(\aheap) = \emptyset$.  Since
  $x \not\in \dom(\aistore')$ and $(\aistore',\aheap)
  \models_{\coreset{\asys}} \varphi$, we have $x \not\in
  \fv{\varphi}$.  By $\alpha$-renaming if necessary, we can assume
  w.l.o.g. that $x \not\in \vec{x} \cup \vec{y}$ ($\dagger$). This is
  possible since $x \in \allvars{\aprob}^1$, hence if $x\in \vec{x}
  \cup \vec{y}$ then by definition of
  $\core{\aprob}$ it cannot occur in $\roots{\phi}$; it can therefore be renamed by a variable not occurring
  in $\allvars{\aprob}^1$. 
  We distinguish the following
  cases. \begin{compactitem}
  \item If $\aistore(x) \neq \aaistorep(x')$ for all $x' \in
    \vec{x}$, then $\aaistorep[x \leftarrow \aistore(x)]$ is an
    injective associate of $\aaistorep$: indeed, by hypothesis,
    $\aistore(x) \in \loc(\aheap)$ and $\aaistorep(\vec{y}) \cap
    \loc(\aheap) = \emptyset$, thus $\aistore(x) \not\in
    \aaistorep(\vec{y})$. Since $\phi$ is quantifier-free and
    $\aaistorep$ agrees with $\aaistorep[x \leftarrow \aistore(x)]$ on
    $\fv{\phi}$, we obtain $(\aaistorep[x \leftarrow \aistore(x)],
    \aheap) \models_{\coreset{\asys}} \phi$. We now prove that
    $\aaistorep[x \leftarrow \aistore(x)] \in
    \witset{\asys}{\aistore}{\aheap}{\varphi}$, which suffices to show
    $\varphi \in \coreabs{\aistore,\aheap}{\aprob}$, by the definition
    of the latter set: \begin{compactitem}
    \item $\aaistorep[x \leftarrow \aistore(x)](\vec{x}) \subseteq
      \aaistorep(\vec{x}) \cup \set{\aistore(x)} \subseteq
      \loc(\aheap)$, because $\aaistorep \in
      \witset{\asys}{\aistore'}{\aheap}{\varphi}$ and $\aistore(x) \in
      \loc(\aheap)$ by hypothesis.
    \item $\aaistorep[x \leftarrow \aistore(x)](\vec{y}) =
      \aaistorep(\vec{y})$ and $\aaistorep(\vec{y}) \cap \loc(\aheap)
      = \emptyset$, because $\aaistorep \in
      \witset{\asys}{\aistore'}{\aheap}{\varphi}$.
    \item $\aaistorep[x \leftarrow \aistore(x)](\lroots{\varphi}) =
      \aaistorep(\lroots{\varphi})$ because $x \not\in \fv{\phi}$, and
      $\aaistorep(\lroots{\varphi}) \cap \dom(\aheap) = \emptyset$
      because $\aaistorep \in \witset{\asys}{\aistore'}{\aheap}{\varphi}$.
    \end{compactitem}
    Consequently we obtain $\varphi \in
    \coreabs{\aistore,\aheap}{\aprob}$, and since $x \not\in
    \fv{\varphi}$, we have $\coreabs{\aistore,\aheap}{\aprob} \subseteq
    \rem{x}{\coreabs{\aistore,\aheap}{\aprob}}$, hence the result.
  \item Otherwise, $\aistore(x) = \aaistorep(x')$ for some $x' \in
    \vec{x}$, hence $\varphi$ is of the form $\heapex x' \heapex \vec{x}' \nheapall
    \vec{y} ~.~ \phi$, where $\vec{x}' \isdef \vec{x} \setminus
    \set{x'}$.  Clearly, the variable $x'$ must be unique, otherwise $\aaistorep$ would not be injective. Let $\aaistore$
    be the injective store obtained from $\aaistorep[x \leftarrow
      \aistore(x)]$ by removing the pair $(x',\aistore(x))$ from
    it. We prove that $\aaistore \in
    \witset{\asys}{\aistore}{\aheap}{\heapex\vec{x}'\nheapall\vec{y}
      ~.~ \phi[x/x']}$: \begin{compactitem}
    \item $(\aaistore,\aheap) \models_{\coreset{\asys}} \phi[x/x']$,
      because $\aaistore$ agrees with
      $\aaistorep[x\leftarrow\aistore(x)]$ on $\fv{\phi[x/x']}$.
    \item $\aaistore(\vec{x}')=\aaistorep(\vec{x}') \subseteq
      \loc(\aheap)$, because $\aaistorep \in
      \witset{\asys}{\aistore'}{\aheap}{\varphi}$. 
    \item $\aaistore(\vec{y})=\aaistorep(\vec{y})$, because $x \not\in
      \vec{y}$ ~($\dagger$) and $\aaistorep(\vec{y}) \cap \loc(\aheap)
      = \emptyset$, because $\aaistorep \in
      \witset{\asys}{\aistore'}{\aheap}{\varphi}$.
    \end{compactitem}
    Furthermore, we have
    $\aaistorep[x\leftarrow\aistore(x)](\lroots{\varphi})=\aaistorep(\lroots{\varphi})$
    because $x \not\in \fv{\varphi}$, hence $x \not\in
    \lroots{\varphi}$. Thus $\aaistore(\lroots{\varphi}) \subseteq
    \aaistorep(\lroots{\varphi})$ and, since
    $\aaistorep(\lroots{\varphi}) \cap \dom(\aheap) = \emptyset$ by
    Definition \ref{def:coreabs}, we deduce that
    $\aaistore(\lroots{\varphi}) \cap \dom(\aheap) = \emptyset$. Still
    by Definition \ref{def:coreabs}, we obtain that $\heapex \vec{x}'
    \nheapall \vec{y} ~.~ \phi[x/x'] \in
    \coreabs{\aistore,\aheap}{\aprob}$ and thus $\heapex x' \heapex
    \vec{x}' \nheapall \vec{y} ~.~ \phi \in
    \rem{x}{\coreabs{\aistore}{\aprob}}$ (with $\hat{x}\isdef x'$).
  \end{compactitem}
  ``$\supseteq$'' Let $\varphi = \heapex \vec{x} \nheapall \vec{y} \,.\,  \phi \in \rem{x}{\coreabs{\aistore,\aheap}{\aprob}}$, for some
  quantifier-free formula $\phi$. We distinguish the following
  cases. \begin{compactitem}
  \item If $\varphi \in \coreabs{\aistore,\aheap}{\aprob}$ and $x
    \not\in \fv{\varphi}$, then for any injective structure
    $(\aistore,\aheap)$ meeting the conditions of Definition
    \ref{def:coreabs}, the structure $(\aistore',\aheap)$ is injective
    and trivially meets the conditions of Definition
    \ref{def:coreabs}, hence $\varphi \in
    \coreabs{\aistore',\aheap}{\aprob}$.
  \item Otherwise, $\varphi = \heapex \hat{x} \heapex \vec{x}'
    \nheapall \vec{y} ~. \phi[\hat{x}/x]$, $x \in \fv{\heapex \vec{x}'
      \nheapall \vec{y} ~. \phi}$ and $\heapex \vec{x}' \nheapall
    \vec{y} ~. \phi \in \coreabs{\aistore,\aheap}{\aprob}$, where
    $\vec{x}' \isdef \vec{x} \setminus \set{x}$. Let $\aaistore$ be an
    injective $(\vec{x}'\cup\vec{y})$-associate of $\aistore$ meeting
    the conditions from Definition \ref{def:coreabs}. It is easy to
    check that $(\aaistore\setminus \{ (x,\aaistore(x) \}) \cup \{
    (\hat{x},\aaistore(x)) \} \in
    \witset{\asys}{\aistore}{\aheap}{\varphi}$, thus $\varphi \in
    \coreabs{\aistore',\aheap}{\aprob}$. \qed
  \end{compactitem}
  }}

\paragraph{The Profile Function}
Let $\profile_\aprob$  be the least relation that
satisfies the constraints (\ref{eq:pto-core}), (\ref{eq:pred-core}),
(\ref{eq:sep-core}) and (\ref{eq:ex-core}).
We prove that $\profile_\aprob$ is a valid profile for
$\aprob$, in the sense of Definition \ref{def:profile}:
\begin{lemma}\label{lemma:profile}
  Given a progressing and normalized entailment problem $\aprob =
  (\asys, \seqset)$, a symbolic heap $\varphi \in \shk{\rank}$ with
  $\fv{\varphi} \subseteq \allvars{\aprob}^1$, a core formula $\phi
  \in \coretrans{\varphi}$ and a set of core formul{\ae} $F \subseteq
  \core{\aprob}$, we have $(\phi, F) \in \profile_\aprob$ iff
  $F=\coreabs{\aistore,\aheap}{\aprob}$, for some injective normal
  $\coreset{\asys}$-model $(\aistore,\aheap)$ of $\phi$, with
  $\dom(\aistore) = \fv{\varphi}\cup \const$.
\end{lemma}
\optionalProof{Lemma \ref{lemma:profile}}{sec:coreabs}{By induction on the
  structure of $\profile_\aprob$, defined as the least set satisfying
  the constraints (\ref{eq:pto-core}), (\ref{eq:pred-core}),
  (\ref{eq:sep-core}) and (\ref{eq:ex-core}), we prove that
  $(\aistore,\aheap)$ is an injective normal $\coreset{\asys}$-model
  of $\phi$ if and only if $(\phi,~ \coreabs{\aistore,\aheap}{\aprob})
  \in \profile_\aprob$. Based on the structure of the core formula
  $\phi \in \coretrans{\varphi}$, for some symbolic heap $\varphi \in
  \shk{\rank}$, we distinguish the following
  cases: \begin{compactitem}
  \item $\phi = t_0 \mapsto (t_1, \ldots, t_\rank)$: because $\asys$
    is progressing, by Lemma \ref{lemma:pto-core}, we obtain that
    $(\phi, F) \in \profile_\aprob$ if and only if $F =
    \coreabs{\aistore,\aheap}{\aprob}$, for some injective
    $\asys$-model $(\aistore,\aheap)$ of $t_0 \mapsto
    (t_1,\ldots,t_\rank)$, such that $\dom(\aistore) = \set{t_0,
      \ldots, t_\rank} \cup \const$. Since any injective $\asys$-model
    $(\aistore,\aheap)$ of $t_0 \mapsto (t_1,\ldots,t_\rank)$ is also
    normal, we conclude this case.
  \item $\phi = \emp \swand p(\vec{t})$: ``$\Rightarrow$'' Since
    $\profile_\aprob$ is the least relation satisfying
    (\ref{eq:pred-core}), ${(\emp \swand p(\vec{t}), F)} \in
    \profile_\aprob$ if and only if ${(\heapex \vec{y} ~.~ \psi, F)}
    \in \profile_\aprob$, for some core unfolding $\emp \swand
    p(\vec{t}) \sunfold{\coreset{\asys}} \psi$, where $\vec{y} =
    \fv{\psi} \setminus \vec{t}$.  By the induction hypothesis, there
    exists an injective normal $\coreset{\asys}$-model
    $(\aistore,\aheap)$ of $\heapex \vec{y} ~.~ \psi$ such that $F =
    \coreabs{\aistore,\aheap}{\aprob}$ and $\dom(\aistore) =
    \fv{\heapex \vec{y} ~.~ \psi} \cup \const$.  Since $\aprob$ is
    normalized, by Condition \ref{it1bis:normalized} in Definition
    \ref{def:normalized} we have $\fv{\heapex \vec{y} ~.~ \psi} =
    \fv{\phi}$. By Lemma \ref{lemma:simple-unfolding-sat},
    $(\aistore,\aheap)$ is an injective $\coreset{\asys}$-model of
    $\emp \swand p(\vec{t})$. Because $\phi$ is quantifier-free,
    $(\aistore,\aheap)$ is also an injective normal
    $\coreset{\asys}$-model of $\phi$. ``$\Leftarrow$'' Let
    $(\aistore,\aheap)$ be an injective normal $\coreset{\asys}$-model
    of $\emp \swand p(\vec{t})$. By Lemma
    \ref{lemma:simple-unfolding-sat}, there exists a core unfolding
    $\emp \swand p(\vec{t}) \sunfold{\coreset{\asys}} \psi$ and an
    injective extension $\aaistore$ of $\aistore$, such that
    $(\aaistore,\aheap)$ is an injective $\coreset{\asys}$-model of
    $\psi$.  Let $\vec{y} \isdef \fv{\psi} \setminus \vec{t}$. Then
    every variable $x \in \vec{y}$ occurs in a points-to or a
    predicate atom, by Definition \ref{def:simple-formulae}. Since
    $\asys$ is normalized, we obtain that $\aaistore(x) \in
    \loc(\aheap)$, by point (\ref{it3:normalized}) of Definition
    \ref{def:normalized}, and therefore $(\aistore,\aheap)$ is an
    injective $\coreset{\asys}$-model of $\heapex \vec{y} ~.~ \psi$.
    Since $\psi$ is satisfiable, it cannot contain two atoms with the
    same root.  We have $\fv{\psi} = \fv{\phi} \subseteq
    \allvars{\aprob}^1$.  Furthermore, since
    $\card{\allvars{\aprob}^2} = \probwidth{\aprob}$ and $\size{\psi}
    \leq \probwidth{\aprob}$, we can assume w.l.o.g. that $\vec{y}
    \subseteq \allvars{\aprob}^2$, hence $\heapex \vec{y} ~.~ \psi$ is
    a core formula.  By the induction hypothesis, we obtain that
    $\tuple{\heapex \vec{y} ~.~ \psi,~
      \coreabs{\aistore,\aheap}{\aprob}} \in \profile_\aprob$, thus
    $\tuple{\emp \swand p(\vec{t}),~
      \coreabs{\aistore,\aheap}{\aprob}} \in \profile_\aprob$ follows,
    by (\ref{eq:pred-core}).
  \item $\phi = \phi_1 * \phi_2$: ``$\Rightarrow$'' Since
    $\profile_\aprob$ is the least set satisfying (\ref{eq:sep-core}),
    $(\phi_1 * \phi_2, F) \in \profile_\aprob$ if and only if
    $(\phi_i, F_i) \in \profile_\aprob$ and $F = \add{X_1}{F_1}
    \coresep{D} \add{X_2}{F_2}$, where $X_i = \fv{\phi_i} \setminus
    \fv{\phi_{3-i}}$, for $i=1,2$, $\alloc{\phi_1}{\coreset{\asys}}
    \cap \alloc{\phi_2}{\coreset{\asys}} = \emptyset$ and $D =
    \alloc{\phi_1*\phi_2}{\asys} \cap (\fv{\phi_1} \cap \fv{\phi_2}
    \cup \const)$. Since $\fv{\phi_i} \subseteq \fv{\phi} \subseteq
    \allvars{\aprob}^1$, by the inductive hypothesis, there exist
    injective normal $\coreset{\asys}$-models $(\aistore_i, \aheap_i)$
    of $\phi_i$, such that $F_i =
    \coreabs{\aistore_i,\aheap_i}{\aprob}$, for $i=1,2$. By renaming
    locations if necessary, we assume w.l.o.g. that $\aistore_1$ and
    $\aistore_2$ agree over $\term{\phi_1} \cap \term{\phi_2}$ and
    that $\aistore_i(\fv{\phi_i} \setminus \fv{\phi_{3-i}}) \cap
    (\aistore_{3-i}(\fv{\phi_{3-i}} \setminus \fv{\phi_i}) \cup
    \loc(\aheap_{3-i})) = \emptyset$, for $i=1,2$ ($\dagger$).  This
    is feasible since the truth value of formul{\ae} does not depend
    on the name of the locations.  Let $\aistore =\aistore_1 \cup
    \aistore_2$. It is easy to check that $\tuple{(\aistore,\aheap_1),
      (\aistore,\aheap_2)}$ is an injective normal
    $\coreset{\asys}$-companion for $(\phi_1, \phi_2)$, by Definition
    \ref{def:companions}. Moreover, by Lemma \ref{lemma:add}, we have
    $\coreabs{\aistore,\aheap_i}{\aprob} = \add{X_i}{F_i}$, for
    $i=1,2$.  Next, we prove that $\aheap_1$ and $\aheap_2$ are
    disjoint heaps. Suppose, for a contradiction, that $\dom(\aheap_1)
    \cap \dom(\aheap_2) \neq \emptyset$. By assumption ($\dagger$),
    there exists a variable $x \in \fv{\phi_1} \cap \fv{\phi_2}$, such
    that $\aistore(x) \in \dom(\aheap_1) \cap \dom(\aheap_2)$. Since
    $\aprob$ is normalized, by Conditions (\ref{it4:normalized}) and
    (\ref{it5:normalized}) in Definition \ref{def:normalized}, the
    only variables that can be allocated by a model of a core formula
    $\phi_i$ are $\alloc{\phi_i}{\coreset{\asys}}$, we must have $x
    \in \alloc{\phi_1}{\coreset{\asys}} \cap
    \alloc{\phi_2}{\coreset{\asys}}$, which contradicts with the
    condition that $\alloc{\phi_1}{\coreset{\asys}} \cap
    \alloc{\phi_2}{\coreset{\asys}} = \emptyset$. We conclude that
    $\aheap_1$ and $\aheap_2$ are disjoint and let $\aheap = \aheap_1
    \uplus \aheap_2$.  By Lemmas \ref{lemma:frontier} and
    \ref{lemma:coresep-d}, we respectively have
    $\front(\aheap_1,\aheap_2) \subseteq \aistore\left(\fv{\phi_1}
    \cap \fv{\phi_2} \cup \const\right) \subseteq
    \aistore(\allvars{\aprob}^1 \cup \const)$ and $\front(\aheap_1,
    \aheap_2) \cap \dom(\aheap) \subseteq \aistore(D) \subseteq
    \dom(\aheap)$. Thus $(\aistore,\aheap)$ is an injective normal
    $\coreset{\asys}$-model of $\phi_1 * \phi_2$ and, by Definition
    \ref{def:coresep}, we have $\coreabs{\aistore,\aheap}{\aprob} =
    \coreabs{\aistore,\aheap_1}{\aprob} \coresep{D}
    \coreabs{\aistore,\aheap_2}{\aprob} = \add{X_1}{F_1} \coresep{D}
    \add{X_2}{F_2}$.

    \noindent''$\Leftarrow$'' Let $(\aistore,\aheap)$ be an injective
    normal $\coreset{\asys}$-model of $\phi_1 * \phi_2$. Note that since $\phi_1 * \phi_2$ is satisfiable 
    we must have $\alloc{\phi_1}{\coreset{\asys}} \cap \alloc{\phi_2}{\coreset{\asys}} = \emptyset$. By Lemma
    \ref{lemma:companions}, there exists an injective
    $\coreset{\asys}$-normal companion $\tuple{(\aistore_1, \aheap_1),
      (\aistore_2, \aheap_2)}$ for $(\phi_1, \phi_2)$, such that
    $\aheap = \aheap_1 \uplus \aheap_2$.  Since $(\aistore_i,
    \aheap_i)$ is an injective normal $\coreset{\asys}$-model of
    $\phi_i$, we have $\tuple{\phi_i,
      \coreabs{\aistore_i,\aheap_i}{\aprob}} \in \profile_\aprob$, by
    the inductive hypothesis, for $i=1,2$. We prove that
    $\aistore(X_i) \cap \loc(\aheap_{i}) = \emptyset$, where $X_i
    \isdef \fv{\phi_i} \setminus \fv{\phi_{3-i}}$, for $i=1,2$. Let
    $i=1$, the case $i=2$ being symmetric, and suppose, for a
    contradiction, that $\aistore(x) \in \loc(\aheap_1)$, for some $x
    \in X_1$. Because $\asys$ is normalized, by point
    (\ref{it3:normalized}) of Definition \ref{def:normalized}, we
    have $\aistore(x) \in \loc(\aheap_2)$, thus $\aistore(x) \in
    \front(\aheap_1,\aheap_2)$. By Lemma \ref{lemma:frontier},
    $\aistore(x) \subseteq \aistore(\fv{\phi_1} \cap \fv{\phi_2} \cup
    \const$ and, since $\aistore$ is injective, we deduce that $x \in
    \fv{\phi_1} \cap \fv{\phi_2} \cup \const$, which contradicts the hypothesis that
    $x \in X_1$. Hence $\aistore(X_i) \cap \loc(\aheap_{i}) =
    \emptyset$ and, by Lemma \ref{lemma:add}, we obtain
    $\coreabs{\aistore,\aheap_i}{\aprob} =
    \add{X_i}{\coreabs{\aistore_i,\aheap_i}{\aprob}}$, for $i=1,2$.
    Moreover, by Lemmas \ref{lemma:frontier} and
    \ref{lemma:coresep-d}, we respectively have $\front(\aheap_1, \aheap_2)
    \subseteq \aistore(\allvars{\aprob}^1 \cup \const)$ and
    $\front(\aheap_1, \aheap_2) \cap \dom(\aheap) \subseteq
    \aistore(D) \subseteq \dom(\aheap)$. By Definition
    \ref{def:coresep}, we have $\coreabs{\aistore,\aheap}{\aprob} =
    \add{X_1}{\coreabs{\aistore_1,\aheap_1}{\aprob}} \coresep{D}
    \add{X_2}{\coreabs{\aistore_2,\aheap_2}{\aprob}}$, thus
    $\tuple{\phi_1 * \phi_2, \coreabs{\aistore,\aheap}{\aprob}} \in
    \profile_\aprob$, by (\ref{eq:sep-core}).
  \item $\heapex x' \,.\, \phi_1'$: By $\alpha$-renaming if necessary,
    we assume that $x' \in \allvars{\aprob}^2$. Note that this is
    possible because $\card{\allvars{\aprob}^2} \geq \size{\phi_1'}$.
    Furthermore, since we also have $\card{\allvars{\aprob}^1} \geq
    \size{\phi_1'}$, we may assume that there exists a variable $x \in
    \allvars{\aprob}^1 \setminus \fv{\phi_1'}$.  It is clear that
    $\phi_1 = \phi_1'[x/x']$ is a core formula. ``$\Rightarrow$''
    Since $\profile_\aprob$ is the least relation satisfying
    (\ref{eq:ex-core}), we have $(\heapex x' ~.~ \phi_1',~ F) \in
    \profile_\aprob$ only if there exists a set of core formul{\ae}
    $F_1 \subseteq \core{\aprob}$, such that $F = \rem{x}{F_1}$ and
    $(\phi_1,F_1) \in \profile_\aprob$. By the inductive hypothesis,
    there exists an injective normal $\coreset{\asys}$-model
    $(\aistore_1,\aheap)$ of $\phi_1$ such that $F_1 =
    \coreabs{\aistore_1,\aheap}{\aprob}$. By Lemma \ref{lemma:rem}, we
    obtain $F = \coreabs{\aistore,\aheap}{\aprob}$, where $\aistore$
    is the restriction of $\aistore_1$ to $\dom(\aistore_1) \setminus
    \set{x}$. Since $\asys$ is normalized and the only occurrences of
    predicate atoms in $\phi_1$ are of the form $\emp \swand
    p(\vec{t})$, we have $\aistore_1(x) \in \loc(\aheap)$. Thus we
    conclude by noticing that $(\aistore,\aheap)$ is an injective
    normal $\coreset{\asys}$-model of $\heapex x' ~.~ \phi_1'$.
    ``$\Leftarrow$'' Let $(\aistore,\aheap)$ be an injective normal
    $\coreset{\asys}$-model of $\heapex x' ~.~ \phi_1'$, with
    $\dom(\aistore) = (\fv{\phi_1} \setminus \set{x}) \cup
    \const$. There exists $\ell \in \loc(\aheap) \setminus
    \img(\aistore)$ such that $(\aistore[x \leftarrow \ell],\aheap)$
    is an injective normal $\coreset{\asys}$-model of $\phi_1$. Since
    $\aistore[x \leftarrow\ell]$ is an injective extension of
    $\aistore$ and $\ell \in \loc(\aheap)$, by Lemma \ref{lemma:rem},
    $\coreabs{\aistore,\aheap}{\aprob} = \rem{x}{\coreabs{\aistore[x
          \leftarrow \ell],\aheap}{\aprob}}$ and $(\heapex x' ~.~
    \phi_1',~ F) \in \profile_\aprob$ follows, by the inductive
    hypothesis. \qed
  \end{compactitem}}

\longVersionOnly{
\subsection{Construction of the Composition Operation}
\label{sec:composition}
}

\longVersionOnly{
As stated by Definition \ref{def:coresep}, a composition operation
combines the core abstractions of two injective structures with
disjoint heaps into a set of core formul{\ae} that is the actual core
abstraction of the disjoint union of the two structures. Since there
are infinitely many structures with the same core abstraction, this
set cannot be computed by enumerating the models of its operands and
computing the core abstraction of their compositions. For this reason,
the construction works symbolically on core formul{\ae}, by saturating
the separating conjunction of two core formul{\ae} via a \emph{modus
  ponens}-style consequence operator. 
}

\shortVersionOnly{The composition operation $\coresep{D}$ works
  symbolically on core formul{\ae}, by saturating the separating
  conjunction of two core formul{\ae} via a \emph{modus ponens}-style
  consequence operator. }
  
\begin{definition}\label{def:conseq}
  Given formul{\ae} $\phi, \psi$, we write $\phi \conseq \psi$ if
  $\phi = \varphi * [\alpha \swand p(\vec{t})] * [(\beta * p(\vec{t}))
    \swand q(\vec{u})]$ and $\psi = \varphi * [(\alpha * \beta) \swand
    q(\vec{u})]$ (up to the commutativity of $*$ and the neutrality of
  $\emp$) for some formula $\varphi$, predicate atoms $p(\vec{t})$ and
  $q(\vec{u)}$ and conjunctions of predicate atoms $\alpha$ and
  $\beta$.
\end{definition}

\begin{example}
Consider the structure $(\astore,\aheap)$ and the rules of Example
\ref{ex:contexts}.  We have $\aheap = \aheap_1 \uplus \aheap_2$, with
$(\astore[y \leftarrow \ell_3],\aheap_1) \models_{\coreset{\asys}}
q(y) \swand p(x)$ and $(\astore[y \leftarrow \ell_3],\aheap_2)
\models_{\asys} q(y)$, i.e., $(\astore[y \leftarrow \ell_3],\aheap_2)
\models_{\coreset{\asys}} \emp \swand q(y)$, thus $(\astore[y
  \leftarrow \ell_3],\aheap) \models_{\coreset{\asys}} q(y) \swand
p(x) * \emp \swand q(y) \conseq \emp \swand p(x)$. \hfill$\blacksquare$
\end{example}

\putInAppendix{

We prove below that $\conseq$ is a logical consequence relation:

\begin{lemma}\label{lemma:conseq}
  If $\phi \conseq^* \psi$ then $\phi \models_{\coreset{\asys}} \psi$.
\end{lemma}
\proof{ The proof is by induction on the length $n \geq 0$ of the derivation
  sequence from $\phi$ to $\psi$. If $n=0$ then $\phi=\psi$ and there
  is nothing to prove. Assume $n = 1$, the case $n > 1$ follows
  immediately by the inductive
  hypothesis. We assume that 
  $\phi = [\alpha \swand p(\vec{t})]
  * [(\beta * p(\vec{t})) \swand q(\vec{u})]$ and $\psi = (\alpha *
    \beta) \swand q(\vec{u})$, for some predicate atoms $p(\vec{t})$
  and $q(\vec{u})$ and some possibly empty conjunctions of predicate
  atoms $\alpha$ and $\beta$. Then there exist two disjoint heaps
  $\aheap_1$ and $\aheap_2$, such that $\aheap = \aheap_1 \uplus
  \aheap_2$, $(\astore,\aheap_1) \models_{\coreset{\asys}} \alpha
  \swand p(\vec{t})$ and $(\astore,\aheap_2) \models_{\coreset{\asys}}
  (\beta * p(\vec{t})) \swand q(\vec{u})$. We prove that $(\astore, \aheap) \models \psi$  by induction on
  $\card{\aheap_2}$. If $\card{\aheap_2}=0$ then $\beta = \emp$ and,
  by Lemma \ref{lemma:context-emp}, we obtain $p=q$ and
  $\astore(\vec{t}) = \astore(\vec{u})$.  Thus $\aheap = \aheap_1$ and
  $(\astore,\aheap) \models (\alpha * \beta) \swand q(\vec{u})$ follows
  trivially. If $\card{\aheap_2} > 0$, then there exists a rule 
  \begin{equation}\label{eq:coreset-rule}
  (\delta * p(\vec{x})) \swand q(\vec{y}) \Leftarrow_{\coreset{\asys}}
  \rho
  \end{equation} 
  and a substitution $\tau$ such that $[(\delta * p(\vec{x})) \swand q(\vec{y})]\tau = (\beta * p(\vec{t})) \swand q(\vec{u})$ and $(\astore', \aheap_2) \models \rho\tau$, where $\astore'$ is an associate of $\astore$. Since $\card{\aheap_2} > 0$, by definition of $\coreset{\asys}$,  rule (\ref{eq:coreset-rule}) must be an instance of  (\ref{infrule:notemp}). Thus $\rho$ is of the form $\exists \vec{v} ~.~ \psi'\sigma * \Asterisk_{j=1}^m (\gamma_j \swand
  p_j(\sigma(\vec{w}_j)))$ for some substitution $\sigma$, where $\gamma_1,
  \ldots, \gamma_m$ are separating conjunctions of predicate atoms such that $\delta * p(\vec{x}) = \Asterisk_{j=1}^m \gamma_j$. Still because (\ref{eq:coreset-rule}) is an instance of (\ref{infrule:notemp}), there exists a rule
    \begin{equation}\label{eq:asys-rule}
  q(\vec{y}) \Leftarrow_\asys \exists \vec{z} ~.~ \psi' * \Asterisk_{j=1}^m p_j(\vec{w}_j)
  \end{equation}
  and we have $\vec{v} = \vec{z} \setminus \dom(\sigma)$. 
  
  Since $(\astore,\aheap_2)
  \models_{\coreset{\asys}} \exists \vec{v} ~.~ \psi'\sigma\tau *
  \Asterisk_{j=1}^m \gamma_j\tau \swand
  p_j(\tau(\sigma(\vec{w}_j)))$, there exists a
  $\vec{v}$-associate $\aastore$ of $\astore$ such that
  $(\aastore,\aheap_2) \models_{\coreset{\asys}} \psi'\sigma\tau *
  \Asterisk_{j=1}^m \gamma_j\tau \swand  p_j(\tau(\sigma(\vec{w}_j)))$. Hence, there exist two disjoint heaps
  $\aheap'_2$ and $\aheap''_2$ such that $\aheap_2 = \aheap'_2 \uplus
  \aheap''_2$, $(\aastore,\aheap'_2) \models \psi'\sigma\tau$ and
  $(\aastore,\aheap''_2) \models_{\coreset{\asys}} \Asterisk_{j=1}^m
  \gamma_j\tau \swand p_j(\tau(\sigma(\vec{w}_j)))$. We deduce that
  \[(\aastore,\aheap_1\uplus\aheap''_2) \models_{\coreset{\asys}} 
  [\alpha \swand p(\vec{t})] * [\Asterisk_{j=1}^m \gamma_j\tau
  \swand p_j(\tau(\sigma(\vec{w}_j)))].\] Since $\delta *
  p(\vec{x}) = \Asterisk_{j=1}^m \gamma_j$, we can assume
  w.l.o.g. that $\gamma_1$ is of the form $p(\vec{x}) * \delta'$, so that
  $\gamma_1\tau = p(\vec{t}) * \delta'\tau$ and
  \[(\aastore,\aheap_1\uplus\aheap''_2) \models_{\coreset{\asys}} 
  [\alpha \swand p(\vec{t})] * [p(\vec{t}) * \delta'\tau \swand
  p_1(\tau(\sigma(\vec{w}_1)))] * [\Asterisk_{j=2}^m \gamma_j\tau
  \swand p_j(\tau(\sigma(\vec{z}_j)))].\] 
  There therefore exist two
  disjoint heaps $\aheap_3$ and $\aheap_4$ such that
  $\aheap_1\uplus\aheap''_2 = \aheap_3\uplus\aheap_4$ and the
  following hold:
  \[\begin{array}{rcl}
  (\aastore,\aheap_3) & \models_{\coreset{\asys}} & [\alpha \swand
  p(\vec{t})] * [p(\vec{t}) * \delta'\tau \swand p_1(\tau(\sigma(\vec{z}_1)))], \\
  (\aastore,\aheap_4) & \models_{\coreset{\asys}} & \Asterisk_{j=2}^m \gamma_j\tau
  \swand p_j(\tau(\sigma(\vec{w}_j))).
  \end{array}\]
   Because $\asys$ is assumed to be progressing, $\psi'$ contains
  exactly one points-to atom, thus $\card{\aheap'_2} = 1$ and
  $\card{\aheap_3} \leq \card{\aheap_1} + \card{\aheap''_2} <
  \card{\aheap_1} + \card{\aheap_2} = \card{\aheap}$. By the
  inductive hypothesis, we deduce that
  $(\aastore,\aheap_3) \models_{\coreset{\asys}} \alpha*\delta'\tau
  \swand p_1(\tau(\sigma(\vec{w}_1)))$. Putting it all together, we obtain
  \begin{eqnarray*}
  	(\aastore,\aheap)& \models_{\coreset{\asys}}& \psi'\sigma\tau * 
  	[\alpha*\delta'\tau \swand p_1(\tau(\sigma(\vec{w}_1)))] *
  	[\Asterisk_{j=2}^m \gamma_j\tau \swand
  	p_j(\tau(\sigma(\vec{w}_j)))], \text{ hence}\\
  	(\astore,\aheap) &\models_{\coreset{\asys}}& \exists \vec{v} ~.~ \psi'\sigma\tau * 
  	[\alpha*\delta'\tau \swand p_1(\tau(\sigma(\vec{w}_1)))] *
  	[\Asterisk_{j=2}^m \gamma_j\tau \swand
  	p_j(\tau(\sigma(\vec{w}_j)))].
  \end{eqnarray*}
  Since $\delta =
  \delta'*\Asterisk_{j=2}^m \gamma_j$, rule
  (\ref{eq:asys-rule}) implies the existence of the following rule that
  is an instance of (\ref{infrule:notemp}):
  \[(\eta * \delta) \swand q(\vec{y}) \Leftarrow_{\coreset{\asys}} \exists \vec{v} ~.~ 
  \psi'\sigma * [\eta * \delta' \swand p_1(\sigma(\vec{w}_1))] *
  [\Asterisk_{j=2}^m \gamma_j \swand
  p_j(\sigma(\vec{w}_j))],\] where $\eta$ is a
  separating conjunction of predicate atoms, such that
  $\eta\tau = \alpha$. Thus we obtain $(\astore,\aheap)
  \models_{\coreset{\asys}} (\eta\tau * \delta\tau)
  \swand q(\tau(\vec{y}))$ and $(\astore,\aheap)
  \models_{\coreset{\asys}} \alpha * \beta \swand
  q(\vec{u})$ follows. \qed
}
}

\longVersionOnly{ The composition operator is defined as follows.}
\shortVersionOnly{\noindent} We define a relation on the set of core
formul{\ae} $\core{\aprob}$, parameterized by a set $D \subseteq
\allvars{\aprob}^1 \cup \const$:
\begin{eqnarray}
&&\heapex \vec{x}_1 \nheapall \vec{y}_1 ~.~ \psi_1, \heapex \vec{x}_2
\nheapall \vec{y}_2 ~.~ \psi_2 \deriv{D} \heapex \vec{x} \nheapall
\vec{y} ~.~ \psi \label{eq:deriv} \\[-1mm]
&&\text{if } \psi_1 * \psi_2
\conseq^* \psi, \vec{x}_1 \cap \vec{x}_2 = \emptyset, \vec{x} =
(\vec{x}_1 \cup \vec{x}_2) \cap \fv{\psi}, 
\vec{y} = ((\vec{y}_1 \cup \vec{y}_2) \cap \fv{\psi}) \setminus \vec{x}, 
\lroots{\psi} \cap D = \emptyset. \nonumber
\end{eqnarray}
\vspace*{-.5\baselineskip}
The composition operator is defined by lifting the $\deriv{}$ relation to sets
of core formul{\ae}:
\begin{equation}\label{eq:coresep}
F_1 \coresep{D} F_2 \isdef \set{\psi \mid \phi_1 \in F_1, \phi_2 \in F_2, \phi_1,\phi_2 \deriv{D} \psi}
\end{equation}
\noindent We show that $\coresep{D}$ is indeed a
composition, in the sense of Definition \ref{def:coresep}:
\begin{lemma}\label{lemma:coresep}
  Let $\asys$ be a normalized, progressing, connected and \restricted
  set of rules, $D \subseteq \allvars{\aprob}^1 \cup \const$ be a set
  of terms and $(\aistore, \aheap)$ be an injective structure, with
  $\dom(\aistore) \subseteq \allvars{\aprob}^1 \cup \const$. Let
  $\aheap_1$ and $\aheap_2$ be two disjoint heaps, such
  that:\begin{inparaenum}[(1)]
  \item\label{it1:coresep} $\aheap = \aheap_1 \uplus \aheap_2$,
  \item\label{it2:coresep} $\front(\aheap_1, \aheap_2) \subseteq
    \aistore(\allvars{\aprob}^1 \cup \const)$ and 
  \item\label{it3:coresep} $\front(\aheap_1, \aheap_2) \cap
    \dom(\aheap) \subseteq \aistore(D) \subseteq
    \dom(\aheap)$.
  \end{inparaenum} 
  Then, we have $\coreabs{\aistore,\aheap}{\aprob} =
  \coreabs{\aistore,\aheap_1}{\aprob} \coresep{D}
  \coreabs{\aistore,\aheap_2}{\aprob}$.
\end{lemma}
\optionalProof{Lemma \ref{lemma:coresep}}{sec:coreabs}{ 
  ``$\subseteq$'' Let $\psi \in
  \coreabs{\aistore,\aheap}{\aprob}$ be a core formula. By equation
  (\ref{eq:coresep}), it is sufficient to show the existence of core
  formul{\ae} $\psi_i \in \coreabs{\aistore,\aheap_i}{\aprob}$, for
  $i=1,2$, such that $\psi_1 * \psi_2 \deriv{D} \psi$. 

  \noindent ({\bf A}) First, we proceed under the following
  assumptions: \begin{compactenum}
  \item $\psi$ is quantifier-free thus, by Definition
    \ref{def:core-formulae}, it is of the form: \[\psi =
    \Asterisk_{i=1}^n \underbrace{\left(\Asterisk_{j=1}^{k_i}
      q_j^i(\vec{u}^i_j) \swand p_i(\vec{t}_i)\right)}_{\isdef\lambda_{i \in
        \interv{1}{n}}} * \Asterisk_{i=n+1}^m \underbrace{x_i \mapsto
      (t^i_1,\ldots,t^i_\rank)}_{\isdef\lambda_{i \in \interv{n+1}{m}}},
    \text{ for some } 0 \leq n \leq m,\] 
  \item $\aistore$ is bijective, i.e.\ $\img(\aistore) = \locs$;
  \item $(\aistore,\aheap)
  \models_{\coreset{\asys}} \psi$ and $\aistore(\lroots{\psi}) \cap
  \dom(\aheap) = \emptyset$ ($\dagger$)
  \end{compactenum}
  We show the existence of two quantifier-free core formul{\ae}
  $\psi_1,\psi_2$ with $\psi_1, \psi_2 \deriv{D} \psi$, $\aistore \in
  \witset{\asys}{\aistore}{\aheap_i}{\psi_i}$ and $\roots{\psi_i}
  \subseteq \allvars{\aprob}^1 \cup \const \cup \roots{\psi}$, for $i
  = 1,2$. By definition, there exist $m$ disjoint heaps $\aheap'_1,
  \ldots, \aheap'_m$, such that $\aheap = \aheap_1 \uplus \aheap_2 =
  \biguplus_{i=1}^m \aheap'_i$ and $(\aistore,\aheap'_i)
  \models_{\coreset{\asys}} \lambda_i$, for all $i \in \interv{1}{m}$.
  First, we prove that: \[\lroots{\psi} \cap D = \emptyset.
  ~(\dagger\dagger)\] Suppose, for a contradiction, that there exists
  a variable $x \in \lroots{\psi} \cap D$. Then $\aistore(x) \in
  \aistore(\lroots{\psi})$, leading to $\aistore(x) \not\in
  \dom(\aheap)$, by ($\dagger$). But we also have $\aistore(x) \in
  \aistore(D)$, hence $\aistore(D) \not\subseteq \dom(\aheap)$, which
  contradicts the hypothesis $\aistore(D) \subseteq
  \aistore(\dom(\aheap)$ from the statement of the Lemma. Second, we
  build $\psi_1$ and $\psi_2$, distinguishing the following cases:

  \noindent ({\bf A.1}) If for all $i\in \interv{1}{m}$, either
  $\aheap'_i \subseteq \aheap_1$ or $\aheap'_i \subseteq \aheap_2$,
  then we let $\psi_i \isdef \Asterisk\{\lambda_j \mid j \in
  \interv{1}{m},~ \aheap'_j \subseteq \aheap_i\}$, for $i = 1,2$ (note
  that we may have $\psi_i = \emp$, if $\aheap_i$ is empty). It is
  clear that the formula $\psi$ can be written in the form $\psi_1 *
  \psi_2$, up the commutativity of $*$ and neutrality of $\emp$ for
  $*$.  Since $\lroots{\psi} \cap D = \emptyset$ by
  ($\dagger\dagger$), we deduce that $\psi_1, \psi_2 \deriv{D} \psi$
  (\ref{eq:deriv}) trivially, since $\psi = \psi_1 * \psi_2$.
   
  \noindent ({\bf A.2}) Otherwise, there exists $i \in \interv{1}{m}$
  such that $\aheap'_i \not\subseteq \aheap_1$ and $\aheap'_i
  \not\subseteq \aheap_2$. Thus, necessarily, $\card{\aheap'_i} >
  1$. Furthermore, since $\card{\aheap'_j} = 1$ for all $j \in
  \interv{n+1}{m}$, we must have $i \in \interv{1}{n}$. For the sake
  of readability we drop all references to $i$ and write $\lambda_i =
  \Asterisk_{j=1}^{k} q_j(\vec{u}_j) \swand p(\vec{t})$ instead of
  $\lambda_i = \Asterisk_{j=1}^{k_i} q^i_j(\vec{u}^i_j) \swand
  p_i(\vec{t}_i)$. Since $\aistore$ is bijective by assumption, by
  Lemma \ref{lemma:bijective-unfolding-sat}, there exists a core
  unfolding $\lambda_i \sunfold{\coreset{\asys}} \varphi_i$, such that
  $(\aistore,\aheap) \models_{\coreset{\asys}} \varphi_i$. Because
  $\card{\aheap'_i}>1$, entails that $\varphi_i\neq \emp$, the rule
  used to obtain this core unfolding (see Definition
  \ref{def:simple-formulae}) must have been generated by inference
  rule (\ref{infrule:notemp}). Since $\asys$ is progressing, we deduce
  that $\varphi_i$ is of the form $t_0 \mapsto (t_1, \ldots, t_\rank)
  * \Asterisk_{j=1}^{k'} (\gamma_j \swand p'_j(\vec{t}_j))$, for some
  separating conjunctions of predicate atoms $\gamma_1, \ldots,
  \gamma_{k'}$ such that $\Asterisk_{j=1}^{k'} \gamma_j =
  \Asterisk_{j=1}^k q_j(\vec{u}_j)$, and that $t_0 =
  \aroot(p(\vec{t}))$.  Then $\aistore(t_0) \in \dom(\aheap'_i)
  \subseteq \dom(\aheap)$ and assume that $\aistore(t_0) \in
  \dom(\aheap_1)$ (the case $\aistore(t_0) \in \dom(\aheap_2)$ is
  symmetric). We construct a sequence of formul{\ae} by applying the
  same process to each occurrence of a subformula of the form $\alpha'
  \swand p'(\vec{t}')$ such that $\aistore(\aroot(p'(\vec{t}'))) \in
  \dom(\aheap_1)$, leading to \(\Asterisk_{j=1}^k q_j(\vec{u}_j)
  \swand p(\vec{t}) \sunfold{\coreset{\asys}}^* \alpha *
  \Asterisk_{j=1}^h \delta_j \swand r_j(\vec{v}_j)\),
  where:\begin{compactitem}
  \item $\alpha$ is a separating conjunction of points-to atoms,
  \item $\delta_1, \ldots, \delta_h$ are separating conjunctions of
    predicate atoms, such that $\Asterisk_{j=1}^h \delta_j =
    \Asterisk_{j=1}^k q_j(\vec{u}_j)$,
  \item $(\aistore,\aheap'_i) \models_{\coreset{\asys}} \alpha *
    \Asterisk_{j=1}^h \delta_j \swand r_j(\vec{v}_j)$,
  \item $\aistore(\aroot(r_j(\vec{v}_j))) \in \dom(\aheap_2)$, for all
    $j \in \interv{1}{h}$.
  \end{compactitem}
  Let $\lambda_{i,1}^1 \isdef \Asterisk_{j=1}^h r_j(\vec{v}_j) \swand
  p(\vec{t})$. By definition $\aheap'_i = \aheap^1_{i,1} \uplus
  \aheap'_{i,1}$, with $(\aistore,\aheap^1_{i,1})
  \models_{\coreset{\asys}} \alpha$ and $(\aistore,\aheap'_{i,1})
  \models_{\coreset{\asys}} \Asterisk_{j=1}^h \delta_j \swand
  r_j(\vec{v}_j)$.  Note that by construction $\aheap^1_{i,1}
  \subseteq \aheap^1$ (but we do not necessarily have $\aheap'_{i,1}
  \subseteq \aheap_2$).  Furthermore, it is easy to check that $\alpha
  \models_{\coreset{\asys}} \lambda_{i,1}^1$ (indeed, by construction,
  $\alpha$ is obtained by starting from $p(\vec{t})$ and repeatedly
  unfolding all atoms not occurring in $\Asterisk_{j=1}^h
  r_j(\vec{v}_j)$), hence $(\aistore,\aheap^1_{i,1})
  \models_{\coreset{\asys}} \lambda_{i,1}^1$.  By Definition
  \ref{def:conseq}, we have $\lambda_{i,1}^1 * \left(\Asterisk_{j=1}^h
  \delta_j \swand r_j(\vec{v}_j)\right) \conseq^* \lambda_i$. We now
  prove that: \[\aroot(r_j(\vec{v}_j)) \in \allvars{\aprob}^1 \cup
  \const, \text{ for each } j \in \interv{1}{h}. ~~ (\star) \enspace\]
  Since $(\aistore,\aheap^1_{i,1}) \models_{\coreset{\asys}}
  \lambda_{i,1}^1$, by Lemma \ref{lemma:lhs-root}, we have
  $\aistore(\aroot(r_j(\vec{v}_j))) \in \loc(\aheap_1) \cup
  \aistore(\const)$. If $\aistore(\aroot(r_j(\vec{v}_j))) \in
  \aistore(\const)$, we obtain $\aroot(r_j(\vec{v}_j)) \in \const$ by
  injectivity of $\aistore$. Otherwise
  $\aistore(\aroot(r_j(\vec{v}_j))) \in \loc(\aheap_1)$, and since
  $\aistore(\aroot(r_j(\vec{v}_j))) \in \dom(\aheap_2) \subseteq
  \loc(\aheap_2)$ by construction, we obtain
  $\aistore(\aroot(r_j(\vec{v}_j))) \in \front(\aheap_1,\aheap_2)
  \subseteq \aistore(\allvars{\aprob}^1 \cup \const)$, by hypothesis
  (\ref{it3:coresep}) of the Lemma. Since $\aistore$ is injective, we
  deduce that $\aroot(r_j(\vec{v}_j)) \in \allvars{\aprob}^1 \cup
  \const$.
  
 We repeat the entire process until we get a formula that satisfies
 Condition ({\bf A.1}). Note that the unfolding terminates because at
 each step we increase the number $h$ of separating conjunctions
 $\delta_1, \ldots, \delta_h$ and $\Asterisk_{j=1}^h \delta_j =
 \Asterisk_{j=1}^k q_j(\vec{u}_j)$, where $k \geq h$ is fixed. If we
 denote by $s$ the number of unfolding steps, and by $\psi(i)$
 the formula obtained after step
 $i$, we eventually obtain a sequence of formul{\ae} $\psi(s)
 \conseq^* \dots \conseq^* \psi(0) = \psi$ where $\psi(s)$ satisfies
 Condition ({\bf A.1}), and $(\aistore,\aheap) \models \psi(i)$, for
 all $i = 0,\ldots,s$.  By Point ({\bf A.1}), we therefore obtain
 formul{\ae} $\psi_j$ such that $(\aistore,\aheap_j)
 \models_{\coreset{\asys}} \psi_j$, for $j=1,2$ and $\psi_1 * \psi_2
 \conseq^* \psi$, which, by ($\dagger\dagger$), leads to $\psi_1,
 \psi_2 \deriv{D} \psi$ (\ref{eq:deriv}).
  
 We prove that $\aistore(\lroots{\psi_i}) \cap \dom(\aheap_i) =
 \emptyset$, for $i = 1,2$. Let $i=1$ and $x \in \lroots{\psi_1}$ (the
 proof is identical for the case $i=2$). If $x \in \lroots{\psi}$ then
 $\aistore(x) \not\in \dom(\aheap)$, by ($\dagger$). Otherwise, $x
 \not\in \lroots{\psi}$ was introduced during the unfolding, hence
 $\aistore(x) \in \dom(\aheap_2)$, by the construction of $\psi_1$. In
 both cases, we have $\aaistore(x) \not\in \dom(\aheap_1)$. Since
 $(\aistore,\aheap_1) \models_{\coreset{\asys}} \psi_1$ and $\psi_1$
 is quantifier-free, by construction, we have $\aistore \in
 \witset{\asys}{\aistore}{\aheap_1}{\psi_1}$, thus $\psi_1 \in
 \coreabs{\aistore,\aheap_1}{\aprob}$, as required.

 Next, we show that for $i= 1, 2$, each root in $\psi_i$ is contained
 in $\allvars{\aprob}^1 \cup \const \cup \roots{\psi}$, and that it
 occurs with multiplicity one. We give the proof when $i = 1$, the
 proof for $i=2$ is symmetric.  First, each $x \in \roots{\psi_1}$ is
 either a root of $\psi$ or it is introduced by the unfoldings
 described above. In the second case we have $x \in \allvars{\aprob}^1
 \cup \const$ by ($\star$). Second, we show that all variables from
 $\roots{\psi_1}$ occur with multiplicity one.  Suppose, for a
 contradiction, that $x$ occurs twice as a root in $\psi_1$.  If both
 occurrences of $x$ are in points-to atoms $x \mapsto (t_1, \ldots,
 t_\rank)$ or in a predicate atom $\delta \swand p(\vec{t})$ with $x =
 \aroot(p(\vec{t}))$, then since all atoms are conjoined by separating
 conjunctions, $\phi_1$ is unsatisfiable, which contradicts the fact
 that $(\aistore,\aheap_1) \models_{\coreset{\asys}} \psi_1$.  If one
 occurrence of $x$ occurs in $\lroots{\psi_1}$ then we have shown that
 $\aistore(x) \not\in \dom(\aheap_1)$, thus the other occurrence of
 $x$ cannot occur in $\rroots{\psi_1}$, which entails that it also
 occurs in $\lroots{\psi_1}$.  Finally, assume that both occurrences
 of $x$ occur in $\lroots{\psi_1}$. Because $\psi \in \core{\aprob}$,
 it must be the case that at least one occurrence of $x$ was
 introduced during the unfolding. This entails that
 $\aistore(x)\in\dom(\aheap)$ thus $x$ cannot occur in
 $\lroots{\psi}$, because $\psi \in \coreabs{\aistore,\aheap}{\aprob}$
 (Definition \ref{def:coreabs}), hence both occurrences of $x$ have
 been introduced during the unfolding. But each time a variable $x$ is
 introduced in $\lroots{\psi_1}$, there is another occurrence of the
 same variable $x$ that is introduced in $\rroots{\psi_2}$, hence
 $\psi_2$ is unsatisfiable, which contradicts the fact that
 $(\aistore,\aheap_2) \models_{\coreset{\asys}} \psi_2$.

 \noindent({\bf B}) Let $\psi = \heapex \vec{x} \nheapall \vec{y}
 \,.\, \varphi$, where $\varphi$ is a quantifier-free core formula in
 $\core{\aprob}$ and let $\aistore$ be an injective store.  Note that,
 since $\psi\in \core{\aprob}$ and $\dom(\aistore) \subseteq
 \allvars{\aprob}^1$, we have $(\vec{x} \cup \vec{y}) \cap
 \dom(\aistore) = \emptyset$. Because $\psi \in
 \coreabs{\aistore,\aheap}{\aprob}$, by Definition \ref{def:coreabs},
 there exists a witness $\aaistore \in
 \witset{\asys}{\aistore}{\aheap}{\psi}$, satisfying the three points
 of Definition \ref{def:coreabs}, and such that: 
  \[\aaistore(\lroots{\psi}) \cap \dom(\aheap) = \emptyset.~(\ddagger)\]
  Note that $\aaistore$ is injective by Definition \ref{def:coreabs},
  and we can assume w.l.o.g. that it is bijective. 

  To this aim, we consider any bijection $\ell \mapsto x_{\ell}$
  between $\locs \setminus \img(\aaistore)$ and $\vars \setminus
  \dom(\aaistore)$. Such a bijection exists because both $\locs
  \setminus \img(\aaistore)$ and $\vars \setminus \dom(\aaistore)$ are
  infinitely countable. Let $\aaistorep$ be the extension of
  $\aaistore$ with the set of pairs $\set{(x_\ell, \ell) \mid \ell
    \mapsto x_\ell}$. It is easy to check that $\aaistorep$ is
  bijective.

  Since $(\aaistore,\aheap) \models_{\coreset{\asys}} \varphi$ by
  point \ref{it1:witset} of Definition \ref{def:coreabs} and $\varphi$
  is quantifier-free, we have $\aaistore \in
  \witset{\asys}{\aaistore}{\aheap}{\varphi}$, hence $\varphi \in
  \coreabs{\aaistore,\aheap}{\aprob}$, because $\lroots{\varphi} =
  \lroots{\psi}$ and $\aaistore(\lroots{\varphi}) \cap \dom(\aheap) =
  \emptyset$ follows from ($\ddagger$). By case ({\bf A}), there exist
  quantifier-free core formul{\ae} $\varphi_1, \varphi_2$, such that
  $\varphi_1, \varphi_2 \deriv{D} \varphi$, $\aaistore \in
  \witset{\asys}{\aistore}{\aheap_i}{\varphi_i}$ and
  $\roots{\varphi_i} \subseteq \allvars{\aprob}^1 \cup \const \cup
  \roots{\varphi}$, for $i = 1,2$. Let $\aistore_i$ be the restriction
  of $\aaistore$ to $\fv{\varphi_i} \cup \const$ and define the
  following sets, for $i=1,2$:
  \[\vec{x}_i \isdef \set{x \in \dom(\aaistore) \setminus \dom(\aistore)\mid 
    \aaistore(x) \in \loc(\aheap_i)} \hspace*{0.5cm} \vec{y}_i \isdef
  \set{x \in \dom(\aistore_i) \setminus \dom(\aistore) \mid
    \aaistore(x) \not\in \loc(\aheap_i)}\] Note that we do not know at
  this point whether $\vec{x}_i \subseteq \dom(\aistore_i)$ (this will be
  established later), while $\vec{y}_i \subseteq \dom(\aistore_i)$
  holds by definition.
    
  We prove that for all variables $x \in \vec{x}_i$, there exists a
  subformula $\delta$ occurring in $\varphi_i$ such that $x \in
  \fv{\delta}$, and either $\delta$ is a points-to atom or $\delta =
  \alpha \swand \beta$ with $x \in \fv{\beta} \setminus
  \fv{\alpha}$. To this aim, we begin by proving that if some formula
  $\varphi'$ is obtained from the initial formula $\varphi$ by a
  sequence of unfoldings as defined in Part ({\bf A}) and if $x \in
  \fv{\varphi'}$, then $\varphi'$ contains a formula of the form
  above. The proof is by induction on the length of the
  unfolding: \begin{compactitem}
  \item{If $\varphi = \varphi'$, then by the hypothesis $x \not \in \dom(\aistore)$ and
    $x \in \fv{\varphi}$, thus $x\in \vec{x} \cup \vec{y}$. Since
    $\aistore_i(x) \in \loc(\aheap_i) \subseteq \loc(\aheap)$, we have
    $\aaistore(x) \in \loc(\aheap)$, hence by Condition
    (\ref{it3:witset}) of Definition \ref{def:coreabs}, necessarily $x
    \in \vec{x}$. Then the proof follows immediately from Condition
    (\ref{core-formulae:exists}) in Definition
    \ref{def:core-formulae}.}
  \item{Otherwise, according to the construction above, $\varphi'$ is
    obtained from an unfolding $\varphi''$ of $\varphi$, by replacing
    some formula $\lambda_i = \Asterisk_{j=1}^{k_i} q_j^i(\vec{u}^i_j)
    \swand p_i(\vec{t}_i)$ in $\varphi''$ by $\lambda_{i,1}^1 *
    \left(\Asterisk_{j=1}^h (\delta_j \swand r_j(\vec{v}_j))\right)$,
    with $\lambda_{i,1}^1 = \Asterisk_{j=1}^h \left(r_j(\vec{v}_j)
    \swand p_i(\vec{t}_i)\right)$, and all atoms in $\delta_j$ occur in
    $\Asterisk_{j=1}^{k_i} q_j^i(\vec{u}^i_j)$. \begin{compactitem}
      \item{ If $x$ occurs in $\varphi''$, then by the induction
        hypothesis $\varphi''$ contains a formula $\delta$ satisfying
        the condition above.  If $\delta$ is distinct from $\lambda_i$
        then $\delta$ occurs in $\varphi'$ and the proof is completed.
        Otherwise, we have $\delta = \alpha \swand \beta$ with $\beta
        = p_i(\vec{t}_i)$, $\alpha = \Asterisk_{j=1}^{k_i}
        q_j^i(\vec{u}^i_j)$ and $x \in \fv{\beta} \setminus
        \fv{\alpha}$.  We distinguish two cases: If $x \in
        \fv{r_j(\vec{v}_j)}$, for some $j \in \interv{1}{h}$, then $x
        \in \fv{r_j(\vec{v}_j)} \setminus\fv{\delta_j}$ (since $x\not
        \in \fv{\alpha}$ and $\fv{\delta_j} \subseteq \fv{\alpha}$),
        thus the formula $\Asterisk_{j=1}^h \delta_j \swand
        r_j(\vec{v}_j)$ fulfills the required property.  Otherwise,
        $x \in \fv{p_i(\vec{t}_i)} \setminus \fv{\Asterisk_{j=1}^h
          r_j(\vec{v}_j)}$ and $\lambda_{i,1}^1$ fulfills the
        property.}
      \item{ Now assume that $x$ does not occur in $\varphi''$.  This
        necessarily entails that $x \in \fv{r_j(\vec{v}_j)}$, for some
        $j \in \interv{1}{h}$, and that $x\not \in \fv{\delta_j}$,
        thus $x \in \fv{r_j(\vec{v}_j)} \setminus \fv{\delta_j}$ and
        the formula $\delta_j \swand r_j(\vec{v}_j)$ fulfills the
        required property.}
  \end{compactitem}}
  \end{compactitem}   
  We show that such a formula $\delta$ cannot occur in
  $\varphi_{3-i}$, hence necessarily occurs in $\varphi_i$, which
  entails that $x \in \dom(\aistore_i)$, and also that $\vec{x}_1 \cap
  \vec{x}_2 = \emptyset$.  This is the case because if $\delta$ occurs
  in $\varphi_{3-i}$, then there exists a subheap $\aheap_{3-i}'$ of
  $\aheap_{3-i}$ such that $(\aaistore,\aheap_{3-i}') \models \delta$.
  By Lemma \ref{lemma:varrightonly}, since $x \in \fv{\beta} \setminus
  \fv{\alpha}$ when $\delta$ is of the form $\alpha \swand\beta$, we
  have $\aaistore(x) \in \loc(\aheap_{3-i})$.  Furthermore, by
  hypothesis $x \in \vec{x}_i$, hence $\aaistore(x)\in
  \loc(\aheap_i)$. Therefore $\aaistore(x) \in \front(\aheap_1,
  \aheap_2) \subseteq \img(\aistore)$ by the hypothesis
  (\ref{it2:coresep}) of the Lemma.  Since $\aaistore$ is injective,
  this entails that $x \in \dom(\aistore)$, which contradicts the
  definition of $\vec{x}_i$.
            
  Let $\psi_i \isdef \heapex\vec{x}_i \heapex\vec{y}_i ~.~ \varphi_i$,
  for $i = 1,2$. Due to the previous property, $\psi_i$ satisfies
  Condition (\ref{core-formulae:exists}) of Definition
  \ref{def:core-formulae}. By definition of $\vec{y}_i$, we have
  $\vec{y}_i \subseteq \dom(\astore_i)$ and by definition of
  $\astore_i$, we have $\dom(\astore_i) \subseteq \fv{\varphi_i} \cup
  \const$, thus $\psi_i$ also fulfills Condition
  (\ref{core-formulae:no_useless_var}) of the same definition. By part
  ({\bf A}) $\varphi_i$ is a core formula, hence Condition
  (\ref{core-formulae:roots_are_distinct}) is satisfied, which entails
  that $\psi_i$ is a core formula.  Still by part ({\bf A}) of the
  proof, $(\aaistore,\aheap_i) \models_{\coreset{\asys}} \varphi_i$,
  thus we also have also $(\aistore_i,\aheap_i)
  \models_{\coreset{\asys}} \varphi_i$, by the definition of
  $\aistore_i$, for $i=1,2$.  By the definition of $\vec{x}_i$ and
  $\vec{y}_i$, we have $\aistore_i \in
  \witset{\asys}{\aistore}{\aheap_i}{\psi_i}$ and since
  $\aistore_i(\lroots{\varphi_i}) = \aaistore(\lroots{\varphi_i})$ and
  $\aaistore(\lroots{\varphi_i}) \cap \dom(\aheap_{i}) = \emptyset$,
  we obtain $\psi_i \in \coreabs{\aistore, \aheap_i}{\aprob}$, for
  $i=1,2$.

  Since $\varphi_1 * \varphi_2 \deriv{D} \varphi$ and
  $\varphi_1,\varphi_2$ are quantifier-free, we have, by definition of
  $\deriv{D}$:
  \[\psi_1 * \psi_2 \deriv{D}
  \heapex\vec{x}'\nheapall\vec{y}'\varphi, \text{ where } \vec{x}' =
  (\vec{x}_1 \cup \vec{x}_2) \cap \fv{\varphi} \text{ and } \vec{y}' =
  ((\vec{y}_1 \cup \vec{y}_2) \cap \fv{\varphi}) \setminus \vec{x}'.\]

  To complete the proof, it is sufficient to show that $\vec{x}' =
  \vec{x}$ and that $\vec{y} = \vec{y}'$, so that
  $\heapex\vec{x}'\nheapall\vec{y}'\varphi = \psi$. 
  \begin{compactitem}
  \item[$\vec{x}' = \vec{x}$]{``$\subseteq$'' Let $x \in \vec{x}'$. We
    have $x \in \vec{x}_i$, for some $i = 1,2$, and $x\in
    \fv{\varphi}$.  By definition of $\vec{x}_i$, this entails that $x
    \in \dom(\aaistore) \setminus \dom(\aistore)$ and that
    $\aaistore(x) \in \loc(\aheap_i) \subseteq \loc(\aheap)$. Since $x
    \in \fv{\varphi}$ and $x \in \dom(\aaistore) \setminus
    \dom(\aistore)$, necessarily $x \in \vec{x} \cup \vec{y}$, and because of
    Condition (\ref{it3:witset}) in Definition \ref{def:coreabs},
    we have $x \not \in \vec{y}$. Hence $x \in
    \vec{x}$. 
    
    ``$\supseteq$'' Let $x \in \vec{x}$.  We have $x \in
    \fv{\varphi}$ by Definition \ref{def:core-formulae}
    (\ref{core-formulae:exists}), and $\aaistore(x) \in \loc(\aheap)$
    by Definition \ref{def:coreabs} (\ref{it1:witset}), thus
    $\aaistore(x) \in \loc(\aheap_i)$, for some $i = 1,2$, so that $x
    \in \vec{x}_i$. Consequently $x \in \vec{x}'$.}
  \item[$\vec{y} = \vec{y}'$]{``$\subseteq$'' Let $y \in \vec{y}'$. By
    definition, we have $y \in \vec{y_i}$ for some $i = 1,2$, $y \in
    \fv{\varphi}$, and $y \not \in \vec{x} = \vec{x'}$.  Since $y \in
    \vec{y_i}$, we have $y \not \in \dom(\aistore)$, thus $y \not \in
    \fv{\psi}$, hence $y \in \vec{x} \cup \vec{y}$. Since $y \not \in
    \vec{x}$, we deduce that $y \in \vec{y}$. 
    
     ``$\supseteq$'' Let
    $y\in \vec{y}$. By definition, $y \not \in \dom(\aistore)$ and $y
    \not \in \vec{x}$, moreover $y \in \fv{\varphi}$, by Definition
    \ref{def:core-formulae} (\ref{core-formulae:no_useless_var}).  By
    Definition \ref{def:coreabs} (\ref{it3:witset}), we have
    $\aaistore(y) \not \in \loc(\aheap)$.  By definition of
    $\deriv{D}$, since $y \in \fv{\varphi}$, necessarily $y \in
    \fv{\varphi_i}$, for some $i=1,2$.  Since $y \not \in
    \dom(\aistore)$, we deduce that $y \in \vec{x}_i \cup
    \vec{y}_i$. Since $\aaistore(y) \not \in \loc(\aheap)$, we have
    $\aaistore(y) \not \in \loc(\aheap_i)$, hence $y \in \vec{y}_i$.
    Consequently, $y \in \vec{y}'$. }
  \end{compactitem}

  \noindent``$\supseteq$'' Let $\psi \in
  \coreabs{\aistore,\aheap_1}{\aprob} \coresep{D}
  \coreabs{\aistore,\aheap_2}{\aprob}$ be a core formula. By the
  definition of $\coresep{D}$ (\ref{eq:coresep}), there exists $\psi_i
  \in \coreabs{\aistore,\aheap_i}{\aprob}$, for $i = 1,2$, such that
  $\psi_1, \psi_2 \deriv{D} \psi$. By the definition of $\deriv{D}$
  (\ref{eq:deriv}), we have $\psi_i = \heapex \vec{x}_i \nheapall
  \vec{y}_i ~.~ \phi_i$, for $i = 1,2$, with $\vec{x} = (\vec{x}_1
  \cup \vec{x}_2) \cap \fv{\phi}$, $\vec{y} = ((\vec{y}_1 \cup
  \vec{y}_2) \cap \fv{\phi}) \setminus \vec{x}$, $\vec{x}_1 \cap
  \vec{x}_2 = \emptyset$ and $\psi = \heapex \vec{x} \nheapall \vec{y}
  ~.~ \phi$, where $\phi_1$, $\phi_2$ and $\phi$ are quantifier-free
  core formul{\ae} and $\lroots{\phi} \cap D = \emptyset$. Since
  $\psi_i \in \coreabs{\aistore,\aheap_i}{\aprob}$, by Definition
  \ref{def:coreabs}, there exist witnesses $\aaistore_i \in
  \witset{\asys}{\aistore}{\aheap_i}{\nheapall \vec{y}_i ~.~ \phi_i}$,
  such that $\aaistore_i(\vec{x}_i) \subseteq \loc(\aheap_i)$ and
  $\aaistore_i(\lroots{\phi_i}) \cap \dom(\aheap_i) = \emptyset$, for
  $i=1,2$. W.l.o.g. we can choose these witnesses such that
  $\dom(\aaistore_i) = \vec{x}_i \cup \dom(\aistore)$. Let $\aaistore$
  be any extension of $\aaistore_1 \cup \aaistore_2$ such that
  $\vec{y}_1 \cup \vec{y}_2 \subseteq \dom(\aaistore)$ and
  $\aaistore(y_1) \neq \aaistore(y_2) \not\in \loc(\aheap) \cup
  \img(\aaistore_1) \cup \img(\aaistore_2)$, for all variables $y_1
  \neq y_2 \in \vec{y}_1 \cup \vec{y}_2$. Note that such an extension
  exists, because $\locs$ is infinite and $\vec{y}_1$, $\vec{y}_2$ are
  finite. Moreover, $\aaistore$ is a well-defined store, because
  $\aaistore_1$ and $\aaistore_2$ both agree over $\dom(\aistore)$ and
  $\vec{x}_1 \cap \vec{x}_2 = \emptyset$.

  We prove that $\aaistore$ is injective. Suppose, for a
  contradiction, that $\aaistore(x_1) = \aaistore(x_2)$, for some
  variables $x_1 \neq x_2 \in \dom(\aaistore)$. By the definition of
  $\aaistore$, since $\aaistore_1$ and $\aaistore_2$ are injective,
  the only possibility is $x_i \in \dom(\aaistore_i) \setminus
  \dom(\aaistore_{3-i})$, for $i=1,2$ (hence $x_i \not \in
  \dom(\aistore)$). Then $x_i \in \vec{x}_i$ must be the case, thus
  $\aaistore_i(x_i) \in \loc(\aheap_i)$, leading to $\aaistore_1(x_1)
  \in \front(\aheap_1,\aheap_2) \subseteq \img(\aistore)$, by the
  hypothesis of the Lemma, hence $x_i \in \dom(\aistore)$, by
  injectivity of $\aistore$, which yields a contradiction.
    
  We prove next that $\aaistore \in
  \witset{\asys}{\aistore}{\aheap}{\phi}$. Since $\aaistore_i \in
  \witset{\asys}{\aistore}{\aheap_i}{\nheapall \vec{y}_i ~.~ \phi_i}$,
  we have $(\aaistore_i, \aheap_i) \models_{\coreset{\asys}} \nheapall
  \vec{y}_i ~.~ \phi_i$, for $i = 1,2$. We show that
  $\aaistore(\vec{y}_i) \cap \loc(\aheap_i) = \emptyset$ for $i =
  1,2$.  Suppose, for a contradiction, that $i=1$ and $\aaistore(x)
  \in \loc(\aheap_1)$, for some $x \in \vec{y}_1$ (the proof when
  $i=2$ is symmetric). By definition of $\aaistore$, this is possible
  only if $x \in \vec{x}_2$, and this entails that $\aaistore(x) \in
  \loc(\aheap_2)$,thus $\aaistore(x) \in \front(\aheap_1,\aheap_2)
  \subseteq \aistore(\allvars{\aprob}^1 \cup \const)$, by the
  hypothesis of the Lemma.  By the injectivity of store $\aistore$,
  this entails that $x\in \dom(\aistore)$, which contradicts the fact
  that $x \in \vec{x}_2$ (since, by definition of $\vec{x}_2$, we have
  $\vec{x}_2 \cap \dom(\aistore) = \emptyset$). Then
  $\aaistore(\vec{y}_i) \cap \loc(\aheap_i) = \emptyset$, hence,
  $(\aaistore,\aheap_i) \models_{\coreset{\asys}} \phi_i$, for $i =
  1,2$. Then $(\aaistore,\aheap) \models_{\coreset{\asys}} \phi_1 *
  \phi_2$, leading to $(\aaistore,\aheap) \models_{\coreset{\asys}}
  \phi$, by Lemma \ref{lemma:conseq}, since $\phi_1 * \phi_2 \conseq^*
  \phi$.  Moreover, $\aaistore(\vec{x}_1 \cup \vec{x}_2) =
  \aaistore_1(\vec{x}_1) \cup \aaistore_2(\vec{x}_2) \subseteq
  \loc(\aheap_1) \cup \loc(\aheap_2) = \loc(\aheap)$ and
  $\aaistore(\vec{y}_i) \cap \loc(\aheap) = \emptyset$, for $i=1,2$,
  by the definition of $\aaistore$. Then $\aaistore \in
  \witset{\asys}{\aistore}{\aheap}{\phi}$, by Definition
  \ref{def:coreabs}.

  Finally, we prove that $\aaistore(\lroots{\phi}) \cap \dom(\aheap) =
  \emptyset$. Suppose, for a contradiction, that there exists $x \in
  \lroots{\phi}$ such that $\aaistore(x) \in \dom(\aheap)$. By
  Definition \ref{def:conseq}, we have $\lroots{\phi} \subseteq
  \lroots{\phi_1} \cup \lroots{\phi_2}$ and we assume that $x \in
  \lroots{\phi_1}$ (the case $x \in \lroots{\phi_2}$ is
  symmetrical). Since $(\aaistore,\aheap_1) \models_{\coreset{\asys}}
  \phi_1$, we obtain $\aaistore(x) \in \loc(\aheap_1) \cup
  \aaistore(\const)$, by Lemma \ref{lemma:lhs-root}, and since $x
  \not\in \const$ and $\aaistore$ is injective, we obtain
  $\aaistore(x) \in \loc(\aheap_1)$. Moreover, we have $\aaistore_1(x)
  \not\in \dom(\aheap_1)$, hence $\aaistore_1(x) \in \dom(\aheap_2)
  \subseteq \loc(\aheap_2)$. Thus $\aaistore(x) \in
  \front(\aheap_1,\aheap_2) \cap \dom(\aheap) \subseteq \aaistore(D)$,
  leading to $x \in D$, by the injectivity of $\aaistore$. This
  contradicts the hypothesis $\lroots{\phi} \cap D = \emptyset$
  (\ref{eq:deriv}). We obtain that $\aaistore(\lroots{\phi}) \cap
  \dom(\aheap) = \emptyset$, thus $\phi \in
  \coreabs{\aistore,\aheap}{\aprob}$.  
\qed}

\section{Main Result}
\label{sec:complexity}

In this section, we state the main complexity result of the paper. As
a prerequisite, we prove that the size of the core formul{\ae} needed
to solve an entailment problem $\aprob$ is polynomial in
$\probwidth{\aprob}$ and the number of such formul{\ae} is simply
exponential in $\probwidth{\aprob} + \log(\size{\aprob})$.

\begin{lemma}\label{lem:card}
Given an entailment problem $\aprob$, for every formula $\phi \in
\core{\aprob}$, we have $\size{\phi} = \bigO(\probwidth{\aprob}^2)$
and $\card{\core{\aprob}} = 2^{\bigO(\probwidth{\aprob}^3 \times
  \log(\size{\aprob}))}$.
\end{lemma}
\optionalProof{Lemma \ref{lem:card}}{sec:complexity}{ Let $\phi \in \core{\aprob}$ be
  a core formula. Then $\phi$ can be viewed as a formula built over
  atoms of the form $p(\vec{t})$ and $t_0 \mapsto (t_1, \ldots,
  t_\rank)$ using the connectives $*$, $\swand$ and the quantifiers
  $\heapex$ and $\nheapall$. By Definition \ref{def:core-formulae}
  (\ref{core-formulae:roots_are_distinct}), $\phi$ contains at most
  $\card{\allvars{\aprob}}$ occurrences of such atoms.  Since, by
  points (\ref{core-formulae:no_useless_var}) and
  (\ref{core-formulae:exists}) of Definition \ref{def:core-formulae},
  all the variables in $\phi$ necessary occur in an atom, this entails
  that $\phi$ contains at most $\card{\allvars{\aprob}}\times \alpha$
  (bound or free) variables, where $\alpha = \max(\{ \#p \mid p \in
  \preds \} \cup \{ \rank+1 \})$ denotes the maximal arity of the
  relation symbols (including $\mapsto$) in $\phi$. Since each atom is
  of size at most $\alpha+1$ and since there is at most one connective
  $*$ or $\swand$ for each atom, we deduce that $\size{\phi} \leq
  2\times\card{\allvars{\aprob}}\times \alpha +
  \card{\allvars{\aprob}}\times (\alpha + 2)$. By definition, we have
  $\alpha \leq \probwidth{\aprob}$, and $\allvars{\aprob}$ is chosen
  is such a way that $\card{\allvars{\aprob}}=
  2\times\probwidth{\aprob}$, thus $\size{\phi} =
  \bigO(\probwidth{\aprob}^2)$.  The symbols that may occur in the
  formula include the set of free and bound variables, the predicate
  symbols and the symbols $\mapsto$, $*$, $\swand$ $\nheapall$,
  $\heapex$, yielding at most $(\card{\allvars{\aprob}} \times \alpha)
  + \size{\aprob} + 5 \leq \probwidth{\aprob}^2 + \size{\aprob} + 5$
  symbols.  Thus there are at most $(\probwidth{\aprob}^2 +
  \size{\aprob} + 5)^{\bigO(\probwidth{\aprob}^2)} =
  2^{\bigO(\probwidth{\aprob}^3 \times \log(\size{\aprob}))}$ core
  formul{\ae} in $\core{\aprob}$. \qed}

\begin{theorem}\label{thm:main}
Checking the validity of progressing, connected and \restricted
entailment problems is \twoexptime-complete.
\end{theorem}
\proof{ \twoexptime-hardness follows from
  \cite{DBLP:conf/lpar/EchenimIP20}; since the reduction in
  \cite{DBLP:conf/lpar/EchenimIP20} involves no (dis-)equality, the
  considered systems are trivially \restricted. We now prove
  \twoexptime-membership. Let $\aprob$ be \arestricted 
  problem.  By Lemma \ref{lemma:normalized}, we compute, in time
  $\size{\aprob} \cdot 2^{\bigO(\probwidth{\aprob}^2)}$, an equivalent
  normalized \restricted problem $\aprob_n$ of $\size{\aprob_n} =
  \size{\aprob} \times 2^{\bigO(\probwidth{\aprob}^2)}$ and
  $\probwidth{\aprob_n} = \bigO(\probwidth{\aprob}^2)$.  We fix an
  arbitrary set of variables $\allvars{\aprob_n} =
  \allvars{\aprob_n}^1 \uplus \allvars{\aprob_n}^2$ with
  $\card{\allvars{\aprob_n}^i} = \probwidth{\aprob_n}$, for $i = 1,2$
  and we compute the relation $\profile_{\aprob_n}$, using a Kleene
  iteration, as explained in Section \ref{sec:coreabs} (Lemma
  \ref{lemma:profile}). By Lemma \ref{lem:card}, if $\psi \in
  \core{\aprob_n}$ then $\size{\psi}= \bigO(\probwidth{\aprob}^2)$ and
  if $(\psi,F) \in \profile_{\aprob_n}$ then $\card{F} =
  2^{\bigO(\probwidth{\aprob_n}^3 \times \log(\size{\aprob_n}))} =
  2^{\bigO(\probwidth{\aprob}^8 \times \log(\size{\aprob}))}$, hence
  $\profile_\aprob$ can be computed in
  $2^{2^{\bigO(\probwidth{\aprob}^8 \times \log(\size{\aprob}))}}$
  steps. It thus suffices to check that each of these steps can be
  performed in polynomial time w.r.t.\ $\core{\aprob_n}$ and
  $\size{\aprob_n}$. This is straightforward for points-to atoms,
  predicate atoms and existential formul{\ae}, by iterating on the
  rules in $\aprob_n$ and applying the construction rules
  (\ref{eq:pto-core}), (\ref{eq:pred-core}) and (\ref{eq:ex-core})
  respectively. For the disjoint composition, one has to compute the
  relation $\conseq^*$, needed to build the operator $\coresep{D}$,
  according to (\ref{eq:deriv}) and (\ref{eq:coresep}). We use again a
  Kleene iteration. It is easy to check that $\phi \conseq \psi
  \Rightarrow \size{\psi} \leq \size{\phi}$, furthermore, one only
  needs to check relations of the form $\phi_1 * \phi_2 \conseq \psi$
  with $\phi_1, \phi_2, \psi \in \core{\aprob_n}$.  This entails that
  the number of iteration steps is $2^{\bigO(\probwidth{\aprob}^8
    \times \log(\size{\aprob}))}$ and, moreover, each step can be
  performed in time polynomial w.r.t.\ $\core{\aprob_n}$.  Finally, we
  apply Lemma \ref{lemma:entailment} to check that all the entailments
  in $\aprob_n$ are valid. This test can be performed in time
  polynomial w.r.t.\ $\card{\profile_{\aprob_n}}$ and
  $\size{\aprob_n}$. \qed}
  
\section{Conclusion and Future Work}

We presented a class of $\seplog$ formul{\ae} built from a set of
inductively defined predicates, used to describe pointer-linked
recursive data structures, whose entailment problem is
\twoexptime-complete. This fragment, consisting of so-called
\restricted formul\ae, is a strict generalization of previous work
defining three sufficient conditions for the decidability of
entailments between $\seplog$ formul{\ae}, namely progress,
connectivity and establishment
\cite{IosifRogalewiczSimacek13,KatelaanMathejaZuleger19,PMZ20}. On one
hand, every progressing, connected and established entailment problem
can be translated into \arestricted problem. On the other hand, the
models of \restricted formul{\ae} form a strict superset of the models
of established formul{\ae}. The proof for the \twoexptime\ upper bound
for \restricted entailments leverages from a novel technique used to
prove the upper bound of established entailments
\cite{KatelaanMathejaZuleger19,PMZ20}. A natural question is whether
the \restrictedness condition can be dropped. We conjecture that this
is not the case, and that entailment is undecidable for progressing,
connected and non-\restricted sets. Another issue is whether the
generalization of symbolic heaps to use guarded negation, magic wand
and septraction from \cite{PZ20} is possible for \restricted
entailment problems. The proof of these conjectures is on-going work.

\longVersionOnly{ Future work focuses on finding efficient ways to
  implement the algorithm in this paper, such as relationships with
  SMT solving and the application of these techniques to combinations
  of symbolic heaps with SMT-supported theories of data (integers,
  real numbers, strings, sets, etc.).  As evidenced by Example
  \ref{ex:acyclic-lists}, non-\restricted rules can sometimes be
  transformed into \restricted ones by replacing variables with
  constants (and propagating these replacements into the rules). It
  would be interesting from a practical point of view to automate this
  transformation and identify syntactic conditions ensuring that it is
  applicable.  }

\bibliography{refs}

\end{document}